\documentclass[preprint, 10pt]{elsarticle}

\newif\ifInputs
\Inputstrue

\usepackage{palatino}
\usepackage[usenames]{color}
\usepackage{epsfig}
\usepackage[fleqn,reqno]{amsmath}
\usepackage{amsfonts,amsthm,bm}
\usepackage{amssymb}
\usepackage{mathrsfs}
\usepackage{stmaryrd}
\usepackage{subfigure}
\usepackage[titletoc]{appendix}
\usepackage{tabularx,booktabs}
\usepackage{graphics,caption}
\usepackage{fancybox}
\usepackage[top=1.2in,bottom=1.2in,left=1in, right=1in]{geometry}
\usepackage{array}
\usepackage{multirow}
\usepackage{wrapfig}
\usepackage{algorithmic,algorithm}
\usepackage{paralist}
\usepackage{ifthen}
\usepackage{enumitem}
\usepackage{tikz,pgfplots,filecontents}
\usepackage{float}
\usepackage{mdframed}
\usepackage{todonotes}
\usepackage{xcolor,colortbl}

\graphicspath{{./figs/}}

\usepackage[pagebackref=false,bookmarks=false]{hyperref} 

\definecolor{dblue}{rgb}{0.03,0.3,0.62}
\definecolor{dorange}{rgb}{1,0.55,0}

\hypersetup{
  bookmarksnumbered=true,
  bookmarksopen=false,
  hypertexnames=false,     
  breaklinks=true,          
  unicode=false,
  pdffitwindow=true,        
  pdfnewwindow=true,        
  colorlinks=true,         
  linkcolor=dblue,
  anchorcolor=red,
  citecolor=dorange,
  filecolor=magenta,
  urlcolor=dblue,
  pdfstartview = FitH,
  pdfkeywords = {},
  pdfcreator = {LaTeX with hyperref package}
}

\definecolor{sblue}{cmyk}{0.98,0.13,0,0.43} 
\definecolor{sblue}{cmyk}{0.98,0.13,0,0.43} 

\newcommand{\bigO}{\mathcal{O}}

\newcommand{\figref}[1]{Figure~\ref{#1}}
\newcommand{\tabref}[1]{Table~\ref{#1}}
\newcommand{\algref}[1]{Algorithm~\ref{#1}}

\newcommand{\xx}{{\mathbf{x}}}

\newcommand{\yy}{{\mathbf{y}}}
\newcommand{\zz}{{\mathbf{z}}}
\newcommand{\ff}{{\mathbf{f}}}

\newcommand{\BB}{{\mathcal{B}}}

\newcommand{\uu}{{\mathbf{u}}}

\newcommand{\rr}{{\mathbf{r}}}
\newcommand{\nn}{{\mathbf{n}}}
\newcommand{\eeta}{{\boldsymbol\eta}}

\newcommand{\llambda}{{\boldsymbol\lambda}}
\newcommand{\grad}{{\nabla}}
\DeclareMathOperator*{\argmin}{\arg\!\min}

\newcolumntype{C}{>{\centering\arraybackslash} m{2.5cm}}

\usepackage{lineno}

\newcommand{\mcaption}[2]{\caption{\small \em #1}\label{#2}}
\newcommand{\secref}[1]{Section \ref{#1}}
\newcommand{\aref}[1]{Algorithm \ref{#1}}

\begin{document}

\title{Low-resolution simulations of vesicle suspensions in 2D}

\author[utMe]{G\"{o}kberk Kabacao\u{g}lu} \ead{gokberk@ices.utexas.edu}
\author[fsu]{Bryan Quaife} \ead{bquaife@fsu.edu}
\author[utMe,ut]{George Biros}\ead{gbiros@acm.org}
\address[utMe]{Department of Mechanical Engineering, \\The University 
of Texas at Austin, Austin, TX, 78712, United States}
\address[fsu]{Department of Scientific Computing, \\ Florida State
University, Tallahassee, FL 32306, United States}
\address[ut]{Institute for Computational Engineering and Sciences,\\
The University of Texas at Austin, Austin, TX, 78712, United States}

\begin{abstract} 
Vesicle suspensions appear in many biological and industrial applications. 
These suspensions are characterized by rich and complex dynamics of vesicles due to
their interaction with the bulk fluid, and their large deformations and nonlinear elastic
properties.   Many existing
state-of-the-art numerical schemes can resolve such complex vesicle
flows. However, even when using provably optimal algorithms, these
simulations can be computationally expensive, especially  for
suspensions with a large number of vesicles.  These high computational costs can
limit the use of simulations for parameter exploration, optimization,
or uncertainty quantification. One way to reduce the cost is to use
low-resolution discretizations in space and time.
However, it is well-known that simply reducing the resolution results
in vesicle collisions, numerical instabilities, and often in erroneous results.

In this paper, we investigate the effect of a number of algorithmic
empirical fixes (which are commonly used by many groups) in an attempt to
make low-resolution simulations more stable and more predictive. Based
on our empirical studies for a number of flow configurations, we
propose a scheme that attempts to integrate these fixes in a
systematic way.  This low-resolution scheme is an extension of our
previous work~\cite{quaife-biros14,quaife-biros16}.  Our
low-resolution correction algorithms (LRCA) include anti-aliasing and membrane reparametrization for avoiding spurious oscillations in vesicles' membranes, 
adaptive time stepping and a repulsion force for handling vesicle collisions and, 
correction of vesicles' area and arc-length for maintaining physical vesicle shapes. 
We perform a systematic error analysis by
comparing the low-resolution simulations of dilute and dense
suspensions with their high-fidelity, fully resolved, counterparts. We
observe that the LRCA enables both efficient and statistically
accurate low-resolution simulations of vesicle suspensions, while it
can be 10$\times$ to 100$\times$ faster.  
\end{abstract}


\maketitle


\section{Introduction\label{s:intro}}
Vesicle suspensions are deformable capsules filled with and submerged
in an incompressible fluid.  Their simulation plays an important role
in many biological applications~\cite{kraus-lipowsky-e96, seifert97},
such as biomembranes~\cite{sackmann96} and red blood cells (RBCs)~\cite{ghigliotti-misbah-e10, kaoui-misbah-e11, misbah06, noguchi-gompper05,pozrikidis90}.

Here we discuss the  numerical simulations of vesicle suspensions; specifically, algorithms that 
enable stable and accurate simulations at \textbf{\emph{low-resolution spatio-temporal discretization}}. 
Although many algorithmically optimal methods exist (see below), the costs remain prohibitively expensive for
large vesicle suspensions. So, the basic question we try to address in this paper is the following. What is the minimum resolution required to
recover different quantities of interest in the context of boundary integral equation methods for vesicle suspensions?

Understanding and improving low-resolution simulations will enable parametric studies and optimization (e.g., phase diagrams and design of microfluidic devices).  Also many boundary integral equation codes  use the empirical corrections  we investigate here
because convergence studies and high-resolution simulations are not possible. Further understanding these corrections and reducing the number of simulation parameters will be valuable for the community. 

In our group, we have capability for both 2D and 3D
simulations~\cite{quaife-biros14,rahimian-biros-e10}.  We have opted
to study two-dimensional Stokesian suspensions since convergence
studies in three dimensions for suspensions with a large number of
vesicles can be extremely expensive~\cite{rahimian-biros-e10}. In
addition, two dimensional simulations are valuable on their own since they can reproduce
experimentally observed flow physics in many regimes (e.g., motion of red blood cells in
microchannels~\cite{kaoui-misbah-e11,fedosov-gompper-e14}, margination
of white blood cells in blood flow~\cite{freund07,fedosov-gompper-e12,fedosov-gompper-e14}, and
sorting of rigid particles and RBCs using deterministic lateral
displacement technique~\cite{quek-chiam-e11,ye-yu-e14,zhang-fedosov-e15}).

\paragraph{Background} Vesicle flows are characterized by large deformations, local
inextensibility of a vesicle's membrane, conservation of enclosed
area due to the incompressibility of the fluid inside the vesicle, and
stiffness related to tension and bending forces.  These
features make suspensions at low resolutions a challenging
problem. In line with our previous work~\cite{shravan-biros-e09,
shravan-zorin-e11, rahimian-biros-e10, quaife-biros14,
quaife-biros15}, and work of others~\cite{freund-zhao10,
zhao-shaqfeh09, zhao-freund-e10, freund-orescanin11, zhao-shaqfeh13a,
zhao-shaqfeh13b, marple-shravan-e15, rahimian-biros-e15}, we use an
integral equation formulation for the viscous interfacial
flow~\cite{pozrikidis92}.  Our previous results for simulating
high-concentration vesicle suspensions in two
dimensions~\cite{quaife-biros14,quaife-biros15} focus on accurate
quadrature and high-order semi-implicit time stepping. The results in
those papers rely on sufficient resolution and provide a robust
framework for simulations. For example, vesicles do not collide because
all hydrodynamic interactions are resolved with spectral accuracy.
Thus, there is no need to introduce artificial repulsion forces
between vesicles. We can accurately resolve long time horizon
simulations for concentrated suspensions with roughly 96 or 128 points
per vesicle.  But in three dimensions such a resolution is
prohibitively expensive.  For example, a similar resolution using the
3D version of these algorithms~\cite{malhotra-biros-e17} would require
over 10,000 points per vesicle.  Therefore, there is a need to use some
empirical fixes to maintain stability in simulations, all the
while accurately capturing the statistics of the underlying flow using
as coarse discretization as possible.  To measure the accuracy of the
physics and statistics, we develop the algorithms in two dimensions so that we can
compare with "ground truth" simulations performed at an adequate
resolution.  Demonstrating the effectiveness of these algorithms at
low resolutions is the first step towards extending them to three
dimensions.

\paragraph*{Contributions} Low-resolution simulations of vesicle
suspensions can become unstable as a result of spurious oscillations
in vesicles' shapes due to computing nonlinear terms, non-physical
changes in vesicles' areas and arc-lengths, and vesicle collisions.
We address these issues and develop a robust method by implementing
some standard techniques and also introducing new schemes.  We calibrate
the parameters for these algorithms heuristically.  We, then,  
investigate accuracy of our low-resolution simulations compared
to the ground truth solutions. We also report the self-convergence of
the low-resolution simulations without the ground truth.
The numerical experiments  help us develop a black-box solver
that can capture underlying physics accurately using as coarse discretization as possible without having to adjust parameters
other than the spatial and temporal resolution. 

We summarize these contributions and our conclusions as follows:
\begin{itemize}
\item We introduce an efficient algorithm for determining an upsampling
rate that is sufficient for controlling the aliasing errors caused by
nonlinear terms, but not too large so that the computational costs are
not unnecessarily inflated.  Additionally, we formulate the
reparametrization algorithm in~\cite{shravan-zorin-e11} into two dimensions,
which is necessary for low-resolution stability.

\item Our previous adaptive time stepping work~\cite{quaife-biros16}
relied on asymptotic assumptions of the truncation error, which are not
valid at the low resolutions.   Since this result breaks down, we
present a new variation of this scheme that can be used at all
resolutions. 

\item A vesicle's area and arc-length are invariant in two-dimensional
vesicle simulations (their counterparts are volume and surface area in 
the three-dimensional simulations). However, at low resolutions the errors can be
extensive and hence result in unstable and non-physical flows in time
scales much shorter than the target time horizons. Therefore, we
present an efficient scheme to correct those errors without modifying
the governing equations. 

\item Near-field (lubrication like) hydrodynamic interactions cannot be resolved
accurately at low resolutions.  This leads to non-physical collisions
between vesicles.  We detect collisions with spectral
accuracy~\cite{quaife-biros14} and implement a short range repulsion
force~\cite{harmon-grinspun-e09,vouga-grinspun-e11} to keep vesicles
sufficiently separated. Unlike many other repulsion models requiring two 
parameters, our scheme is parameter-free, i.e., the repulsion length 
scale is set beforehand based on numerical experiments and the strength of the force is 
adaptive that guarantees no collision.

\item We calibrate all the parameters of the LRCA heuristically and thereby 
develop a black-box solver with a single parameter. We test the solver 
in a real-world application of a microfluidic cell sorting device.

\end{itemize}

Summary of conclusions:
\begin{itemize}
\item {\bf Corrections:} All empirical fixes (anti-aliasing, reparametrization, repulsion, adaptive time
  stepping, area-length correction) are necessary to stabilize low-resolution simulations.
  Dropping one can result in failure.

\item {\bf Parameters:} The main parameters are the spatial resolution $N$, the
  temporal resolution $\rho_{AL}$, and a time budget $T_\mathrm{comp}$ so the solver can automatically set the minimum time steps.
  Overall, the simulations are quite sensitive to time-discretization. 

\item {\bf Failure modes:} If $T_\mathrm{comp}$ is not sufficient the code will terminate early. 
  This is because the required time-step size is too small or equivalently the time per time step is 
  too large (for example, the suspensions has too many vesicles).

\item {\bf Convergence:} We don't have a way to guarantee convergence. 
  Goal-oriented error estimation requires adjoints and we don't have this capability. 
  The only way to check for convergence is to start with a coarse $N$ and $\rho_{AL}$ and 
  refine until the results do not change significantly. Notice that this is also true for the 
  fine-resolution simulations. Notice even in this scenario in which we 
  compare simulations at different resolutions, the error metric matters a lot. 
  If we're interested in convergence of individual trajectories, very refined simulations are necessary,
  especially for dense suspensions. But for error metrics that look at average quantities, 
  (e.g., effective viscosity) convergence is faster and less sensitive to the details of the simulation. 
\end{itemize}

\paragraph*{Limitations} 
One limitation is that our results are entirely empirical. In general,
there is very little work on theoretical results for general vesicles.
Indeed the only results are for vesicles that are small perturbations
of a disc and thus resemble rigid spheres.  Another limitation is that
the methods are implemented in two dimensions.  However, the algorithms
can be naturally extended to three dimensions: e.g. 
local area and length correction can be extended
to a volume and surface area correction~\cite{malhotra-biros-e17}, and
a surface reparameterization has already been implemented in three
dimensions~\cite{shravan-zorin-e11, rahimian-biros-e15,
malhotra-biros-e17}.  Another limitation is that our methods do not
allow for spatial adaptivity.  However, upsampling is utilized to avoid
aliasing that would otherwise be unavoidable at low frequencies.

Our methods allow for a viscosity contrast between the interior and
exterior of the vesicles, and several numerical examples are presented.
But the methods are not directly applicable to suspensions in which the
bulk fluid is non-Newtonian or inertial flows.

\paragraph*{Related work} 

This paper is an extension of our work for high-concentration
suspensions~\cite{quaife-biros14} and for high-order adaptive time
stepping with spectral deferred correction
(SDC)~\cite{quaife-biros16}.  That's why, we refer the reader 
to~\cite{quaife-biros14,quaife-biros16} for the review of the 
literature on the numerical methods for Stokesian particulate flows. 
Here, we only review the literature on
anti-aliasing techniques, surface reparametrization algorithms,
area-length correction methods, repulsion models, and error measures
for vesicle dynamics and rheology. 

\textbf{Anti-aliasing.} Classical works in aliasing
include~\cite{canuto-zang-e87, patterson-orszag71, kravchenko-moin97}.
In~\cite{orszag71} and~\cite{orszag74}, if the discretization is with $N$ 
points, the nonlinear terms are computed
at the higher resolution $1.5N$ and filtered back to $N$ points.
While this removes aliasing errors due to quadratic operations, the
nonlinearities in the vesicle model, such as roots and inverses, are
much stronger.  Therefore, it is essential to find appropriate
upsampling rates.  In~\cite{rahimian-biros-e15}, an algorithm that
automatically adjusts the upsampling rate for differentiation is based
on the mean curvature of the three-dimensional vesicles; our upsampling
scheme is similar.  It efficiently determines the sufficient upsampling
rate for each vesicle to compute the force due to bending while we
always upsample to $N^{3/2}$ to compute the layer potentials.

\textbf{Reparametrization.} By using reparametrization, the grid
quality of the vesicle membrane is preserved and this also helps
control aliasing errors.  An algorithm for distributing grid points
equally in arc-length for two-dimensional membranes is presented
in~\cite{baker-shelley90} and implemented in~\cite{hou-shelley-e94,
marple-shravan-e15}.  Additionally,~\cite{shravan-zorin-e11,
rahimian-biros-e15} present a reparametrization scheme for
three-dimensional vesicles which redistributes points so that
high-frequency components of the spectral discretization are
minimized.  Our reparametrization scheme is based on the latter works and 
smooths vesicle shapes by penalizing its high frequencies. We have 
observed that this provides better grid quality than equally spacing 
the points in arc-length.

\textbf{Local correction to area and arc-length.} Despite the local
inextensibility and incompressibility conditions, errors in the area and
length of a vesicle can become large because of error accumulating at
each time step.  This not only results in non-physical vesicle shapes,
but can also lead to instabilities.  In~\cite{marple-shravan-e15}, this
issue is addressed by performing an area-length correction after each
time step.  The length is corrected by adding a correction term to the
inextensibility condition and the area correction requires solving a
quadratic equation.  In~\cite{biben-misbah03, beaucourt-misbah-e04,
aland-voigt-e14}, area and length errors are corrected by adding
artificial forces. Unlike those techniques, our area-length correction
scheme does not modify the governing equations. We correct area and
length after each time step by solving a constrained optimization
problem. This scheme is also extended to three-dimensions
in~\cite{malhotra-biros-e17}. 

\textbf{Repulsion.} There is extensive work on repulsion force models
for avoiding collisions in particulate
flows~\cite{glowinski-periaux-e01, feng-michaelides02,
feng-michaelides04, pan-glowinski-e10}.  These models are
in either polynomial or exponential form. They 
have two parameters: One is the repulsion length scale where the force is
non-zero and the other is the strength of the force.  However these two parameters
are set a priori and the cannot be adapted during the simulation. In
our scheme we employ a state-of-the-art scheme from computer 
graphics~\cite{harmon-grinspun-e09, vouga-grinspun-e11}. This model is 
in a polynomial form which performs well in dense suspension simulations because 
it is developed for simulations with objects coming close frequently with
low velocities in the context of contact mechanics. The length scale is the only 
parameter of the model, which we calibrate heuristically. 
The strength of the repulsion is determined adaptively, therefore, 
no vesicle collision is guaranteed.  
 
\textbf{Error measures.} A significant question that arises in these
low-resolution calculations is an appropriate definition of the error.
Obviously one has to give up on capturing individual trajectories
accurately and look at appropriate statistics that should depend on the
particular application in dense suspensions. By contrast, there are 
applications such as cell sorting in which the trajectories 
are of interest. Since we do not have a particular goal in mind
and we consider this coarsening problem generically, we quantify the 
error in terms of individual trajectories in dilute suspensions and of
upscaled quantities or statistics in dense suspensions. 

The dynamics and rheology of vesicle suspensions have been investigated
widely and various error measures have been introduced. 
For dilute suspensions, local 
error measures such error in the vesicles' inclination angles, centers and 
proximity to other vesicles are frequently used.
In~\cite{kantsler-steinberg-e08, lamura-gompper13, kaoui-harting-e14},
the error is quantified using the vesicles' inclination angles and
centers in dilute suspensions. In~\cite{rahimian-biros-e10} 
distance between two vesicles in a shear flow, i.e. error in proximity. 
For dense suspensions, it is typical to consider collective dynamics rather than
the behavior of each vesicle. For instance, effective viscosity of a 
suspension is an upscaling measure which is equivalent to 
the viscosity of a homogeneous Newtonian fluid having the same energy 
dissipation as the suspension~\cite{jeffrey-acrivos76,rahimian-biros-e10}. 
Additionally, 
in \cite{eckstein-shapiro-e77,
leighton-acrivos87}, the so-called shear-induced diffusion, that is, the
evolution of probability distributions of vesicles' centers is
investigated. This phenomenon is studied both
computationally~\cite{narsimhan-shaqfeh-e13, muller-gompper-e14} and
experimentally~\cite{podgorski-misbah-e11}. We also studied mixing in 
vesicle suspension in~\cite{kabacaoglu-biros-e16}, where we need 
accurate averages of velocity field. In this study, 
we quantify the error based on those quantities of interest.

\paragraph*{Outline of the paper} 
In Section~\ref{s:formulation} we summarize the formulation of our
problem.  In Section~\ref{s:algos} we introduce the LRCA
including anti-aliasing, a new adaptive time stepping method, 
area-length correction, reparametrization, repulsion and alignment of 
shapes.  In Section~\ref{s:experiments} we test the stability of the low-resolution
simulations with the LRCA in various confined and
unconfined flows, and we report accuracy in terms of different error
measures.

\section{Formulation\label{s:formulation}} 
In this section, we summarize the formulation and discretization
algorithm from~\cite{quaife-biros14} (see~\cite{pozrikidis92} for a
detailed derivation). 

\subsection{Governing equations}
In the length and velocity scales of vesicle flows, the inertial forces
are often negligible so we use the quasi-static incompressible Stokes
equations. The dynamics of the flow is fully characterized by the
position of the interface $\mathbf{x}(s,t) \in {\gamma}_i$, where $s$
is arc-length, $t$ is time, and ${\gamma}_i$ is the membrane of the
$i^{th}$ vesicle.  Given $M$ vesicles, we define $\gamma =
\bigcup_{i = 1}^M {\gamma}_i$.  The interior of the $i^{th}$ vesicle is
denoted by $\omega_{i}$, and we define $\omega = \omega_{1} \cup \cdots
\cup \omega_{M}$.  Let $\Omega$ be the $m$-ply connected domain
containing the vesicles, and $\Gamma = {\Gamma}_0 \cup {\Gamma}_1 \cup
\cdots \cup {\Gamma}_m$ be its boundary.  The interior connected
components of $\Gamma$ are ${\Gamma}_i$, $i = 1, \ldots, m$, and
${\Gamma}_0$ is the connected component containing all other connected
components.  See \figref{f:schematicDomain} for the schematic.

\begin{figure}[htb]
\begin{center}
\includegraphics[scale=0.53]{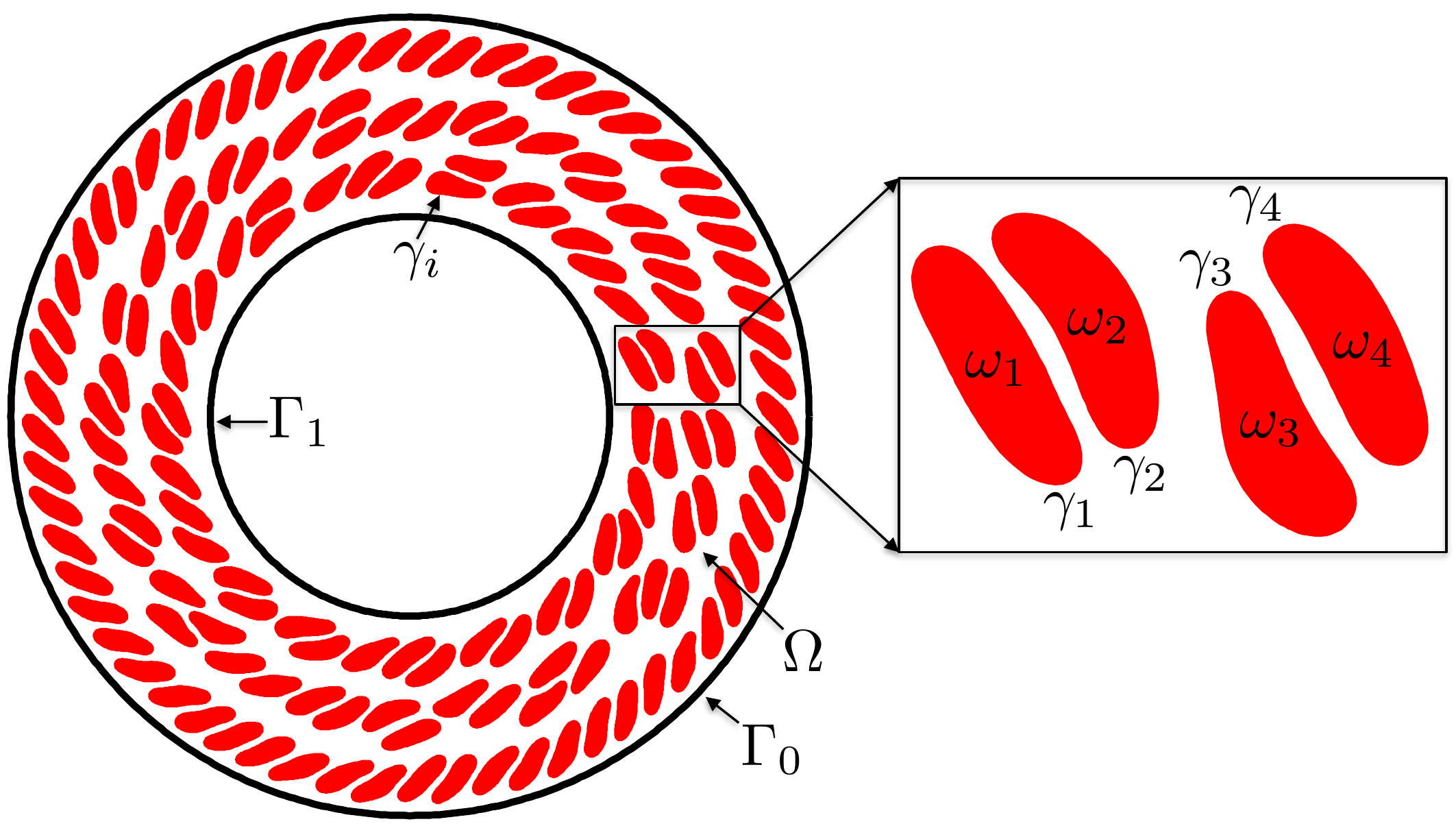} \mcaption{A vesicle
suspension in a Couette apparatus.  $\Omega$ is the fluid domain
between the walls (both inside and outside the vesicles), its
boundary is denoted by $\Gamma$, ${\gamma}_i$ is the boundary of the
$i^{th}$ vesicle whose interior is ${\omega}_i$, $\omega =
{\bigcup}_{i}{\omega}_i$ is the red area, and $\gamma = {\bigcup}_i
{\gamma}_i$.}{f:schematicDomain}
\end{center}
\end{figure}

Let $\mu$ and ${\mu}_i$ be the viscosities of the bulk fluid and the
interior fluid of the $i^{th}$ vesicle, respectively.  The position of
the vesicle is determined by the moving interface problem modeling the
mechanical interactions between the viscous incompressible fluids and
the vesicles' boundaries.  The equations governing the motion of
vesicles are 
\begin{subequations} \label{e:model}
\begin{alignat}{4}
\mu \nabla \cdot \left(\nabla\mathbf{u}(\mathbf{x}) + \nabla {\mathbf{u}}^T(\mathbf{x})   \right) - \nabla p(\mathbf{x}) & = 0, \quad & \mathbf{x} & \in   \Omega \setminus \gamma, \,\, & \text{conservation of momentum,} \\
\nabla \cdot \mathbf{u}(\mathbf{x}) & = 0, \quad  & \mathbf{x} & \in  \Omega \setminus \gamma, \,\, & \text{conservation of mass,}\\
{\mathbf{x}}_s \cdot {\mathbf{u}}_s & = 0, \quad  & \mathbf{x} & \in
\gamma, \,\, & \text{vesicle inextensibility,} \\
\mathbf{u}(\mathbf{x},t) & = \dot{\mathbf{x}}(t), \quad  & \mathbf{x} & \in \gamma, \,\, & \text{velocity continuity,}\\
-{\kappa}_b {\mathbf{x}}_{ssss} +
\left(\sigma(\mathbf{x}){\mathbf{x}}_s\right)_s & = \llbracket
T\mathbf{n} \rrbracket,  \quad &  \mathbf{x} & \in \gamma, \,\, & \text{traction jump,} \label{e:interfacialForce}\\
\mathbf{u}(\mathbf{x},t) & = \mathbf{U}(\mathbf{x},t), \quad  & \mathbf{x} & \in \Gamma, \,\, & \text{no-slip boundary condition.}
\end{alignat}%
\end{subequations}
Here $T = -pI + \mu \left(\nabla \mathbf{u} + \nabla
{\mathbf{u}}^T\right)$ is the Cauchy stress tensor and $\mathbf{n}$ is
the outward normal vector to the membrane $\gamma$ at point
$\mathbf{x}$. $\llbracket \cdot \rrbracket$ denotes the jump across the
interface, $\mathbf{x}_s$ is arc-length derivative of $\mathbf{x}$,
$\kappa_{b}$ is bending stiffness of a membrane, and $\sigma$ is tension 
of a membrane. Here, the right-hand side of \eqref{e:interfacialForce} 
is the interfacial force applied by the membrane to the fluid due to 
bending and tension. $\mathbf{U}$ is
velocity on the boundary $\Gamma$. 

There exist several methods for solving interface evolution equations
similar to~\eqref{e:model}.  In line with our previous
work~\cite{quaife-biros14, quaife-biros16, rahimian-biros-e10,
shravan-biros-e09, shravan-zorin-e11, quaife-biros15}, we use an
integral equation formulation which naturally handle the piecewise
constant viscosity and the discontinuity along the interface.

\subsection{Integral equation formulation}

We present an integral equation formulation of~\eqref{e:model} with a
viscosity contrast $\nu_{p} = \mu_{p}/\mu$ between the interior
fluid with viscosity $\mu_p$ and the exterior fluid with viscosity
$\mu$.  The single and double layer potentials for Stokes
flow (${\mathcal{S}}_{pq}$ and ${\mathcal{D}}_{pq}$, respectively)
denote the potential induced by hydrodynamic densities of 
the interfacial force $\mathbf{f}$ 
and velocity $\mathbf{u}$ on vesicle $q$ and evaluated on vesicle $p$: 
\begin{subequations} \label{e:potentials}
\begin{alignat}{3}
{\mathcal{S}}_{pq}[\mathbf{f}](\mathbf{x}) & := \frac{1}{4\pi \mu}
\int_{{\gamma}_q} \left(-\mathbf{I} \log \rho + \frac{\rr \otimes
\rr}{{\rho}^2}\right) \mathbf{f}(\yy) ds_{\mathbf{y}}, \quad &
\mathbf{x} & \in {\gamma}_p, \label{e:singleLayer} \\
{\mathcal{D}}_{pq}[\mathbf{u}](\mathbf{x}) & := \frac{1-{\nu}_q}{\pi}
\int_{{\gamma}_q} \frac{\rr \cdot \nn}{{\rho}^2} \frac{\rr \otimes
\rr}{{\rho}^2} \mathbf{u}(\yy) ds_{\mathbf{y}}, \quad & \mathbf{x} & \in {\gamma}_p, \label{e:doubleLayer}
\end{alignat}%
\end{subequations}
where $\rr = \xx - \yy$ and $\rho = \|\rr\|_2$. Let ${\mathcal{S}}_p := {\mathcal{S}}_{pp}$ and ${\mathcal{D}}_p := {\mathcal{D}}_{pp}$ denote vesicle self-interactions. We, then, define
\begin{alignat*}{3}
{\mathcal{E}}_{pq}[\ff, \uu](\xx) & = {\mathcal{S}}_{pq}[\ff](\xx) + {\mathcal{D}}_{pq}[\uu](\xx), \quad & \xx & \in {\gamma}_p,\\
{\mathcal{E}}_p [\ff, \uu](\xx) & = \sum_{q=1}^M {\mathcal{E}}_{pq} [\ff, \uu](\xx), \quad & \xx & \in {\gamma}_p.
\end{alignat*}%
For confined flows, we use the completed double layer potential due to
a density function $\eeta$ defined on the solid walls
\begin{align*}
\mathcal{B}[\eeta](\xx) = {\mathcal{D}}_{\Gamma}[\eeta](\xx) + \sum_{q=1}^M R\left[ {\xi}_q(\eeta), {\mathbf{c}}_q\right](\xx) + \sum_{q=1}^M S\left[ {\llambda}_q(\eeta), {\mathbf{c}}_q\right](\xx), \quad \xx \in \gamma \cup \Gamma.
\end{align*}
The Stokeslets and rotlets are
\begin{align*}
S\left[ {\llambda}_q(\eeta), {\mathbf{c}}_q\right](\xx) =
\frac{1}{4\pi\mu} \left(-\log \rho + \frac{\rr \otimes \rr}{{\rho}^2}\right){\llambda}_q(\eeta) \quad \text{and} \quad R\left[ {\xi}_q(\eeta), {\mathbf{c}}_q\right](\xx) = \frac{{\xi}_q(\eeta)}{\mu}\frac{{\rr}^{\perp}}{{\rho}^2},
\end{align*}
where ${\mathbf{c}}_q$ is a point inside ${\omega}_q$, $\rr = \xx -
{\mathbf{c}}_q$, and ${\rr}^{\perp} = (r_2, -r_1)$. The size of the
Stokeslets and rotlets are 
\begin{align*} 
{\llambda}_{q,i} = \frac{1}{2\pi} \int_{{\gamma}_q} {\eeta}_i (\yy)
ds_{\yy}, \,\, i = 1,2 \quad \text{and} \quad {\xi}_q =
\frac{1}{2\pi}\int_{{\gamma}_q} {\yy}^{\perp} \cdot \eeta (\yy) ds_{\yy}.
\end{align*}
If $\xx \in {\Gamma}_0$, we add the rank one modification
${\mathcal{N}}_0[\eeta](\xx) = \int_{{\Gamma}_0}
\left(\nn(\xx)\otimes\nn(\yy)\right)\eeta(\yy)ds_{\yy}$ to
$\mathcal{B}$ to remove a one-dimensional null space.  Finally, by
expressing the inextensibility constraint in operator form as 
\begin{align*} 
\mathcal{P}[\uu](\xx) = {\xx}_s \cdot {\uu}_s,
\end{align*}
the integral equation formulation of~\eqref{e:model} is
\begin{subequations} \label{e:integEquation}
\begin{alignat}{4}
 \left(1+{\nu}_p\right) \uu(\xx) & = {\mathcal{E}}_p [\ff,\uu](\xx) + {\mathcal{B}}_p [\eeta](\xx), \quad & \xx  & \in {\gamma}_p, \,\, & \text{vesicle evolution,}\\
\left(1+{\nu}_p\right) \mathbf{U}(\xx) & = -\frac{1}{2} \eeta(\xx) + {\mathcal{E}}_{\Gamma} [\ff,\uu](\xx) + \mathcal{B}[\eeta](\xx), \quad & \xx & \in \Gamma, \,\, & \text{fixed boundaries,}\\
\quad \mathcal{P}[\uu](\xx)  & = 0, \quad & \xx & \in {\gamma}_p, \,\, & \text{vesicle inextensibility.}
\end{alignat}%
\end{subequations}
Since the velocity
$\uu = d\xx/dt$ and the interfacial force $\ff$ depend on $\sigma$ and
$\xx$,~\eqref{e:integEquation} is a system of
integro-differential-algebraic equations for $\xx, \sigma$, and $\eeta$.

\subsection{Temporal discretization}
We discretize~\eqref{e:integEquation} in time with a first-order
IMEX~\cite{ascher-wetton-e95} time stepping method. We
linearize~\eqref{e:integEquation} and treat the stiff terms, such as
the bending, implicitly, while treating nonlinear terms, such as the
layer potential kernel, explicitly.  In particular, an approximation
for the position $\xx$ and tension $\sigma$ of vesicle $p$ at time
$n+1$ is computed by solving
\begin{subequations} \label{e:temporalDiscretization}
\begin{alignat}{3}
\frac{{\alpha}_p}{\Delta t} \left({\xx}_p^{n+1} - {\xx}_p^n\right) & = {\mathcal{S}}_p^n {\ff}_p^{n+1} + {\mathcal{D}}_p^n{\uu}_p^{n+1} + {\mathcal{B}}_p[{\eeta}^{n+1}] + \sum_{\substack{q = 1 \\ q \neq p}}^M \, {\mathcal{E}}_{pq}^n [{\ff}_q^{n+1}, {\uu}_q^{n+1}], \,\, & \xx & \in {\gamma}_p \\
\mathbf{U}^{n+1}(\xx) & = -\frac{1}{2} {\eeta}^{n+1}(\xx) + {\mathcal{E}}_{\Gamma}^n [{\ff}^{n+1},{\uu}^{n+1}](\xx) + \mathcal{B}[{\eeta}^{n+1}](\xx) + {\mathcal{N}}_0 [{\eeta}^{n+1}](\xx), \,\, & \xx & \in \Gamma, \\
{\mathcal{P}}^n {\xx}_p^{n+1}  & = {\mathcal{P}}^n {\xx}_p^n, \,\, & \xx &  \in {\gamma}_p, \\
{\uu}_p^{n+1} & = \frac{{\xx}_p^{n+1} - {\xx}_p^n}{\Delta t}, \,\, & \xx & \in {\gamma}_p,
\end{alignat}
\end{subequations}
where ${\alpha}_p = (1+{\nu}_p)/2$, and operators with a superscript
$n$ are discretized at ${\xx}^n$.
Although~\eqref{e:temporalDiscretization} is fully coupled, it is more
stable method than methods that treat vesicle-vesicle and
vesicle-boundary interactions explicitly~\cite{quaife-biros14}.

\subsection{Spatial discretization}

Let $\xx(\theta)$, $\theta \in (0, 2\pi]$ be a parametrization of the
interface ${\gamma}_p$, and let$\{{\xx(\theta}_k) = 2k\pi/N\}_{k=1}^N$
be $N$ uniformly distributed discretization points. Then, a spectral
representation of the vesicle membrane is given by
\begin{equation*} \label{e:spatialDiscretization}
  \xx (\theta) = \sum_{k = -N/2 + 1}^{N/2} \hat{\xx}(k)e^{ik\theta}.
\end{equation*}
We use the fast Fourier transform to compute $\hat{\xx}$, and
arc-length derivatives are computed pseudospectrally.  Nearly singular
integrals are computed with an interpolation
scheme~\cite{quaife-biros14}.  Finally, we use a Gauss-trapezoid
quadrature rule~\cite{alpert99} with accuracy $\mathcal{O}(h^8 \log h)$
to evaluate the single layer potential and the spectrally accurate
trapezoid rule for the double layer potential.

We build and factorize a block-diagonal preconditioner introduced
in~\cite{quaife-biros14}.  This preconditioner removes the stiffness
due to the self-interactions of vesicles but does nothing for the
inter-vesicle and inter-wall interactions. As a result, the number of
preconditioned GMRES iterations depends mostly on the magnitude of the
inter-vesicle interactions which is a function of the vesicles'
proximity. As we will see later, we upsample vesicles' boundaries to
avoid aliasing. Thus, we construct the preconditioner on the upsampled
grid. Although this increases the cost of building the preconditioner,
the cost is offset by a significant reduction in the number of GMRES
iterations.

\section{Algorithms for low-resolution simulations\label{s:algos}}
In this section, we present our low-resolution correction algorithms 
(LRCA) for simulations of vesicle suspensions: anti-aliasing in
\secref{s:antialias}, adaptive time stepping in
\secref{s:timeStepAlgo}, local correction to area and length in
\secref{s:correctShapeAlgo}, reparametrization in
\secref{s:reparamAlgo}, alignment of shapes in \secref{s:alignment},
and repulsion force in \secref{s:repulsion}. In
\algref{a:mainAlgorithm}, we list the order that these algorithms are
called in conjuction with the advancing the vesicles forward one time
step.  

\begin{algorithm}[H]
\begin{algorithmic} 
\STATE $ [{\mathbf{x}}_{n+1},{\sigma}_{n+1},{\eeta}_{n+1}] =
\mathtt{timeStep}({\xx}_{n},{\sigma}_n,{\eeta}_n,{\Delta t}_n)$
\COMMENT{Solve the system of equations~\eqref{e:temporalDiscretization}}
\STATE $[\mathtt{accept},{\Delta t}_{\mathrm{new}}] = \mathtt{newTimeStepSize}({\xx}_{n+1},{\xx}_{n},{\Delta t}_n, T_{\mathrm{CPU}})$
\COMMENT{Choose the new time step size}
\IF[If solution is accepted]{$\mathtt{accept}$}
\STATE ${\tilde{\xx}}_{n+1} = \mathtt{correctShape}({\xx}_{n+1},A_0,L_0)$
\COMMENT{Correct errors in area and length of vesicles}
\STATE ${\tilde{\xx}}_{n+1} = \mathtt{reparametrize}({\tilde{\xx}}_{n+1})$
\COMMENT{Reparametrize vesicles' membranes}
\STATE ${\xx}_{n+1} = \mathtt{alignShape}({\tilde{\xx}}_{n+1},{\xx}_{n+1})$
\COMMENT{Align reparametrized shapes with the original ones}
\STATE $ t = t + {\Delta t}_n $
\STATE $\Delta t_{n+1} = \Delta t_{\mathrm{new}}$ 
\COMMENT{Set the time step size for the next time step}
\ELSE[If solution is not accepted]
\STATE $[{\xx}_{n+1},{\sigma}_{n+1},{\eeta}_{n+1}] \leftarrow
[{\xx}_{n},{\sigma}_{n},{\eeta}_{n}]$
\COMMENT{Reject solution and try again with smaller time step}
\STATE $\Delta t_{n} = \Delta t_{\mathrm{new}}$ 
\COMMENT{Set the new time step size for the subsequent attempt}
\ENDIF
\end{algorithmic}
\caption{$\mathtt{Ves2D}$: Main stages in one time step of vesicle flows} \label{a:mainAlgorithm}
\end{algorithm}
At every time step, we solve~\eqref{e:temporalDiscretization} with our
anti-aliasing algorithm to update the vesicles' position $\mathbf{x}$,
tension $\sigma$, and density function $\eeta$ (if the flow is
confined). After solving the evolution equation, given a tolerance 
$\rho_{\mathrm{AL}}$ $\mathtt{newTimeStepSize}$ determines if the solution
${\mathbf{x}}_{n+1}$ is accepted or rejected, and chooses a new time
step size, $\Delta t_{\mathrm{new}}$.  If the solution ${\xx}_{n+1}$ is
accepted, we correct the errors in area and length of every vesicle.
We, then, reparametrize the vesicles' boundaries to redistribute points
such that high frequency components of the surface parametrization are
minimized.  The reparametrization and the area-length correction cause
vesicles to translate and rotate, so we align their centers and
inclination angles with those of the original ones.  Finally, if we
detect that too much error has been committed, then the solution
${\xx}_{n+1}$ is rejected and a time step is taken with a smaller time
step size.

We list and comment on the parameters required by the algorithms under
the pertinent sections.  As a result of numerical experiments we
heuristically decide on the values of these parameters.  
There are two main parameters setting resolution of a simulation: 
Spatial resolution is determined by numbers of 
points per vesicle $N$ and per
wall $N_{\mathrm{wall}}$ and the tolerance for the error in area and
length at each time step, ${\rho}_{\mathrm{AL}}$, sets the temporal
resolution. \cite{quaife-biros16} introduced new higher-order adaptive 
time integrators based on spectral deferred corrections (SDC). 
The number of SDC sweeps $n_{\text{sdc}}$
determines the time stepping order of accuracy.  At low resolutions,
we have observed that SDC does not achieve high-order accuracy unless
a very small time step is taken meaning that a small tolerance
${\rho}_{\mathrm{AL}}$ is requested.  Since we are not interested in
taking small time step sizes, we do not use SDC sweeps for low-
resolution simulations, but they are used for our ground truth high-
resolution simulations.

We propose a black-box solver using~\algref{a:mainAlgorithm} which
requires a single parameter: allocated CPU time $T_{\mathrm{CPU}}$ in
which a simulation is desired to be completed. Our experiments in
\secref{s:experiments}  show that the temporal resolution
$\rho_{\mathrm{AL}}$ required for accurate and  efficient simulations
does not vary much at  low spatial resolutions. The low-resolution
simulations can be successfully completed using $\rho_{\mathrm{AL}}$ =
1E-2 or 1E-3. Since the errors in  area and length are large 
at the coarse spatial
resolutions, the smaller temporal resolutions result in excessive
computing times at the coarse spatial resolutions, i.e. $N \leq 24$.  
This renders the low-resolution simulations impractical. Therefore, we do not require the
temporal resolution to be defined in our solver and instead use the
tolerances we consider workable at low resolutions.

Our solver starts with a coarse spatial discretization $N = 8$ points
per  vesicle and a high tolerance $\rho_{\mathrm{AL}} =$ 1E-2. Then it
indicates possible refinement of the resolutions
to provide an accurate physics or to avoid the failure of the simulation
due to a computation time going beyond the allocated time
$T_{\mathrm{CPU}}$.  We summarize the scheme as follows:
\begin{enumerate}
\item First, the solver runs the simulation with $N_0 = 8$ and $\rho_{\mathrm{AL}} = $ 1E-2 and monitors on-the-fly if the simulation can be completed within $T_{\mathrm{CPU}}$. 
\item If the estimated CPU time goes beyond the allocated time $T_{\mathrm{CPU}}$, the solver terminates the simulation and increases the temporal resolution, first. The next simulation is run with $\rho_{\mathrm{AL}} = $ 1E-3.
\item If the estimated CPU time again exceeds the allocated time $T_{\mathrm{CPU}}$, it increases the spatial resolution to $1.5N_0$ and uses $\rho_{\mathrm{AL}} = $ 1E-2 for the next simulation. 
\item The last two steps are repeated until the simulation is completed within $T_{\mathrm{CPU}}$. If this is not possible, it seems that $T_{\mathrm{CPU}}$ is not achievable at the low resolutions. 
\item Once the solver finds a resolution $N$ and $\rho_{\mathrm{AL}}$, it then checks the accuracy of the simulation. To do so, it runs two more simulations: one with $1.5N$ and $0.1\rho_{\mathrm{AL}}$, and the other with $2N$ and $0.1\rho_{\mathrm{AL}}$. 
\item The self-error is computed with respect to these higher resolution simulations in terms of the quantity of interest. If the self-convergence is achieved, the simulation is terminated. If not, then the procedure above is repeated. 
\end{enumerate}

This scheme can guarantee the accuracy of the physics in terms of the
quantity of interest using as coarse discretization as possible.
But it may not find the simulation which takes the shortest CPU
time. However, it is expected to be faster than to simulate using some high
spatial and temporal resolutions at which it is still unknown if the 
simulation is stable or not beforehand. Additionally, another
simulation of a similar CPU time is still needed to estimate the accuracy of
that solution. We test the proposed solver with an example of a
microfluidic device for cell sorting in \secref{s:dld}.

\subsection{Anti-aliasing} \label{s:antialias}

When representing periodic functions at $N$ grid points, only $N$
frequencies can be represented.  Therefore, if a certain operation such
as the multiplication of two periodic functions is performed, new
high-frequency components are formed and can not be represented with $N$
points.  These newly introduced high-frequency components are identical
to one of the low-frequency components, and the result is that the
high-frequency components are aliased as one of the $N$ frequencies.

In vesicle suspensions, two operations that result in aliasing errors,
especially at low resolutions, are computing the traction jump
$-\kappa_b\xx_{ssss} + (\sigma \xx_{s})_{s}$, and computing the single
and double layer potentials~\eqref{e:potentials}.  The bending term
$\xx_{ssss}$ is especially susceptible to aliasing errors since it
requires multiplication by the Jacobian four times.  We control the
aliasing error by upsampling (uniformly).  But how much should we
upsample?  We adjust the upsampling rate using the decay of the spectrum
of $\xx_{ssss}$.  First, we upsample the $N$ point vesicle to $16N$
points and compute the fourth derivative of this upsampled shape.  Then,
we systematically compare the high-frequency and low-frequency energy
using a growing number of points of this upsampled shape.  We start by
considering the first $1.5N$ Fourier modes.  If the low-frequency energy
exceeds the high-frequency energy, then we use $1.5$ as the upsampling
rate.  Otherwise, we continue by comparing the low-frequency and
high-frequency energy of the first $2N$ Fourier modes.  This algorithm
is continued until the low-frequency energy exceeds the high-frequency
energy, or the maximum upsampled rate of $16$ is reached.  The algorithm
is outlined in \aref{a:upsampleRate}.

\begin{algorithm}[!htb]
\begin{algorithmic} 
\REQUIRE{$\xx$}
\STATE {// Input current configuration $\xx$}
\STATE {$\xx \leftarrow \text{upsample } \xx$}
\COMMENT {Upsample by a pre-specified rate of 16}
\STATE {$\BB\xx = \mathtt{fourthDeriv}(\xx)$}
\COMMENT {Compute the fourth arc-length derivative of the upsampled shape}
\STATE {$\widehat{\BB\xx} = \mathtt{fft}(\BB\xx)$}
\COMMENT {Compute the FFT of the arc-length derivative}
\STATE {$\alpha = 1.5$}
\COMMENT {Upsample by at least $1.5$}
\STATE {$\mathtt{low\_energy} = \|\widehat{\BB\xx}(1:\alpha \,
N/2)\|$}
\COMMENT {Energy in low frequencies}
\STATE {$\mathtt{high\_energy} = \|\widehat{\BB\xx}(\alpha \, N/2+1:
\alpha \, N)\|$}
\COMMENT {Energy in high frequencies}
\WHILE {($\mathtt{high\_energy} > \mathtt{low\_energy} \: \& \:
  \alpha \leq 16$)}
\STATE {$\alpha = \alpha + 0.5$}
\COMMENT {Increase the upsampling rate}
\STATE {$\mathtt{low\_energy} = \|\widehat{\BB\xx}(1:\alpha \, N/2)\|$}
\COMMENT {Energy in low frequencies}
\STATE {$\mathtt{high\_energy} = \|\widehat{\BB\xx}(\alpha \, N/2+1:
    \alpha \, N)\|$}
\COMMENT {Energy in high frequencies}
\ENDWHILE

\RETURN {$\alpha$}
\end{algorithmic}
\caption{Choose upsampling rate for computing traction jump}
\label{a:upsampleRate}
\end{algorithm}

While the upsampling rate may be as large as 16, the vesicle shape is
only tracked at the low resolutions with $N$ points.  Therefore, the
additional cost of computing the traction jump with our anti-aliasing
algorithm is proportional to the upsampling rate.  In addition, our
numerical examples never required an upsampling rate larger than $10$,
and, at most time steps, they do not exceed $3$. 

In \figref{f:derivativeAlias} we use \aref{a:upsampleRate} to compute
the aliasing error in the traction jump of a single elliptical vesicle.
To compute the error, we first compute a reference traction jump with
$1024$ points.  Then, we compute the traction jump, but with $N=12, 16,
24$, and, $32$ points both with (red) and without (blue) anti-aliasing.
As expected, smaller values of $N$ require a larger upsampling rate.  In
addition, the error of the Fourier modes of the traction jump when our
upsampling algorithm is applied is bounded in the interval
$[10^{-6},10^{-4}]$ for all four values of $N$; in contrast, when no
upsampling is applied, the error decays in the low frequencies as $N$ is
increased, but remains large in the high frequencies.  Finally, even
when a high resolution such as $N=32$ is used, we see that it is
important to upsample by at least $1.5$ to control the aliasing error.

\begin{figure}[!htb]
\begin{center}
\begin{tabular}{cc}
  \begin{tikzpicture}

\pgfplotsset{every tick label/.append style={font=\small}}

\begin{axis}[
  width = 0.4\textwidth,
  xmin = -6,
  xmax = 6,
  xtick = {-6,-3,0,3,6},
  xlabel = {$k$},
  xlabel style = {yshift = 5pt},
  ymin = 1.0e-6,
  ymax = 1.0e-0,
  ytick = {1e-6,1e-4,1e-2,1e0},
  ymode = log,
  ylabel = {Error},
  ylabel style = {yshift = 0pt},
  legend entries = {Aliased, Anti-Aliased},
  legend style = {draw=none},
  legend cell align = left,
  legend style = {at={(0.75,0.7)},anchor=north},
  title = {{$N=12$ \quad $\alpha=4$}}
  ]

\addplot [color=blue,only marks,mark=*] table{
-5 4.147060413725352e-02
-3 3.321552634563704e-02
-1 3.308187302019867e-02
+1 6.616374604039738e-02
+3 1.602121069748798e-01
+5 3.775533174666825e-01
};

\addplot [color=red,only marks,mark=*] table{
-5 8.265165353726008e-06
-3 4.762296123364542e-06
-1 5.825256355491739e-06
+1 1.165051271047608e-05
+3 2.539160500212286e-05
+5 5.504874225853487e-05
};

\end{axis}

\end{tikzpicture} &
  \begin{tikzpicture}

\pgfplotsset{every tick label/.append style={font=\small}}

\begin{axis}[
  width = 0.4\textwidth,
  xmin = -8,
  xmax = 8,
  xtick = {-8,-4,0,4,8},
  xlabel = {$k$},
  xlabel style = {yshift = 5pt},
  ymin = 1.0e-6,
  ymax = 1.0e-0,
  ytick = {1e-6,1e-4,1e-2,1e0},
  ymode = log,
  ylabel = {Error},
  ylabel style = {yshift = 0pt},
  title = {{$N=16$ \quad $\alpha=3$}}
  ]

\addplot [color=blue,only marks,mark=*] table{
-7 1.073076422103312e-02
-5 2.488129844563628e-02
-3 1.625390453613296e-02
-1 1.678351321171157e-02
+1 3.356702642342332e-02
+3 7.934365457872801e-02
+5 1.844891429647213e-01
+7 4.099069816787672e-01
};

\addplot [color=red,only marks,mark=*] table{
-7 1.856529124647493e-05
-5 8.265165351809227e-06
-3 4.762296120572701e-06
-1 5.825256352194450e-06
+1 1.165051270711469e-05
+3 2.539160499898798e-05
+5 5.504874225657407e-05
+7 1.180326563790985e-04
};

\end{axis}

\end{tikzpicture} \\
  \begin{tikzpicture}

\pgfplotsset{every tick label/.append style={font=\small}}

\begin{axis}[
  width = 0.4\textwidth,
  xmin = -12,
  xmax = 12,
  xtick = {-12,-6,0,6,12},
  xlabel = {$k$},
  xlabel style = {yshift = 5pt},
  ymin = 1.0e-6,
  ymax = 1.0e-0,
  ytick = {1e-6,1e-4,1e-2,1e0},
  ymode = log,
  ylabel = {Error},
  ylabel style = {yshift = 0pt},
  title = {{$N=24$ \quad $\alpha=2$}}
  ]

\addplot [color=blue,only marks,mark=*] table{
-11 4.912331336074843e-02
-9 4.704664988831683e-03
-7 8.100217338797537e-03
-5 4.858752348276967e-03
-3 2.855200742958988e-03
-1 3.132825647901532e-03
+1 6.265651295803466e-03
+3 1.432120213334397e-02
+5 3.245701594225098e-02
+7 7.177941154043017e-02
+9 1.547853351850682e-01
+11 3.251661260797103e-01
};

\addplot [color=red,only marks,mark=*] table{
-11 7.704452184619908e-05
-9 4.006620782938741e-05
-7 1.856529123640851e-05
-5 8.265165347771424e-06
-3 4.762296119662681e-06
-1 5.825256352753742e-06
+1 1.165051270792047e-05
+3 2.539160499966036e-05
+5 5.504874225629755e-05
+7 1.180326563769799e-04
+9 2.509719550407907e-04
+11 5.300302774969602e-04
};

\end{axis}

\end{tikzpicture} &
  \begin{tikzpicture}

\pgfplotsset{every tick label/.append style={font=\small}}

\begin{axis}[
  width = 0.4\textwidth,
  xmin = -16,
  xmax = 16,
  xtick = {-16,-8,0,8,16},
  xlabel = {$k$},
  xlabel style = {yshift = 5pt},
  ymin = 1.0e-6,
  ymax = 1.0e-0,
  ytick = {1e-6,1e-4,1e-2,1e0},
  ymode = log,
  ylabel = {Error},
  ylabel style = {yshift = 0pt},
  title = {{$N=32$ \quad $\alpha=1.5$}}
  ]

\addplot [color=blue,only marks,mark=*] table{
-15 6.390186377580721e-02
-13 8.547610927201169e-03
-11 1.582958542880021e-03
-9 2.230909802116491e-03
-7 1.358212233963117e-03
-5 6.767801586656984e-04
-3 3.886917450420155e-04
-1 4.464785364574605e-04
+1 8.929570729147161e-04
+3 1.997226009386603e-03
+5 4.439872827107595e-03
+7 9.705085519953511e-03
+9 2.089246391823021e-02
+11 4.432580180657873e-02
+13 9.265080616121177e-02
+15 1.911437821323900e-01
};

 Anti-Aliased error
\addplot [color=red,only marks,mark=*] table{
-15 1.152273160812851e-04
-13 1.222818418070630e-04
-11 7.704452188829601e-05
-9 4.006620786301955e-05
-7 1.856529125557067e-05
-5 8.265165357579124e-06
-3 4.762296124437676e-06
-1 5.825256355621073e-06
+1 1.165051271090840e-05
+3 2.539160500370958e-05
+5 5.504874226329137e-05
+7 1.180326563900547e-04
+9 2.509719550682934e-04
+11 5.300302775543883e-04
+13 1.112363635429560e-03
+15 2.319877774136744e-03
};

\end{axis}

\end{tikzpicture}
\end{tabular}
\end{center}
\mcaption{Aliasing error of the traction jump both with and without
upsampling at different resolutions.  With our anti-aliasing algorithm, the
aliasing error is controlled and mesh-independent.  Because of symmetry
in the geometry, all the even indexed Fourier modes
vanish.}{f:derivativeAlias}
\end{figure}

For the layer potentials, applying \aref{a:upsampleRate} is too
expensive.  Even if we used a low resolution such as $N=12$, this would
require a dense matrix-vector multiplication with $192$ points.
Therefore, we simply fix an upsampling rate that is used at all
resolutions.  We have experimented with upsampling by a factor of $2$
and upsampling by a factor of $\lceil \sqrt{N} \rceil$.  We use the
latter value since we have found that the additional cost is offset by
the number of rejected time steps in some of our numerical examples.  In
\figref{f:layerAlias}, we plot aliasing errors with and without
upsampling, again for an ellipse, and the density function is the
vesicle shape.  By upsampling to $N^{3/2}$, the error is controlled at
all frequencies for all the resolutions.  Moreover, the upsampling rate
used is less than 6 for the four small values of $N$ that we will be
considering.

\begin{figure}[!htb]
\begin{center}
\begin{tabular}{ccc}
  \begin{tikzpicture}

\pgfplotsset{every tick label/.append style={font=\small}}

\begin{axis}[
  width = 0.4\textwidth,
  xmin = -6,
  xmax = 6,
  xtick = {-6,-3,0,3,6},
  xlabel = {$k$},
  xlabel style = {yshift = 5pt},
  ymin = 1.0e-16,
  ymax = 1.0e0,
  ytick = {1e-16,1e-8,1e-0},
  ymode = log,
  ylabel = {Error},
  ylabel style = {yshift = 0pt},
  legend entries = {Aliased, Anti-Aliased},
  legend style = {draw=none},
  legend cell align = left,
  legend pos = south east,
  title = {{$N=12$ \quad $\alpha = 4$}}
  ]

\addplot [color=blue,only marks,mark=*] table{
-5 1.907397994023494e-03
-3 4.991538938668244e-04
-1 2.153774560247649e-03
+1 1.453796692976720e-03
+3 1.069594622552648e-03
+5 1.756772046913773e-03
};

\addplot [color=red,only marks,mark=*] table{
-5 1.175544731900866e-10
-3 1.924129968750502e-10
-1 7.815024450929604e-10
+1 4.020081525912802e-10
+3 1.885114708802130e-10
+5 4.215821200550322e-10
};

\end{axis}

\end{tikzpicture} &
  \begin{tikzpicture}

\pgfplotsset{every tick label/.append style={font=\small}}

\begin{axis}[
  width = 0.4\textwidth,
  xmin = -8,
  xmax = 8,
  xtick = {-8,-4,0,4,8},
  xlabel = {$k$},
  xlabel style = {yshift = 5pt},
  ymin = 1.0e-16,
  ymax = 1.0e0,
  ytick = {1e-16,1e-8,1e-0},
  ymode = log,
  ylabel = {Error},
  ylabel style = {yshift = 0pt},
  title = {{$N=16$ \quad $\alpha=4$}}
  ]

\addplot [color=blue,only marks,mark=*] table{
-7 4.839363075241133e-04
-5 2.050525186009074e-04
-3 9.428873508730829e-05
-1 3.149104866227687e-04
+1 1.886252140516890e-04
+3 1.230595148565249e-04
+5 2.042936106967242e-04
+7 3.853063802630689e-04
};

\addplot [color=red,only marks,mark=*] table{
-7 1.353766735694010e-11
-5 4.563770283063636e-12
-3 1.468195828470418e-12
-1 1.989117824589593e-12
+1 9.703226438373350e-13
+3 1.996416020961174e-12
+5 1.187576432431696e-11
+7 3.986484225911901e-11
};

\end{axis}

\end{tikzpicture} \\
  \begin{tikzpicture}

\pgfplotsset{every tick label/.append style={font=\small}}

\begin{axis}[
  width = 0.4\textwidth,
  xmin = -12,
  xmax = 12,
  xtick = {-12,-6,0,6,12},
  xlabel = {$k$},
  xlabel style = {yshift = 5pt},
  ymin = 1.0e-16,
  ymax = 1.0e-0,
  ytick = {1e-16,1e-8,1e0},
  ymode = log,
  ylabel = {Error},
  ylabel style = {yshift = 0pt},
  title = {{$N=24$ \quad $\alpha=5$}}
  ]

\addplot [color=blue,only marks,mark=*] table{
-11 4.256197554011164e-05
-9 1.930168296628391e-05
-7 8.816083143751436e-06
-5 4.145560293917091e-06
-3 3.028228226814919e-06
-1 9.701055343682879e-06
+1 5.277921944520581e-06
+3 2.808837136125676e-06
+5 4.771429101795270e-06
+7 8.093575299136480e-06
+9 1.566205303949091e-05
+11 3.314621806589346e-05
};

\addplot [color=red,only marks,mark=*] table{
-11 1.458057169654969e-13
-9 7.072053983561286e-14
-7 2.775947277156472e-14
-5 9.097307951939829e-15
-3 2.096777561113693e-15
-1 2.187941802901245e-16
+1 7.588019945828043e-16
+3 1.277373368863105e-15
+5 9.939776059961801e-15
+7 5.042982068707388e-14
+9 1.472260585296673e-13
+11 2.857128224020975e-13
};

\end{axis}

\end{tikzpicture} &
  \begin{tikzpicture}

\pgfplotsset{every tick label/.append style={font=\small}}

\begin{axis}[
  width = 0.4\textwidth,
  xmin = -16,
  xmax = 16,
  xtick = {-16,-8,0,8,16},
  xlabel = {$k$},
  xlabel style = {yshift = 5pt},
  ymin = 1.0e-16,
  ymax = 1.0e-0,
  ytick = {1e-16,1e-8,1e0},
  ymode = log,
  ylabel = {Error},
  ylabel style = {yshift = 0pt},
  title = {{$N=32$ \quad $\alpha=6$}}
  ]

\addplot [color=blue,only marks,mark=*] table{
-15 5.069060139805772e-06
-13 2.309524476914593e-06
-11 1.053284942772067e-06
-9 4.710105434467186e-07
-7 2.064561169421035e-07
-5 1.043996869265469e-07
-3 1.054891408957658e-07
-1 3.795086847586811e-07
+1 1.999634447137742e-07
+3 9.370212704608087e-08
+5 1.615705047538552e-07
+7 2.585539338832250e-07
+9 4.286732831225018e-07
+11 8.695964009070511e-07
+13 1.911950654892104e-06
+15 4.039357114048784e-06
};

\addplot [color=red,only marks,mark=*] table{
-15 4.152150571460241e-13
-13 2.803157212057122e-13
-11 1.592598565158554e-13
-9 7.573867269867849e-14
-7 3.042425521954532e-14
-5 1.043207189266274e-14
-3 2.380504170756641e-15
-1 7.293139343004151e-16
+1 8.513002614336331e-16
+3 6.443706326918795e-16
+5 8.868573061980173e-15
+7 5.105833283369368e-14
+9 1.610353347228538e-13
+11 3.359842670963867e-13
+13 5.183819771131037e-13
+15 6.272702895877702e-13
};

\end{axis}

\end{tikzpicture} &
\end{tabular}
\end{center}
\mcaption{Aliasing error of the single layer potential both with and
without upsampling at different resolutions.  With our anti-aliasing
algorithm, the aliasing error is controlled.  Because of symmetry in the
geometry, all the even indexed Fourier modes vanish.}{f:layerAlias}
\end{figure}

\subsection{Adaptive time stepping} \label{s:timeStepAlgo}

In~\cite{quaife-biros15, quaife-biros16}, we presented an adaptive
high-order time stepping method for vesicle suspensions. The scheme
uses the errors in the vesicles' area and length to estimate the local
truncation error.  This is possible since the area and length are
invariant by the incompressibility and inextensibility conditions,
respectively.  The major advantage is that this estimate can be
computed with spectral accuracy, basically for free, and, in contrast
to many adaptive time stepping methods~\cite{hai-nor-wan1993}, only one
numerical solution is formed.  High order accuracy can be achieved
through spectral deferred correction (SDC) sweeps~\cite{quaife-biros16}.

This algorithm poses two issues that need to be addressed in the
context of the present study.  One issue is that the original proposed
algorithm~\cite{quaife-biros16} uses asymptotic estimates of the error,
so it assumes that the temporal error dominates the spatial error, and
that $\Delta t$ is sufficiently small.  The time stepping error does
not always dominate in low-resolution simulations, and even if it does,
it is possible that a very small $\Delta t$ is necessary to be in the
asymptotic regime.  Therefore, before adjusting the time step size, we
check if we are in the asymptotic regime.  If we are, we use the method
proposed in~\cite{quaife-biros16}, and if not, then we simply increase
or decrease the time step size by a constant factor.  Moreover, we do
not expect to achieve second- or higher-order accuracy in time, and
this must be accounted for when adjusting the time step size.  The
second issue is that the algorithm assumes accumulation of errors in
area and length.  However, to maintain stability, we will be correcting
these errors at every time step.  This is easily resolved by specifying
a error tolerance for each time step rather than for the time horizon 
as done in~\cite{quaife-biros16}.

In \aref{a:newTimeStepSizeAlgo}, we describe our new scheme that
uses errors in area and length to accept or reject a solution and
selects a new time step size. Let ${\rho_{\mathrm{AL}}}$ be the
user-defined tolerance for errors in each vesicle's area and length. The area $A$
and length $L$ of a vesicle at time $t$ whose boundary is
$\mathbf{x}(\theta,t) = (x(\theta,t),y(\theta,t))$ is
\begin{equation*}\label{e:AreaLengthVesicle}
A = \frac{1}{2} \int_0^{2\pi} \left(xy_{\theta} -
yx_{\theta}\right)d\theta, \quad L = \int_0^{2\pi} \sqrt{x_{\theta}^2 +
y_{\theta}^2} d\theta.
\end{equation*}
Shortly we will require $dA/dt$ and $dL/dt$ to adjust the time step. The
time derivatives are given by
\begin{equation*}\label{e:DerivativeAreaLengthVesicle}
\frac{dA}{dt} = \frac{1}{2} \int_0^{2\pi} \left( uy_{\theta} +
xv_{\theta} - vx_{\theta} - yu_{\theta} \right)d\theta, \quad
\frac{dL}{dt} = \int_0^{2\pi} \frac{x_{\theta}u_{\theta} +
y_{\alpha}v_{\theta}}{\sqrt{x_{\theta}^2 + y_{\theta}^2}} d\theta,
\end{equation*}
where $u = \frac{dx}{dt}$ and $v = \frac{dy}{dt}$.  We approximate the
velocities with
\begin{equation*}\label{e:velocities}
u(t) = \frac{x(t+\Delta t)-x(t)}{\Delta t}, \quad v(t) = \frac{y(t+\Delta t)-y(t)}{\Delta t}.
\end{equation*}

Suppose we compute the solution at time $t+\Delta t$ with the
first-order time stepping scheme and the solution $\mathbf{x}(t+\Delta
t)$ has area $A(t+\Delta t)$ and length $L(t+\Delta t)$.  The
errors in area and length are 
\begin{align}
  {\epsilon}_A = \frac{|A(t+\Delta t) - A(t)|}{A(t)}, \quad 
  {\epsilon}_L = \frac{|L(t+\Delta t) - L(t)|}{L(t)}. 
  \label{e:errorInArea}
\end{align}
Assuming ${\epsilon}_A > {\epsilon}_L$ (the same argument holds if the
situation is reversed), we either accept or reject the solution and
choose a new time step size for a single vesicle (we take the maximum
errors over all vesicles if we have multiple vesicles) as follows:
\begin{enumerate}
\item We, first, check for any collisions between different vesicles
and between vesicles and solid walls using the technique presented
in~\cite{quaife-biros14}. If there is a collision, we reject the
solution and decrease the time step size by a factor of two. 

\item We define an interval $[\rho_{\min},\rho_{\mathrm{AL}}]$ where 
$\rho_{\min} = 0.5\rho_{\mathrm{AL}}$. We accept
the solution if ${\rho}_{\min} \leq {\epsilon}_A \leq
{\rho}_{\mathrm{AL}}$, and the time step size is not changed. This step 
helps reduce the number of rejected time steps since it does not increase 
the time step size when the error is close to the tolerance $\rho_{\mathrm{AL}}$.

\item If ${\epsilon}_A < {\rho}_{\min}$, we check if the time step size
is in the asymptotic regime.  This is done by examining the Taylor
series of the area
\begin{equation}\label{e:expandArea}
  A(t + \Delta t) = A(t) + \frac{dA}{dt}\left(t\right) \Delta t +
  \mathcal{O}({\Delta t}^2).
\end{equation}
We check if the right-hand side in~\eqref{e:expandArea} is dominated by
the first two terms by defining~$q_{A}(t) =
\left|\frac{dA}{dt}\left(t\right)/A(t)\right|$ so that
\begin{equation*}\label{e:defineQa}
  {\epsilon}_A \leq q_A(t)\Delta t + |\mathcal{O}({\Delta t}^2)|.
\end{equation*}
Then, we say that $\Delta t$ is in the asymptotic regime if
\begin{equation}\label{e:checkAsymptoticIncrease}
  \frac{\left |{\epsilon}_A - q_A \Delta t\right |}{{\epsilon}_A} \leq 
    {\rho}_{\mathrm{up}},
\end{equation}
and the new time step size is 
\begin{equation}\label{e:DeltaTAsymptoticIncrease}
  {\Delta t}_{\mathrm{new}} = \frac{{\rho}_{\mathrm{AL}}}{q_A(t)}.
\end{equation}
If condition~\eqref{e:checkAsymptoticIncrease} is not satisfied, then
we increase the time step size by a constant factor
${\beta}_{\mathrm{up}}$.  Finally, we do not allow the time step size
to exceed the maximal value $\Delta t_{\max}$, which can be determined 
based on the length $L$ and velocity $U$ scales of a flow, 
i.e. $\Delta t_{\max} \propto L/U$.

\item If ${\epsilon}_A > \rho_{\mathrm{AL}}$, we reject the solution and
decrease the time step size.  Again, we first check if the time step
size is in the asymptotic regime.  If
\begin{equation*}\label{e:checkAsymptoticDecrease}
\frac{\left |{\epsilon}_A - q_A \Delta t\right |}{{\epsilon}_A} \leq {\rho}_{\mathrm{down}},
\end{equation*}
then the new time step size is chosen as
in~\eqref{e:DeltaTAsymptoticIncrease}.  Otherwise, we decrease the time
step size by a constant factor ${\beta}_{\mathrm{down}}$.

\item Once the time step size is chosen, we compute the average of the 
last 10 time step sizes $\overline{\Delta t}$. Then assuming that we 
will keep taking time steps of size $\overline{\Delta t}$ we compute the 
number of remaining time steps to reach the time horizon
${\widetilde{m}} = (T_{\mathrm{h}}-T_{\mathrm{current}})/\overline{\Delta t}$. 
We also compute the average of the CPU times it took in 
the last 10 time steps, $\overline{t}_{\mathrm{CPU}}$. Then assuming that 
each remaining time step will take $\overline{t}_{\mathrm{CPU}}$ on 
average we estimate the remaining CPU time and the total CPU time the 
simulation will take, $\widetilde{T}_{\mathrm{CPU}}$. 
If the total estimated CPU time $\widetilde{T}_{\mathrm{CPU}}$ exceeds 
the allocated time 
$T_{\mathrm{CPU}}$, we terminate the simulation.
\end{enumerate} 
At low resolutions, collisions are likely as the
hydrodynamic forces may not have been resolved sufficiently.  In
addition to the collision detection~\cite{quaife-biros14} in this
scheme, we introduce a repulsion force in~\secref{s:repulsion} to
handle the collisions. However, an imminent collision might require
small time step sizes which result in a computing time exceeding the
allocated time $T_{\mathrm{CPU}}$. This usually occurs when the
vesicles get too close due to large time steps taken before the
repulsion force is activated and once they are too close, the
repulsion force introduces stiffness which requires very small time
step sizes. In those cases we terminate the simulation and take a
finer temporal resolution or maybe a finer spatial resolution so that
the simulation can be completed within the allocated time.

In summary, we have several parameters in our scheme. First, we have a
tolerance~$\rho_{\mathrm{AL}}$ to decide whether the solution is
acceptable.  If it is acceptable, then we need to decide if we should
increase the time step size. We do this by comparing the error with a
tolerance~$\rho_{\min}$.  The tolerance~$\rho_{\mathrm{AL}}$ might be
an input but we observe from our experiments that it should not be
less than 1E-3 at low resolutions to result in reasonable computing
times. The tolerances lower than that requires very small time step
sizes which are needed to keep the errors in area and length below
those tolerances at coarse spatial resolutions. If we are to increase the
time step size, then we need the tolerance~$\rho_{\mathrm{up}}$
in~\eqref{e:checkAsymptoticIncrease} to determine if we can use the
asymptotic assumption to adjust the the time step size
using~\eqref{e:DeltaTAsymptoticIncrease}. If the asymptotic assumption
is not valid, then we need a constant factor~$\beta_{\mathrm{up}}$ by
which we increase the time step size.  If the solution is not
acceptable, then we need to decrease the time step size.  Similarly,
we decide if the asymptotic assumption is valid using a
tolerance~$\rho_{\mathrm{down}}$. If it is not valid, then we need a
constant factor $\beta_{\mathrm{down}}$ by which we decrease the time
step size.  
\begin{table}[htb] \mcaption{List of parameters of the
adaptive time stepping.}{t:adaptiveParams} \centering
\begin{tabular}{l|l|l}  \hline   Symbol & Definition & Value\\
\hline ${\rho}_{\mathrm{AL}}$ & Tolerance for errors in
area-length & [1E-4, 1E-1] \\

${\rho}_{\min}$ & Tolerance that must be reached for time step to be
increased & ${\rho}_{\mathrm{AL}}/2$ \\

${\beta}_{\mathrm{up}}$ & Maximum factor of increment in time step size & $1.2$ \\

${\beta}_{\mathrm{down}}$ & Minimum factor of decrement in time step size & $0.5$ \\

${\rho}_{\mathrm{up}}$ & Tolerance for using the asymptotic assumption
to increase time step size & $10^{-3}$ \\

${\rho}_{\mathrm{down}}$ & Tolerance for using the asymptotic
assumption to decrease time step size & $10^{-2}$ \\

${\Delta t}_{\max}$ & Maximum time step size & $\propto L/U$\\

\hline 
\end{tabular} 
\end{table} 
We list the parameters of the adaptive time stepping and their values in
\tabref{t:adaptiveParams}. Here, $L$ and $U$ are length and velocity scales 
of a flow. We want to be aggressive in decreasing
the time step size but cautious in increasing it. Therefore, we choose
${\rho}_{\mathrm{up}} < {\rho}_{\mathrm{down}}$.  The other parameters
are chosen by running a few experiments and choosing values that
minimize the total number of rejected time steps. The parameter values
in~\tabref{t:adaptiveParams} work very well for a variety of problems we
have tested.  We apply the proposed adaptive time stepping scheme to a
confined and unconfined suspension in \figref{f:timeAdaptiveChokeShear}. 

\begin{figure}[H]
\begin{center}
 \begin{minipage}{\textwidth}
\setcounter{subfigure}{0}
\renewcommand*{\thesubfigure}{(a-1)} 
      \hspace{0cm}\subfigure[Stenosis flow]{\scalebox{0.47}{{\includegraphics{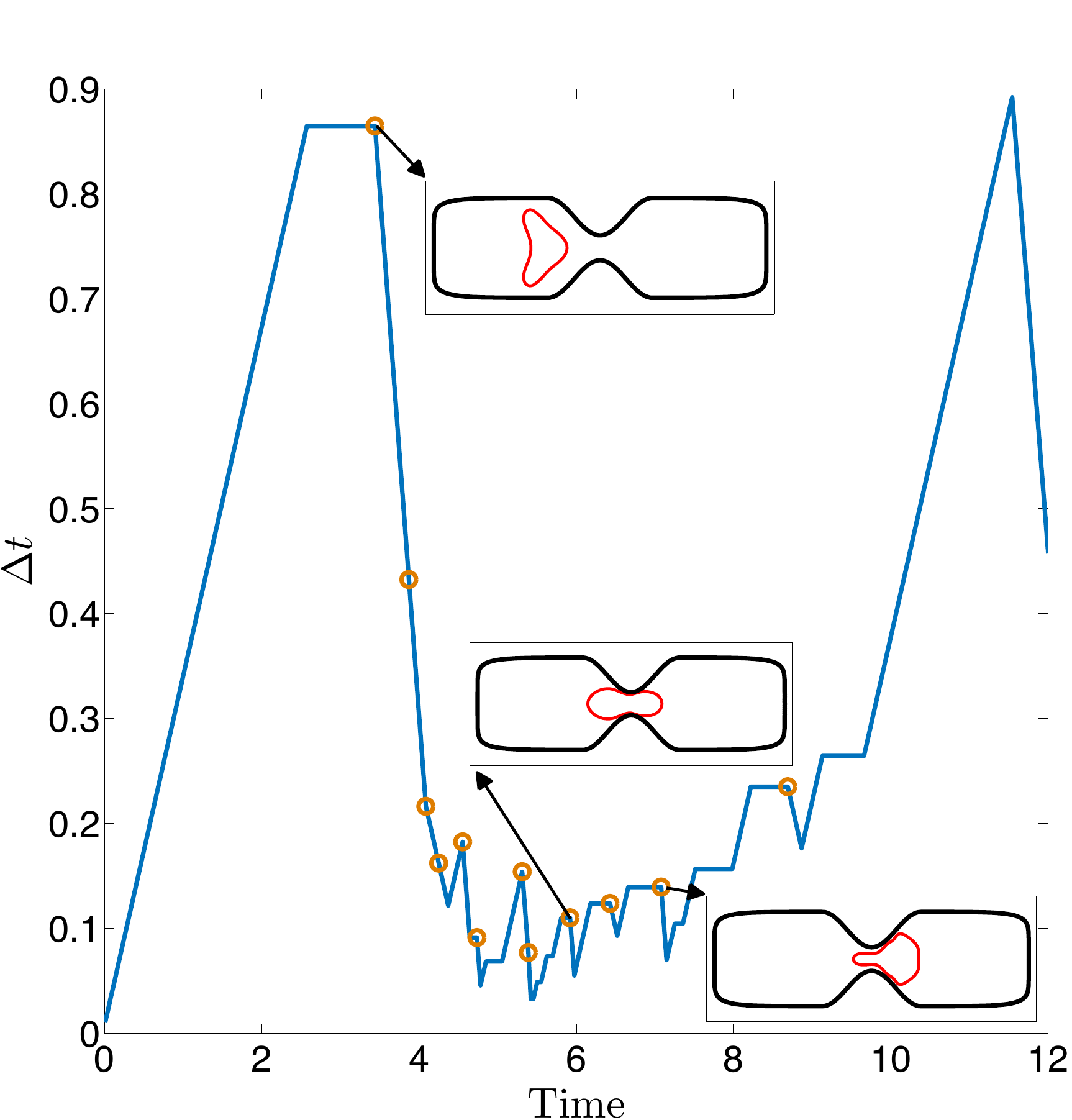}}}	
      \label{f:tstepSizeChoke}}
\setcounter{subfigure}{0}
\renewcommand*{\thesubfigure}{(a-2)} 
      \hspace{-0.2cm}\subfigure[Shear flow]{\scalebox{0.47}{{\includegraphics{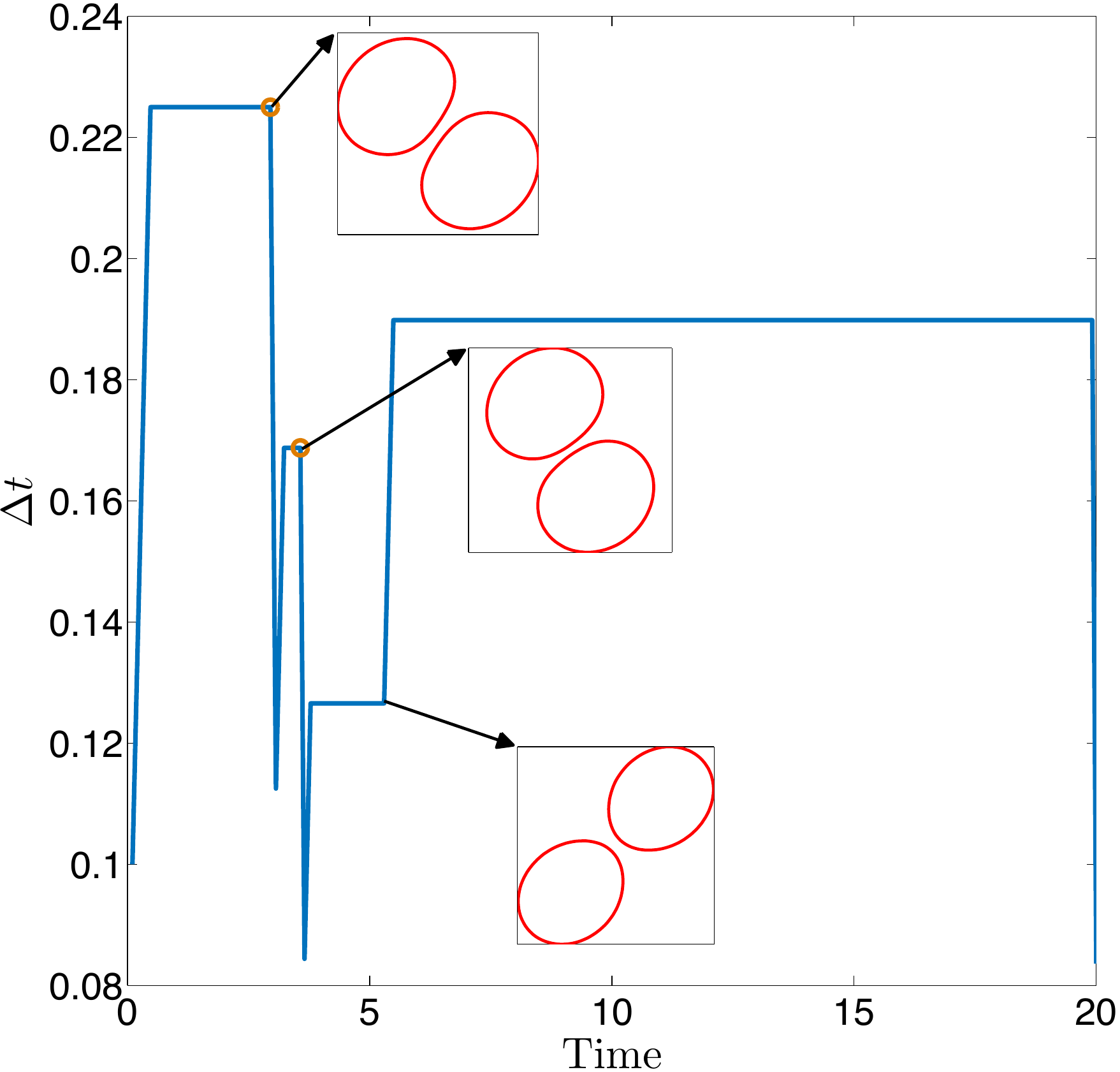}}}
      \label{f:tstepSizeShear}}
\mcaption{We demonstrate how the time step size varies in a {\bf
stenosis flow} (left) and a {\bf shear flow} (right). Open circles
indicate the times when a time step size is rejected. In both
simulations, vesicles are discretized with $N = 16$ points, and the
tolerance is ${\rho}_{\mathrm{AL}} = 10^{-2}$. In the stenosis flow, the
outer wall is discretized with ${N}_{\mathrm{wall}} = 256$ points. There
are 12 rejected and 64 accepted time steps in the stenosis flow, and 2
rejected and 110 accepted time steps in the shear
flow.}{f:timeAdaptiveChokeShear}
\end{minipage}
\end{center}
\end{figure}

\begin{algorithm}[!htb]
\begin{algorithmic} 
\REQUIRE{${\rho}_{\mathrm{AL}}$, ${\rho}_{\min}$, ${\Delta t}_{\max}$,
${\beta}_{\mathrm{up}}$, ${\beta}_{\mathrm{down}}$, ${\rho}_{\mathrm{up}}$, ${\rho}_{\mathrm{down}}$}
\STATE{// Input current and previous configurations ${\xx}_{n+1}$, ${\xx}_n$ and previous time step size ${\Delta t}_n$}
\STATE{// Suppose there is a single vesicle and find the new time step size based on error in area, first.}
\STATE $\mathtt{crossing} = \mathtt{detectCollision}({\xx}_{n+1})$
\COMMENT{Check for collisions}
\IF[If there is a collision]{$\mathtt{crossing}$}
\STATE $\mathtt{accept} = \mathtt{false}$
\COMMENT{Do not accept the solution}
\STATE ${\Delta t}_{\mathrm{new}} = 0.5{\Delta t}_n$
\COMMENT{Decrease ${\Delta t}$}
\ELSE[If there is no collision, check error in area]
\STATE $[A_{n+1},A_n] \leftarrow \mathtt{getArea}({\xx}_{n+1},\xx)$
\COMMENT{Compute area of ${\xx}_{n+1}$ and ${\xx}_n$}
\STATE{$[q_A, q_L] \leftarrow \mathtt{computeTimeDerivaties}({\xx}_{n+1})$}
\COMMENT{Compute $q_A = \left|\frac{dA}{dt}/A\right|$ analytically}
\STATE{${\epsilon}_A \leftarrow \mathtt{computeError}(A_{n+1},A_n)$}
\COMMENT{Compute error in area~\eqref{e:errorInArea}}
\IF[If error falls into the buffer zone]{${\rho}_{\min} \leq {\epsilon}_A \leq {\rho}_{\mathrm{AL}}$}
\STATE $\mathtt{accept} = \mathtt{true}$
\STATE ${\Delta t}_{\mathrm{new}} = {\Delta t}_n$
\COMMENT{Accept the solution and do not change the time step size}
\ELSIF[If error is less than the minimum tolerance]{${\epsilon}_A
< {\rho}_{\min}$}
\STATE $\mathtt{accept} = \mathtt{true}$
\COMMENT{Accept the solution}
\IF[If asymptotic assumption is valid]{$\frac{|{\epsilon}_A -
q_A\Delta t |}{{\epsilon}_A} \leq {\rho}_{\mathrm{up}}$}
\STATE ${\Delta t}_{\mathrm{new}} = {\rho}_{\mathrm{AL}} / q_A$
\COMMENT{Increase $\Delta t$ based on the asymptotic assumption}
\ELSE[If asymptotic assumption is not valid]
\STATE ${\Delta t}_{\mathrm{new}} = {\beta}_{\mathrm{up}} {\Delta t}_n$
\COMMENT{Increase ${\Delta t}$ by a constant factor}
\ENDIF
\STATE${\Delta t}_{\mathrm{new}} = \min ({\Delta t}_{\mathrm{new}},{\Delta t}_{\max})$
\COMMENT{Make sure that $\Delta t$ is not greater than ${\Delta t}_{\max}$}
\ELSIF[If error is greater than tolerance]{${\epsilon}_A > {\rho}_{\mathrm{AL}}$}
\STATE $\mathtt{accept} = \mathtt{false}$
\COMMENT{Do not accept the solution}
\IF[If asymptotic assumption is valid]{$\frac{|{\epsilon}_A-q_A
{\Delta t}_n|}{{\epsilon}_A} \leq {\rho}_{\mathrm{down}}$}
\STATE ${\Delta t}_{\mathrm{new}} = {\rho}_{\mathrm{AL}} / q_A$
\COMMENT{Decrease $\Delta t$ based on the asymptotic assumption}
\ELSE[If asymptotic assumption is not valid]
\STATE ${\Delta t}_{\mathrm{new}} = {\beta}_{\mathrm{down}} {\Delta t}_n$
\COMMENT{Decrease ${\Delta t}$ by a constant factor}
\ENDIF
\ENDIF
\ENDIF
\STATE{// Repeat for ${\epsilon}_L$ to obtain another 
${\Delta t}_{\mathrm{new}}$ and choose the smaller of the two
time step sizes.}
\STATE{// If there are multiple vesicles, then we reject the solution if 
the tolerance is violated by}
\STATE{// at least one vesicle and we choose the minimum of ${\Delta t}_{\mathrm{new}}$ over the vesicles.}
\STATE{$\overline{\Delta t} \leftarrow \mathtt{mean}(\Delta t(\mathtt{end-9:end}))$}
\COMMENT{Average time step size taken in the last 10 time steps}
\STATE{$\widetilde{m} = (T_h-T_{\mathrm{current}})/\overline{\Delta t}$}
\COMMENT{Estimate the remanining number of time steps}
\STATE{$\overline{t}_{\mathrm{CPU}} \leftarrow \mathtt{mean}
(t_{\mathrm{CPU}}(\mathtt{end-9:end})$}
\COMMENT{Average CPU time the last 10 time steps took}
\STATE{$\widetilde{T}_{\mathrm{CPU}} = T_{\mathrm{CPU}}^{\mathrm{sofar}} +
 \widetilde{m}\times \overline{t}_{\mathrm{CPU}}$}
\COMMENT{Estimate how long the simulation is going to take}
\IF[If total estimated CPU time exceeds the allocated time]
{$\widetilde{T}_{\mathrm{CPU}}>T_{\mathrm{CPU}}$}
\STATE{$\mathtt{terminate}$}
\COMMENT{terminate the simulation.}
\ENDIF

\STATE{ }
\RETURN ${\Delta t}_{\mathrm{new}}, \mathtt{accept}$
\end{algorithmic}
\caption{$\mathtt{newTimeStepSize}({\xx}_{n+1},{\xx}_n,{\Delta t}_n,T_{\mathrm{CPU}})$} \label{a:newTimeStepSizeAlgo}
\end{algorithm}

\subsection{Local corrections to area and length}
\label{s:correctShapeAlgo}

The incompressibility and inextensibility conditions guarantee that the
area and length of each vesicle are constant.  However, long time horizon
simulations suffer from the accumulation of errors in area and length
which often leads to instabilities or non-physical simulations.
Therefore, area-length correction is essential in long time horizon
simulations at low resolutions.  One way is to add a correction term to
the inextensibility condition to correct the length and solve a
quadratic equation to correct the area~\cite{marple-shravan-e15}.
Another way is to add a forcing term to the inextensibility
condition~\cite{biben-misbah03,beaucourt-misbah-e04,aland-voigt-e14}.
Here, we introduce a postprocessing technique that maintains the errors
in area and length below a prescribed tolerance without modifying 
the governing equations.  This is done with a
constrained optimization problem where the constraints require the
vesicle's area and length to be fixed.

Suppose that a vesicle initially has area $A_{0}$ and length $L_{0}$,
and that $\xx(t)$ is the solution at time $t$. We make a local
correction to the vesicle's shape by applying sequential quadratic
programming (SQP) to
\begin{equation} \label{e:correctShape}
\argmin_{\substack{A(t) = A_0 \\ L(t) = L_0}} \, \| \tilde{\xx}(t) - \xx(t) \|^2,
\end{equation}
to obtain a new shape $\tilde{\xx}$.  Equation~\eqref{e:correctShape}
is solved iteratively with a MATLAB built-in function,
$\mathtt{fmincon}$, which is used for minimum constrained algebraic
equations (see \aref{a:correctAlgorithm} and
\aref{a:constraintsAlgorithm}). The function requires tolerances for
the objective function ${\rho}_{\mathrm{fun}}$ and for the constraints
${\rho}_{\mathrm{con}}$.  In our low-resolution simulations, both
tolerances are $10^{-3}$.  After correcting the area and length,
it is possible that vesicles are closer than a minimum distance set by
our repulsion force (see \secref{s:repulsion}).   Since we will be
treating repulsion explicitly, the result would be a stiffer system and
a smaller time step size would be required.  To avoid this issue, we
only correct the vesicles shape if the correction does not result in
the distance between any two vesicles decreasing below the repulsion
length scale.

We demonstrate the effectiveness of the local correction in
\figref{f:howCorrectionWorks}.  We consider a single vesicle of reduced
area $0.65$ in a shear flow with no viscosity contrast.  The vesicle
tilts to a certain inclination angle and then undergoes a tank-treading
motion.  We discretize the vesicle with $N = 12$ points and
reparametrize (see \secref{s:reparamAlgo}) its boundary at every time
step. We take a time horizon of $T = 30$ so that the vesicle
tank-treads approximately 1.5 times.  We run the simulation with
various tolerances for errors in area and length
${\rho}_{\mathrm{AL}}$.  We plot the maximum of the errors in area and
length without the correction (top row), 
snapshots of the vesicle configurations without
(middle row) and with (bottom row) the local correction to the
vesicle's shape.  Without correction, the error grows to
$\mathcal{O}(10^{-1})$ at the time horizon, and it is still growing.
However, the simulations remains stable and accurate, even with large
tolerances, when the vesicle's shape is corrected.
\begin{figure}[htb]
\begin{center}
\includegraphics[scale=0.38]{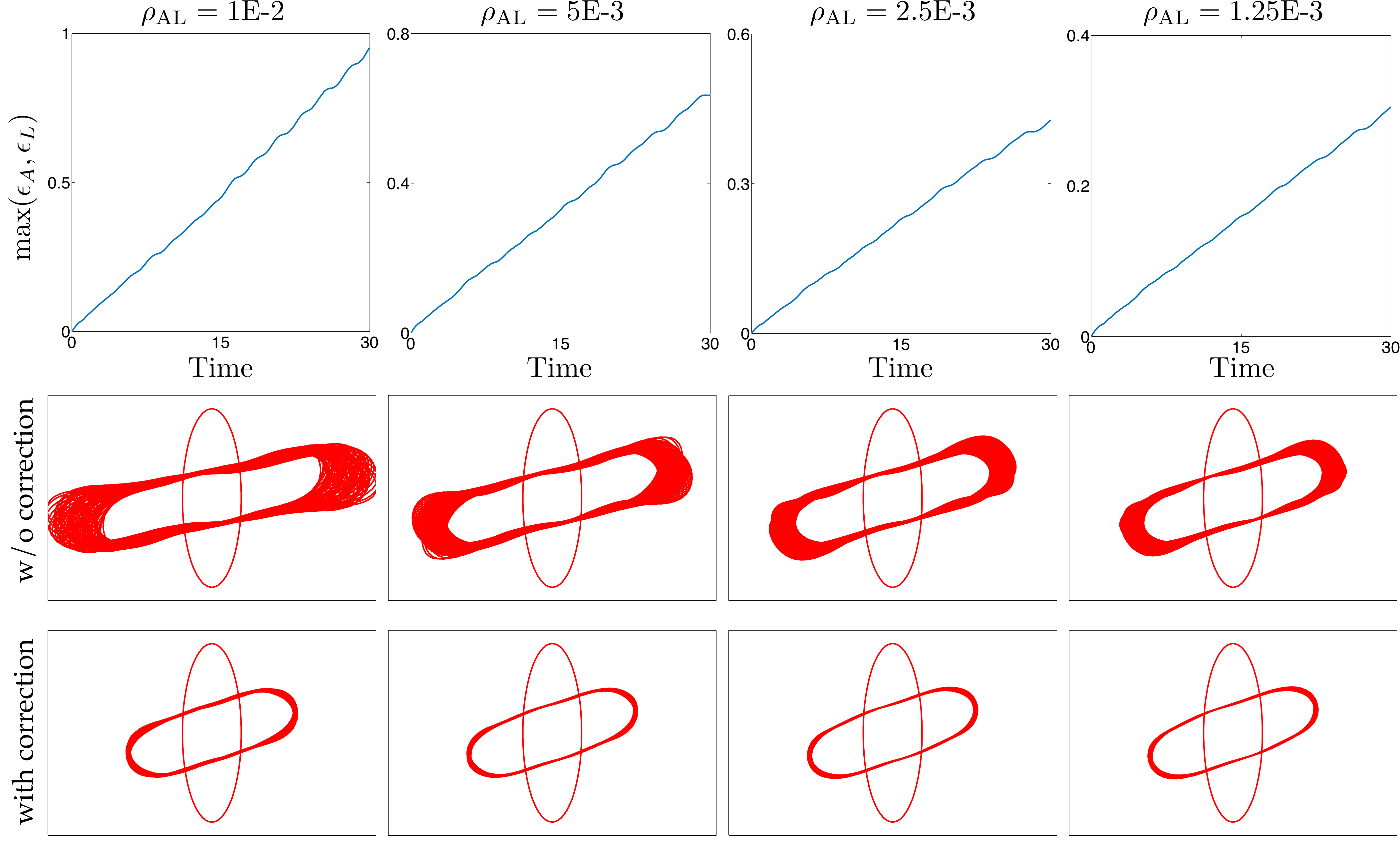} 
\mcaption{The effect of correcting the vesicle's area and length. We
discretize the vesicle with $N = 12$ points.  Each column corresponds to
a simulation with a tolerance for errors in area and length
${\rho}_{\mathrm{AL}}$ indicated at the top. The top plots are the
maximum of the errors in area and length at each time step when the
vesicle's shape is not locally corrected and the middle plots show
superimposed snapshots.  The bottom row are superimposed snapshots of
the vesicles when the shape is corrected at every time
step.}{f:howCorrectionWorks}
\end{center}
\end{figure}

\begin{algorithm}[!htb]
\begin{algorithmic} 
\STATE{// Input the initial area and length of each vesicle $A_0$, $L_0$}
\STATE {// Choose ${\rho}_{\mathrm{con}}$, ${\rho}_{\mathrm{fun}}$}
\COMMENT{Choose tolerances for constraints and function we want to minimize}
\STATE {// Set options of $\mathtt{fmincon}$:} 
\COMMENT{MATLAB's $\mathtt{fmincon}$  finds minimum of constrained function}
\STATE {// Use Sequential Quadratic Programming (SQP) algorithm with tolerances ${\rho}_{\mathrm{con}}$, ${\rho}_{\mathrm{fun}}$}
\FOR[Loop over vesicles]{$k = 1, \ldots, m$}
\STATE $\mathtt{minFun} = @(\zz) \: {\| \zz - {\xx}_k \|}_2$
\COMMENT{Construct the function we want to minimize}
\STATE ${\tilde{\xx}}_k = \mathtt{fmincon}(\mathtt{minFun}, \ldots,@(\zz) \mathtt{constraints}(\zz,A_0,L_0),\text{options})$
\COMMENT{Make a local correction}
\IF{$\mathtt{fmincon}$ fails}
\STATE ${\tilde{\xx}}_k = {\xx}_k$
\COMMENT{If the solver fails, do not correct the $k^{th}$ vesicle}
\ENDIF
\ENDFOR
\STATE $\mathtt{crossing} = \mathtt{detectCollision}({\tilde{\xx}})$
\COMMENT{Check if there is any collision of corrected shapes}
\IF[If there is a collision]{$\mathtt{crossing}$}
\STATE ${\tilde{\xx}} = \xx$
\COMMENT{Do not correct any of the shapes}
\ELSE
\FOR[Loop over vesicles]{$k = 1, \ldots, m$}
\STATE{// Check if there are any vesicles in near zone of $k^{th}$ vesicle approaching to that}
\STATE{// This avoids sudden increases in repulsion forces due to unphysical motions led by correction}
\STATE $\mathtt{approaching} = \mathtt{detectNearCollision}({\tilde{\xx}}_k)$
\IF{$\mathtt{approaching}$}
\STATE ${\tilde{\xx}}_k = {{\xx}_k}$
\COMMENT{Do not correct shape of the $k^{th}$ vesicle}
\ENDIF
\ENDFOR
\ENDIF
\RETURN $\tilde{\xx}$
\end{algorithmic}
\caption{$\mathtt{correctShape}(\xx,A_0,L_0)$} \label{a:correctAlgorithm}
\end{algorithm}
\begin{algorithm}[!htb]
\begin{algorithmic} 
\STATE{$A_z \leftarrow \mathtt{getArea}(\zz), L_z \leftarrow \mathtt{getArcLength}(\zz)$}
\COMMENT{Compute area and arc-length of current shape $\zz$}
\STATE{$\mathtt{cEq} = [(A_z-A_0)/A_0 \quad (L_z-L_0)/L_0]$}
\COMMENT{Non-linear equalities}
\RETURN $\mathtt{cEq}$
\COMMENT{Return constraints of the optimization problem}
\end{algorithmic}
\caption{$\mathtt{constraints}(\zz,A_0,L_0)$} \label{a:constraintsAlgorithm}
\end{algorithm}


\subsection{Reparametrization} \label{s:reparamAlgo} When a vesicle is
discretized at low resolutions, time stepping can quickly distort the
point distribution.  This introduces high frequency components into the
boundary parametrization which leads to aliasing errors and numerical
instabilities.  Therefore, it is essential to redistribute points so
that high-frequency components are minimized.  The reparametrization
algorithm is presented in our previous work~\cite{shravan-zorin-e11,
rahimian-biros-e15} for three-dimensional vesicles.  In
\aref{a:reparamAlgorithm} we mimic this algorithm for two-dimensional
vesicles\footnote{Let us emphasize however that in high spatial
resolutions such a correction is not necessary.}.

Let $\gamma$ be the boundary of a vesicle that is parameterized as
$\xx(s)$ where $s$ is arc-length.  Let $F: {\mathbb{R}}^2 \rightarrow
\mathbb{R}$ denote an implicit representation of the surface such that
$F(\gamma) = 0$ and $\grad F$ does not vanish.  We seek a surface parametrization $\yy(s)$
which minimizes the quality measure $E(\yy) := \sum_{k=1}^{N} a_k
{|{\hat{\yy}}_k|}^2$:
\begin{equation*} \label{e:reparamMinim}
\argmin_{\substack{y \in C^\infty}} E(\yy(s)), \quad \text{subject to}
\quad F(\yy(s)) = 0 \quad \text{for all} \, s,
\end{equation*}
where $a_k$ are attenuation coefficients. By introducing the Lagrangian
$E(\yy) + \int_{\gamma} \lambda F(\yy)$, the optimality condition is
obtained by taking the variation of $E$ with respect to $\yy$ and $\lambda$ (see~\cite{shravan-zorin-e11}): 
\begin{equation} \label{e:optimalityCond}
\left(I - \nn (\yy) \otimes \nn (\yy) \right)\nabla E(\yy) = 0 \quad \text{and} \quad F(\yy) = 0.
\end{equation}
We introduce a parameter $\tau$ and use pseudo-transient continuation to
solve~\eqref{e:optimalityCond}. The discretized equation using an
explicit scheme is
\begin{equation*} \label{e:discreteReparamEqn}
{\yy}_{n+1} = {\yy}_n - \Delta \tau \left(
  I - {\nn}(\yy_n) \otimes {\nn}(\yy_n)\right) \grad E({\yy}_n).
\end{equation*}
Letting $\mathbf{g} = -\left(I - {\nn}(\yy) \otimes {\nn}(\yy)\right) \grad
E({\yy})$, the iteration is continued until the change in $\yy$ or the
gradient $\mathbf{g}$ is sufficiently small.  The parameters
${\rho}_{\yy}$ and ${\rho}_{\mathbf{g}}$ in \aref{a:reparamAlgorithm}
set this stopping criteria.

Since the goal of reparametrization is to smooth the boundary $\gamma$,
the attenuation coefficients $a_k$ should be small for low frequencies
and grow for high frequencies.  We choose $a_k = k^4$ resulting in
$\grad E(\yy) = \sum_{k=1}^{N} k^4 {\hat{\yy}}_k e^{ik\alpha}$. We have
also experimented with $a_k = k^2$, but we found that the resulting
shapes could still have undesirable high frequencies (see
\figref{f:reparamDiffKs}).

\begin{figure}[H]
\begin{center}
 \begin{minipage}{\textwidth}
\setcounter{subfigure}{0}
\centering
\renewcommand*{\thesubfigure}{(a-1)} 
      \hspace{0cm}\subfigure[Vesicle's shape]{\scalebox{0.48}{{\includegraphics{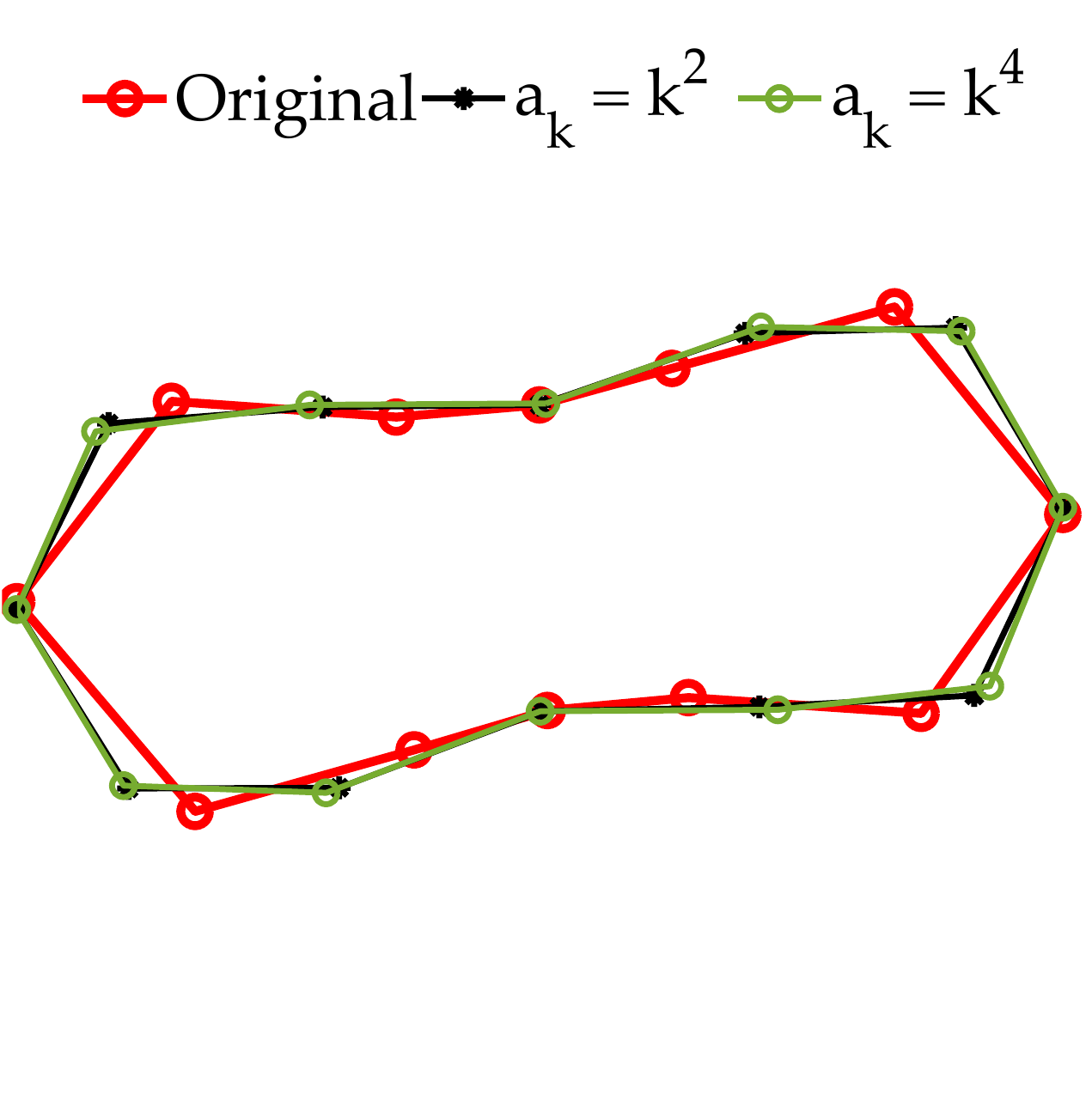}}}	
      \label{f:reparamShapes}}
\setcounter{subfigure}{0}
\centering
\renewcommand*{\thesubfigure}{(a-2)} 
      \hspace{0cm}\subfigure[Energies of the boundaries in \figref{f:reparamShapes}]{\scalebox{0.48}{{\includegraphics{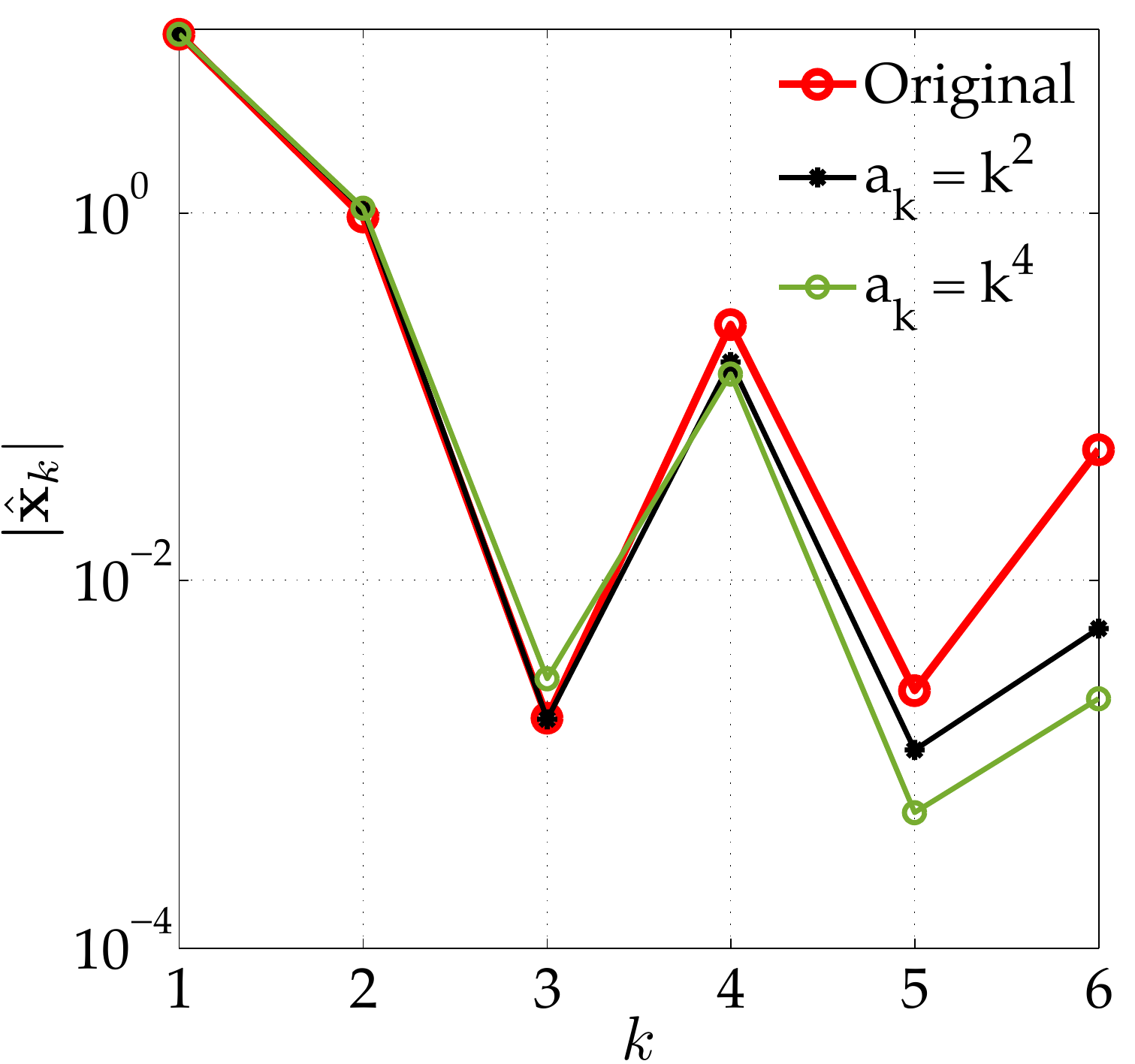}}}
      \label{f:reparamEnergies}}
\mcaption{We reports results for two different choices of the
attenuation coefficient $a_k$. We reparametrize the original shape
(red) discretized by $N = 12$ points in \figref{f:reparamShapes} with
$a_k = k^2$ (black) and $a_k = k^4$ (green).  The corresponding shapes
are in \figref{f:reparamShapes}. \figref{f:reparamEnergies} shows the
absolute values of the shapes' energies.  While the arc-length spacing
turns out to be almost uniform with $a_k = k^2$, the additional
reduction in the high frequencies from using $a_k = k^4$ results in
smoother vesicles and stabler simulations.}{f:reparamDiffKs}
\end{minipage}
\end{center}
\end{figure}

In \figref{f:howReparamWorks}, we compare the simulation of two
vesicles in a shear flow with and without reparametrization.  The
vesicles are discretized with $N=12$ points.  We use our new adaptive time
stepping scheme (\secref{s:timeStepAlgo}) with a tolerance of
${\rho}_{\mathrm{AL}} = 10^{-2}$ and we correct the area and length of
the vesicles after each time step (\secref{s:correctShapeAlgo}).  The
top row does not use reparametrization while the bottom row does. The
grey vesicles are from the ground truth.  The shapes with
reparametrization are significantly smoother and closer to the ground
truth.  The number of required time steps when we reparametrize is
reduced; there are 94 accepted, 4 rejected time steps with
reparametrization and 108 accepted, 11 rejected time steps without
reparametrization.

\begin{figure}[H]
\begin{center}
\includegraphics[scale=0.57]{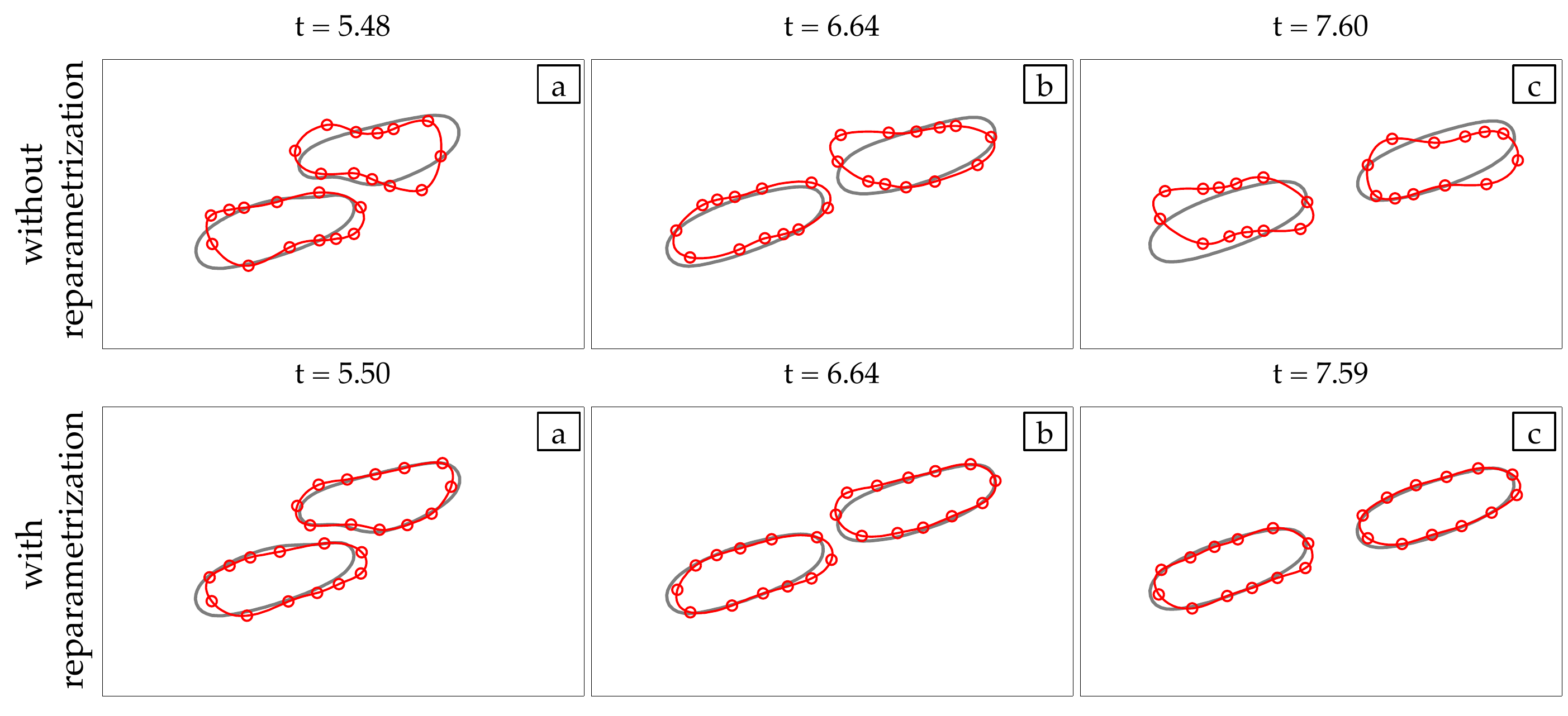}
\mcaption{Our reparametrization algorithm applied to two vesicles of
reduced area $0.65$ in a {\bf shear flow} with no viscosity contrast.
Here, the vesicles are discretized with $N = 12$ points and we set
${\rho}_{\mathrm{AL}} = 10^{-2}$. The top row does not use
reparametrization while the bottom row does.  The grey vesicles are the
ground truth solution, which is computed using the high-fidelity version
of the code.}{f:howReparamWorks}
\end{center}
\end{figure}

\begin{algorithm}[!htb]
\begin{algorithmic} 
\REQUIRE{${\rho}_{\yy}$, ${\rho}_{\mathbf{g}}$, ${\Delta \tau}$,$i_{\max}$}
\STATE{// We choose ${\rho}_{\yy} = 10^{-3}\Delta\xx$ and ${\rho}_{\mathbf{g}} = 10^{-3}$, where $\Delta \xx = \uu \Delta t$; and $i_{\max} = 200$}
\STATE{// We use a line search to find $\Delta \tau$ at every iteration for stability}
\STATE ${\yy}_0 \leftarrow \text{upsample } \xx$
\COMMENT{Upsample to the anti-aliasing frequency} 
\STATE ${\mathbf{g}}_0 = -(I - \nn ({\yy}_0) \otimes \nn ({\yy}_0))\nabla E({\yy}_0)$
\COMMENT{Projected gradient}
\STATE{$i = 0$}
\WHILE{$i < i_{\max}$}
\STATE ${\mathbf{g}} = -\left(I - \nn ({\yy}) \otimes \nn ({\yy})\right) \nabla E({\yy})$
\COMMENT{$\grad E(\yy) = \sum_{k=1}^{N} k^4 {\hat{\yy}}_k e^{ik\alpha}$}
\STATE ${\yy}^+ = {\yy} - \Delta \tau {\mathbf{g}}$
\STATE $\yy \leftarrow {\yy}^+, \, i \leftarrow i+1$
\IF{$\|{\mathbf{g}}\| < \max({\rho}_{\yy} /\Delta \tau,
{\rho}_{\mathbf{g}} \| {\mathbf{g}}_0 \|)$}
\STATE $\mathtt{break}$
\COMMENT{Terminate if the gradient or change in $\yy$ is small}
\ENDIF
\ENDWHILE
\STATE {$\tilde{\xx} \leftarrow \text{downsample } {\yy}$}
\COMMENT{Downsample to the original grid}
\RETURN $\tilde{\xx}$
\end{algorithmic}
\caption{$\mathtt{reparametrize}(\xx)$} \label{a:reparamAlgorithm}
\end{algorithm}

\subsection{Alignment of shapes}\label{s:alignment}
Locally correcting (\secref{s:correctShapeAlgo}) and reparametrizing
(\secref{s:reparamAlgo}) the vesicle shape often results in translations
and rotations.  To remove these errors, we apply a rigid body motion to
the vesicle shape $\tilde{\mathbf{x}} = (\tilde{x},\tilde{y})$ after
each of the algorithms so that the corrected and reparametrized shape
aligns with the original shape.  Given a single vesicle, the rigid body
motion from $\tilde{\mathbf{x}}$ to $\mathbf{x}$ is
\begin{equation*} \label{e:rigidBody}
  \xx = R\tilde{\xx} + \mathbf{t},
\end{equation*}
where $R$ is a rotation matrix ($R^{T}R = I$) and $\mathbf{t}$ is a
translation vector~\cite{besl-mckay92}. To compute $R$, first let
${\mathbf{c}}_{\tilde{\xx}}$ and ${\mathbf{c}}_{\xx}$ be the centers of
the shapes, and then define the $2 \times 2$ matrix
\begin{equation*}\label{e:covariance}
H = \sum_{i=1}^N
\left({\tilde{\mathbf{x}}}_i-{\mathbf{c}}_{\tilde{\mathbf{x}}}\right)
\left({\mathbf{x}}_i-{\mathbf{c}}_{\mathbf{x}}\right)^T,
\end{equation*}
where $\xx_{i}$ are the discretization points of the vesicles.  By
computing the singular value decomposition of $H = U \Sigma V^{T}$, we obtain 
the rotation matrix $R = VU^{T}.$  The translation operator $\mathbf{t}$ is, 
then,
\begin{equation*}\label{e:translation}
  \mathbf{t} = -R{\mathbf{c}}_{\tilde{\xx}} + {\mathbf{c}}_{\xx}.
\end{equation*}

The new shape is ${\mathbf{x}}_{\mathrm{new}} = R{\tilde{\mathbf{x}}}
+ \mathbf{t}$ which, in
addition to having the correct area, length, and a smooth boundary, has
the same center and inclination angle as the shape prior to these local
corrections. Therefore, this algorithm helps minimize the artificial effects of 
the correction algorithms on the dynamics given by the governing equations.

\subsection{Repulsion} \label{s:repulsion} While hydrodynamic
forces do not allow vesicles to cross, these forces are often not
accurately resolved in simulations with low spatial resolutions, and
vesicles may collide.  We introduce a repulsion force to handle
collisions.  We use discrete penalty layers to penalize close proximity
between discretization points on vesicles.  The form of the repulsion we
use has been introduced for contact mechanics~\cite{harmon-grinspun-e09,
vouga-grinspun-e11}.  Letting $h_{\max}$ be the maximum arc-length
spacing and $d_{\min}$ a repulsion length scale, the repulsion
force applies on the points of the vesicles' membranes when they get closer 
than $d_{\min}= {\delta}_{\min} h_{\max}$.  
We define a gap function for discrete layer
$\ell$ between two discretization points $\xx \in \gamma_{p}$ and 
$\yy \in \gamma_{q}$, $p \neq q$
\begin{equation*} \label{e:gapFunction}
  g_{\ell} = \|{\mathbf{r}}\| - \frac{d_{\min}}{\ell},
\end{equation*}
where $\|{\mathbf{r}}\| = \| \xx - \yy \|$.  The gap function measures
the proximity of two points on the vesicles ($\gamma_p$ and $\gamma_q$).  When
$g_{\ell} < 0$, the points are in the proximity of the layer $\ell$. The
repulsion force to penalize being in the proximity of the $\ell^{th}$ discrete
layer is 
\begin{align} \label{e:forceLthDiscLayer}
\mathbf{F}^l = \begin{cases}
      -2W\ell^2\frac{g_{\ell}}{\|\rr\|}
      \rr, & \text{if}\, g_{\ell} < 0,  \\
      0, & \text{otherwise},
      \end{cases}.
\end{align}
where $W$ is the repulsion strength.

The penalty force can be considered as placing a spring between
approaching vesicles. If there is a single spring between them, the
spring will compress fully and eventually fail for sufficiently large
relative velocity.  However, having penalty forces as a function of the
active discrete layers as in~\eqref{e:forceLthDiscLayer} can be
considered as placing an infinite number of springs between approaching
vesicles. This guarantees that two vesicles do not collide, which
makes the method robust.  Although this guarantee is independent of
the repulsion strength $W$, performance of the method and error in
physics depend on the choice of $W$.  

\begin{figure}[H]
 \begin{minipage}{\textwidth}
 \begin{center}
\setcounter{subfigure}{0}
\renewcommand*{\thesubfigure}{(a-1)} 
     \setcounter{subfigure}{0}
     \hspace{0cm}\subfigure[Total number of discrete layers activated vs. distance]{\scalebox{0.6}{{\includegraphics{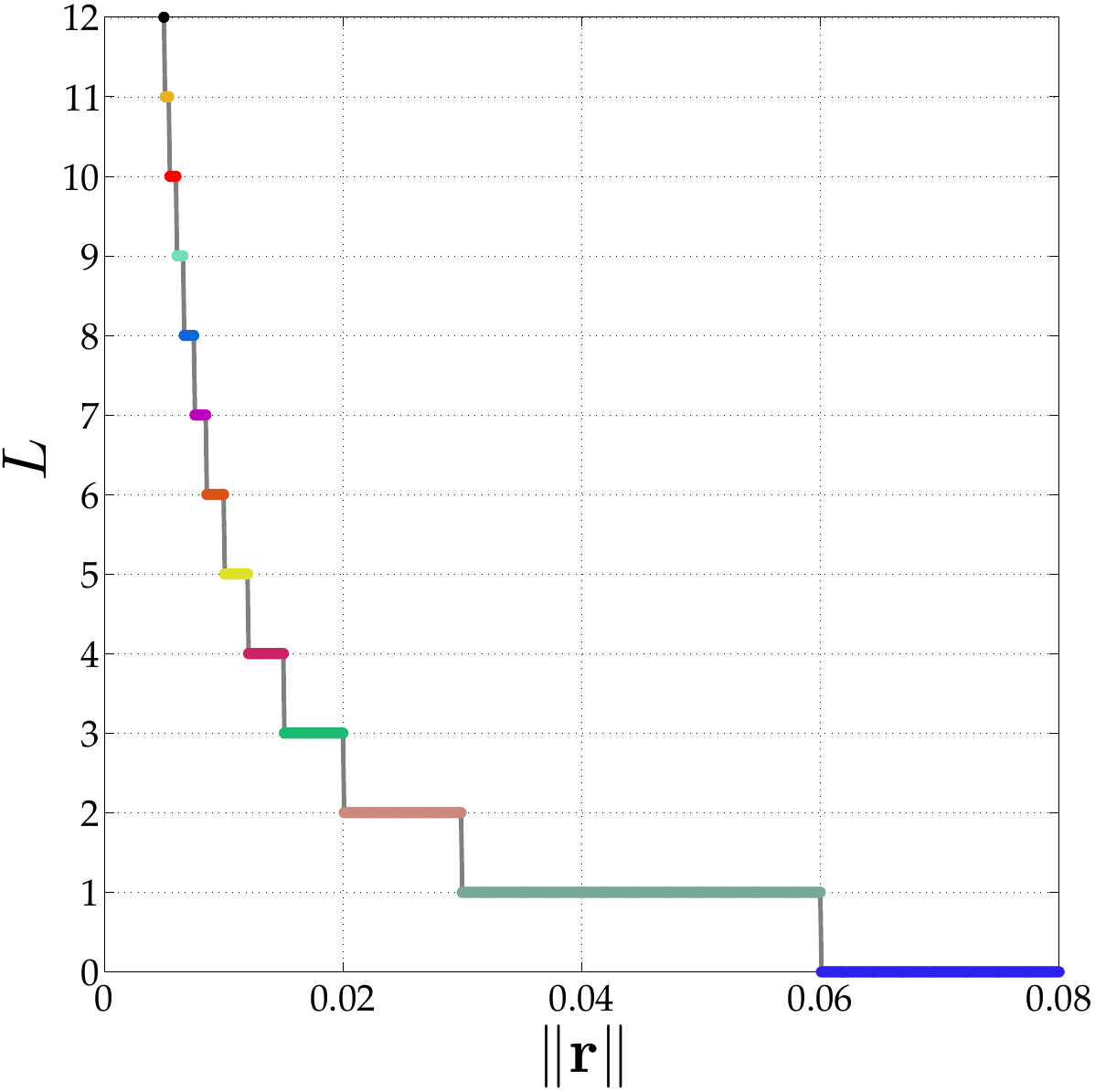}}}
      \label{f:LayersDist}}
\renewcommand*{\thesubfigure}{(a-2)} 
      \hspace{0cm}\subfigure[Repulsion force vs. distance]{\scalebox{0.6}{{\includegraphics{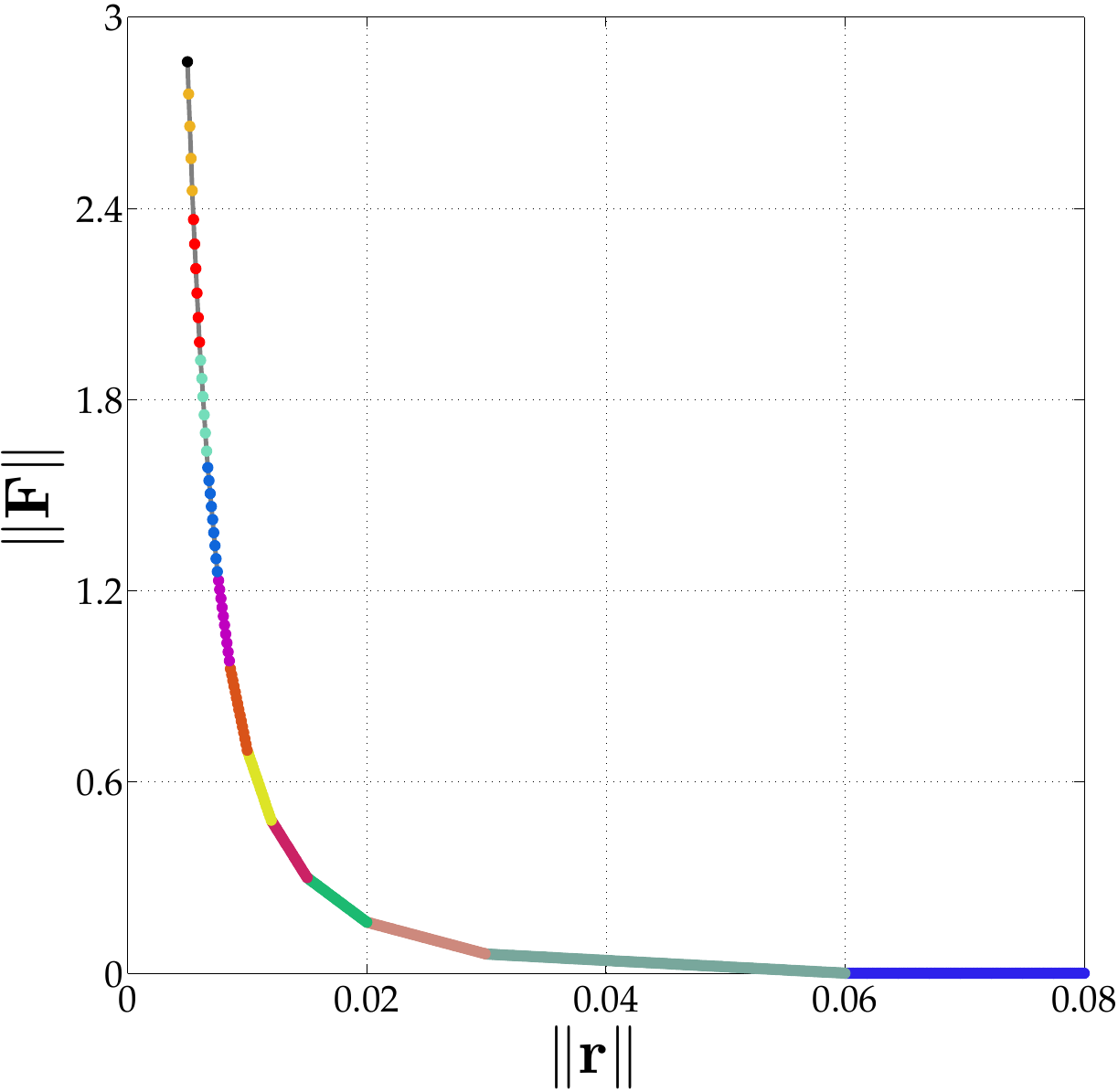}}}	
      \label{f:ForceDist}}
\end{center}
\mcaption{Here we take two approaching points and compute the total
number of activated discrete layers $L$ and the total penalty force
${\mathbf{F}}$ (see~\eqref{e:totalPenaltyForce}). We choose $d_{\min} =
0.06$ and $W = 1$.  We show each $L$ on the left and magnitude of the
corresponding total penalty force $\|{\mathbf{F}}\|$ on the right with
the same color.  The repulsion force increases as the points approach
each other.}{f:repulsionLayersForce}
\end{minipage}
\end{figure}

The total number of activated discrete layers, $L$, is the largest
integer less than $\frac{d_{\min}}{\|{\mathbf{r}}\|}$.  Hence, the
total penalty force on point $\xx \in {\gamma}_p$ due to point $\yy \in
{\gamma}_q$ is 
\begin{equation} \label{e:totalPenaltyForce}
{\mathbf{F}} = \sum_{\ell = 1}^L {\mathbf{F}}^\ell = W
\left(-\frac{L(L+1)(2L+1)}{3} + L(L+1) \frac{d_{\min}}{\|{\mathbf{r}}\|}\right){\mathbf{r}},
\quad L = \left\lfloor \frac{d_{\min}}{\|{\mathbf{r}}\|} \right\rfloor
\end{equation}
In \figref{f:repulsionLayersForce} we plot the total number of discrete
layers activated $L$ and the total penalty force of two approaching
points. We show each $L$ in \figref{f:LayersDist} and the corresponding
total penalty force ${\mathbf{F}}$ in \figref{f:ForceDist} with the
same color.  As the points approach to each other, the number of
activated layers $L$ increases and the color of the curves showing $L$
and $\mathbf{F}$ simultaneously change.   Finally, the repulsion force
at a point $\xx \in {\gamma}_p$ due to all other vesicles is formed by
summing~\eqref{e:totalPenaltyForce} over all discretization points $\yy
\notin \gamma_{p}$.

We treat the repulsion force explicitly.
That is, single layer potentials of the repulsion forces are computed
and placed on the right hand side of the linear system.  That can
introduce stiffness when the vesicles suddenly come too close. 

\paragraph*{Remark} In order to choose the repulsion length scale
$d_{\min}$, we place two vesicles of reduced area 0.65 symmetrically
about the origin in an extensional flow.  This simulation is done at a
low resolution with $N=12$ points.  We examine the energy in the six
lowest frequencies relative to the total energy of the vesicles'
velocities.  This ratio is used to heuristically set $d_{\min}$.  In
this example, when the vesicle separation is less than $0.3 h_{\max}$,
this ratio drops significantly and high-frequency  components appear.
Therefore, we set the repulsion length scale to $d_{\min} =
0.3h_{\max}$ implying that the minimum distance between two points on
the vesicles is  $d_{\min} = 0.3h_{\max}$ after which the repulsion
force is non-zero. However, in our experiments we observed that the
vesicles got so close that an imminent collision required very small
time step sizes in some cases and $d_{\min} = 0.5h_{\max}$ performs
better in those cases. Therefore, we set the repulsion length scale to
$d_{\min} = 0.5h_{\max}$ and never adjusted it again.  We set the
repulsion strength $W$ so that the velocity induced by the repulsion force is
$10\%$ of the velocity due to all hydrodynamic forces in the example above.  
While vesicles can approach one another in various ways, this example
represents one of the worse case scenarios since the proximity between
the vesicles decreases for all time, and we have successfully used
this length scale parameter for all of our experiments in
\secref{s:experiments}.

\section{Numerical experiments\label{s:experiments}} 
In this section, we demonstrate with various examples that the
low-resolution correction algorithms (LRCA) introduced in \secref{s:algos} 
are necessary to maintain
stability and to increase accuracy at low spatial resolution.  We
discuss the accuracy of the low-resolution simulations in terms of
different error measures. The error measures are discussed in
\secref{s:errorMeasures}, and a summary of the numerical experiments are
in \secref{s:experimentsSum}.

\subsection{Error measures}\label{s:errorMeasures} 
We examine the convergence of the method to a ground
truth, where the ground truth is formed at high resolutions with small
error tolerances. We also report self-error of a low-resolution 
simulation with respect to another low but higher resolution simulation. 
We denote the error with respect to a ground truth with $\epsilon^{\mathrm{g}}$ 
and the self-error with $\epsilon^{\mathrm{s}}$. In our previous 
work~\cite{rahimian-biros-e10, shravan-zorin-e11, quaife-biros14,
quaife-biros16}, we use the errors in the area and length of the
vesicles to measure the accuracy.  However, since we correct the area
and length of vesicles at each time step, this error measure becomes
obsolete.  We present two new sets of error measures, one for dilute
suspensions and one for dense suspensions. 

For dilute suspensions, we are interested in the accuracy of the vesicle
configuration. We summarize the error measures
for dilute suspensions in \tabref{t:errorMeasDilute}, and then discuss
details of each measure.
\begin{table}[H]
\mcaption{List of error measures for dilute suspensions.}{t:errorMeasDilute}
\centering
\begin{tabular}{l | l | l}
 \hline 
 Symbol &  Definition & Formulation\\ 
 \hline
  ${\epsilon}_{\mathrm{center}}$ & Error in the center of a vesicle &
  \eqref{e:errorInCenter}\\

  ${\epsilon}_{\mathrm{prox}}$ & Error in the proximity of two vesicles
  & \eqref{e:errorInProx}\\

  ${\epsilon}_{\mathrm{IA}}$ & Error in the inclination angle of a
  vesicle & \eqref{e:errorInIA} \\

  \hline 
\end{tabular} 
\end{table} 
Let $\hat{\mathbf{x}}_{k}$ and $\mathbf{x}_{k}$, $k=1,\ldots,m$, denote
the position of $m$ vesicles formed with a high-resolution simulation
(either ground truth or another low-resolution simulation) and with a low-resolution simulation, respectively.  If
$\hat{\mathbf{c}}_{k}$ and $\mathbf{c}_{k}$ are the centers of the
vesicles, then the error over all time of the center of vesicle $k$,
and the maximum of this error over all vesicles are
\begin{equation} \label{e:errorInCenter}
{\epsilon}_{\mathrm{center}}^{k} = \max_{\substack{t \in [0,T]}}\begin{cases}
      \frac{\|{\mathbf{c}}_k - {\hat{\mathbf{c}}}_k\|_2}{\varepsilon} , & \text{if}\, \|{\hat{\mathbf{c}}}_k\|_2 = 0  \\
      \frac{\|{\mathbf{c}}_k - {\hat{\mathbf{c}}}_k\|_2}{\|{\hat{\mathbf{c}}}_k\|_2} , & \text{otherwise}
      \end{cases}, \quad {\epsilon}_{\mathrm{center}} =
      \max_{\substack{k = 1,\ldots,m}} {\epsilon}_{\mathrm{center}}^k,
\end{equation}
where $\varepsilon$ is MATLAB's floating point relative accuracy
$\mathtt{eps}$.

The error in proximity is used for examples with two vesicles.  Letting
$\mathbf{d}={\mathbf{c}}_{1} - {\mathbf{c}}_{2}$ and $\hat{\mathbf{d}}
= \hat{\mathbf{c}}_{1} - \hat{\mathbf{c}}_{2}$, the error in proximity
of the two vesicles is
\begin{equation} \label{e:errorInProx}
  {\epsilon}_{\mathrm{prox}} = \max_{\substack{t \in [0,T]}}
  \frac{\|\mathbf{d} - \hat{\mathbf{d}}\|_2}{\|\hat{\mathbf{d}}\|_2}.
\end{equation}

The inclination angle ($\mathrm{IA}$) is the angle between the $x$-axis
and the principal axis corresponding to the smallest principal moment of
inertia~\cite{rahimian-biros-e10}.  The moment of inertia tensor is
\begin{equation*} \label{e:momInertiaTensor}
J = \int_{\omega} \left(|\mathbf{r}|^2 I - \mathbf{r} \otimes
\mathbf{r}\right)d\mathbf{x} = \frac{1}{4} \int_{\gamma}
\mathbf{r}\cdot\mathbf{n}\left(|\mathbf{r}|^2 I - \mathbf{r} \otimes
\mathbf{r}\right)ds,
\end{equation*}
where $\mathbf{r} = \xx - \mathbf{c}$, and $\mathbf{c}$ is the center of
the vesicle. Then the error over all time of the inclination angle of
the $k^{th}$ vesicle, and the maximum over all
vesicles are
\begin{equation} \label{e:errorInIA}
{\epsilon}_{\mathrm{IA}}^{k} = \max_{\substack{t \in [0,T]}}\begin{cases}
      \frac{|{\mathrm{IA}}_k - {\widehat{\mathrm{IA}}}_k |}{\varepsilon} , & \text{if}\, |{\widehat{\mathrm{IA}}}_k | = 0  \\
      \frac{|{\mathrm{IA}}_k - {\widehat{\mathrm{IA}}}_k |}{|{\widehat{\mathrm{IA}}}_k| } , & \text{otherwise}
      \end{cases}, \quad {\epsilon}_{\mathrm{IA}} =
      \max_{\substack{k=1,\ldots,m
      }} {\epsilon}_{\mathrm{IA}}^k.
\end{equation}

For dense suspensions, the error in the vesicles' configurations at
low resolutions is large and irrelevant.  However, depending on the
purpose of the simulation, low-resolution simulations can provide
significant information with a considerably low computational cost.  We
consider upscaling measures such as errors in statistics and space-time
averages of physical quantities.  A list of error measures for dense
suspensions are in \autoref{t:errorMeasDense}.
\begin{table}[H]
\mcaption{List of error measures for dense suspensions.}{t:errorMeasDense}
\centering
\begin{tabular}{l | l | l}
 \hline 
 Symbol &  Definition & Formulation \\ 
 \hline
${\epsilon}_{\langle v \rangle}$ & Error in the space-time average of a
velocity field & \eqref{e:spaceAveVel},~\eqref{e:errorInSpaceTime}\\

${\epsilon}_{\langle V \rangle}$ & Error in the time average of the $L^2$
norm of a velocity field &
\eqref{e:L2normVel},~\eqref{e:errorInL2normVel}\\

${\epsilon}_{{\mu}_{\mathrm{eff}}}$ & Error in the effective viscosity
of a suspension &
\eqref{e:effectViscosity},~\eqref{e:perturbedSigma},~\eqref{e:errorInEffVisc}\\
 \hline 
\end{tabular} 
\end{table} 

The velocity field of the fluid bulk plays an important role in many
applications.  For instance, in~\cite{kabacaoglu-biros-e16} we study
mixing in a Couette apparatus containing vesicles (see
\figref{f:ICsCouette}).  We model transport with an advection-diffusion
equation, so capturing the correct averages of the velocity field is
crucial.  We consider the error in space-time average of the velocity
field and the error in time average of the $L^2$ norm of the velocity
field.

The space and space-time averages of a velocity field
$\mathbf{V}(\xx,t)$ are 
\begin{equation} \label{e:spaceAveVel}
v(t) = \frac{1}{| \Omega |} \int_\Omega\mathbf{V}(\xx,t) d\xx, \quad \langle v \rangle = \frac{1}{T} \int_0^T v(t) dt,
\end{equation}
respectively. Letting $\langle \hat{v} \rangle$ and $\langle v \rangle$
denote the space-time averages of velocity fields given by a high-resolution
simulation and its corresponding low-resolution simulations, the error is
\begin{equation} \label{e:errorInSpaceTime}
{\epsilon}_{\langle v \rangle} = \frac{|\langle v \rangle  - \langle \hat{v} \rangle|}{|\langle \hat{v} \rangle|}.
\end{equation}
Additionally, the $L^2$ norm of the velocity field and the time average
of this quantity are 
\begin{equation} \label{e:L2normVel}
V(t) = \frac{1}{| \Omega |} \left(\int_\Omega {\mathbf{V}}^2(\xx,t)
d\xx\right)^{\frac{1}{2}}, \quad \langle V \rangle = \frac{1}{T} \int_0^T V(t) dt,
\end{equation}
respectively.  The error in the time average of the $L^2$ norm of a
velocity field is
\begin{equation} \label{e:errorInL2normVel}
{\epsilon}_{\langle V \rangle} = \frac{|\langle V \rangle  - \langle \hat{V} \rangle|}{|\langle \hat{V} \rangle|}. 
\end{equation}

Another error measure is based on a numerical homogenization for
suspension rheology. The effective viscosity of a suspension is the
viscosity of a homogeneous Newtonian fluid having the same energy
dissipation per macroscopic volume element.  For vesicle suspensions,
it is given by~\cite{rahimian-biros-e10}
\begin{equation} \label{e:effectViscosity}
{\mu}_{\mathrm{eff}} = {\mu}_0 + \phi \frac{1}{T} \int_0^T {\overline{\sigma}}_{12}^p dt
\end{equation}
where
\begin{equation} \label{e:perturbedSigma}
{\overline{\sigma}}^p = \frac{1}{|\Omega|} \int_{\gamma}
\left[-\left({\kappa}_b {\kappa}^2 \mathbf{n} \otimes \mathbf{n} +
\sigma \mathbf{t} \otimes \mathbf{t} \right) + {\mu}_0 (\nu - 1)
\left(\mathbf{u}\otimes\mathbf{n} +
\mathbf{n}\otimes\mathbf{u}\right)\right]ds.
\end{equation}
Here, ${\mu}_0$ is viscosity of the bulk fluid, $\phi$ is the volume
fraction of vesicles, ${\overline{\sigma}}^p$ is the spatial average of the
perturbation in stress $\sigma$ due to the presence of vesicles,
${\kappa}_b$ is the bending stiffness, ${\kappa}$ is the curvature,
$\mathbf{n}$, $\mathbf{t}$ are the unit normal and tangent vectors, and
$\mathbf{u}$ is the velocity.  Letting ${\hat{\mu}}_{\mathrm{eff}}$ and
${\mu}_{\mathrm{eff}}$ be effective viscosities of a suspension obtained
from a high- and a low-resolution simulation, the error in
effective viscosity of a suspension is
\begin{equation} \label{e:errorInEffVisc}
{\epsilon}_{{\mu}_{\mathrm{eff}}} =
\frac{|{{\mu}}_{\mathrm{eff}}  -
{\hat{\mu}}_{\mathrm{eff}}|}{|{\hat{\mu}}_{\mathrm{eff}}|}. 
\end{equation}

For dense suspensions in a Couette apparatus, we also report 
probability distribution functions of the location of each vesicle's
center and the magnitude of the velocity at certain radii.

We report the {\em self-convergence} of the solutions within the low-
resolution simulations in addition to the convergence to a ground
truth solution. The self-convergence is useful to estimate the 
accuracy of a low-resolution
simulation  when a ground truth solution is not
available. In the following sections, the self-errors $\epsilon^{\mathrm{s}}$ 
are reported in a way that they are computed with respect to the simulation which is
reported in one row below on the same table. 

\subsection{Summary of numerical experiments}\label{s:experimentsSum}
We perform numerical experiments of both dilute and dense vesicle 
suspensions in bounded and unbounded domains.  We use our adaptive time
stepping in all runs except when forming the ground truth.  Then, we
compare the simulations with and without the {\em LRCA} 
introduced in \secref{s:algos}.  We
report timings and the (self-) errors defined in Tables~\ref{t:errorMeasDilute}
and~\ref{t:errorMeasDense}.  A simulation is stopped if it takes orders of 
magnitude more computing time than the other simulations of the same example 
with different resolutions. 
The examples we consider are:

\begin{itemize}
\item \textbf{Two vesicles in a shear flow} (\secref{s:shear}): We
simulate a pair of vesicles with viscosity contrasts $\nu = 1$ and $\nu
= 10$ and reduced areas (RA) $0.65$ and $0.99$.  The initial
configurations result in the vesicles nearly touching. The purpose 
of these experiments is to demonstrate errors in average quantities such 
as the proximity between vesicles and the actual trajectories of the vesicles.

\item \textbf{One vesicle in a stenosis flow} (\secref{s:stenosis}): We
simulate a single vesicle of reduced area $0.65$ and without viscosity
contrast $\nu = 1$ in a constricted tube (stenosis) with a parabolic
flow profile at the intake and the outtake. In these experiments, 
the vesicle's initial height is 3.5 times larger than the constriction size. 
As a result of that it highly deforms and gets close to the tube's boundary as 
it passes the constriction. Here, we show that the LRCA are essential to 
avoid the vesicle-solid boundary collisions.

\item \textbf{Four vesicles in a Taylor-Green flow}(\secref{s:taylorGreen}): 
We simulate four vesicles of reduced area
$0.65$ with viscosity contrasts of $\nu = 1$ and $\nu = 10$ in a
periodic Taylor-Green flow.  The vesicles cover approximately 50\% of
the area of a periodic cell $(0,\pi)^2$ making vesicle interactions
stronger and the problem more complicated than the previous examples. 
Here we demonstrate that although the simulations do not converge 
in terms of the local error measures 
such as ${\epsilon}_{\mathrm{center}}$, 
the convergence in the upscaling measures can be achieved at low resolutions.

\item \textbf{Couette apparatus} (\secref{s:couette}):  We simulate
vesicles of reduced area $0.65$ without viscosity contrast in a Couette
apparatus.  Simulations with volume fractions $\phi = 20\%$ and $\phi =
40\%$ are performed.  For these examples, we report errors in the upscaled
quantities (see \tabref{t:errorMeasDense}) and statistics. Similar to the 
experiment with a Taylor-Green flow, many vesicle interactions result in 
large local errors. However, the low-resolution simulations are 100$\times$ 
faster while capturing the upscaled quantitites and statistics accurately.

\item \textbf{Microfluidic device} (\secref{s:dld}): We simulate 
the separation of a healthy red blood cell (RBC) in a microfluidic device 
using deterministic lateral displacement (DLD) technique \cite{huang-sturm-e04}. 
The device we consider here leads the RBC to show no net lateral displacement, 
which is confirmed by the actual and numerical experiments 
\cite{beech-tegenfeldt-e12,kruger-coveney-e14}. The purpose of 
this experiment is to show the ability of our black-box solver to 
deliver the accurate physics using as coarse discretization as possible.

\end{itemize}

\paragraph*{Remark}  For all runs, we fix the bending stiffness to
${\kappa}_b = 10^{-1}$ and the GMRES tolerance to
${\rho}_{\text{GMRES}} = 10^{-10}$. Ground truth solutions computed
with the high-fidelity version of the code are illustrated as grey
vesicles. Additionally, since we use our adaptive time stepping scheme,
simulations are compared at different, but comparable times.

\subsection{Shear flow}\label{s:shear}
\begin{figure}[H]
\begin{minipage}{\textwidth}
  \begin{minipage}[t]{0.55\textwidth}
    \centering
    \setcounter{subfigure}{0}
    \renewcommand*{\thesubfigure}{(a-1)} 
          \hspace{0cm}\subfigure[RA =
          0.65]{\scalebox{0.25}{{\includegraphics{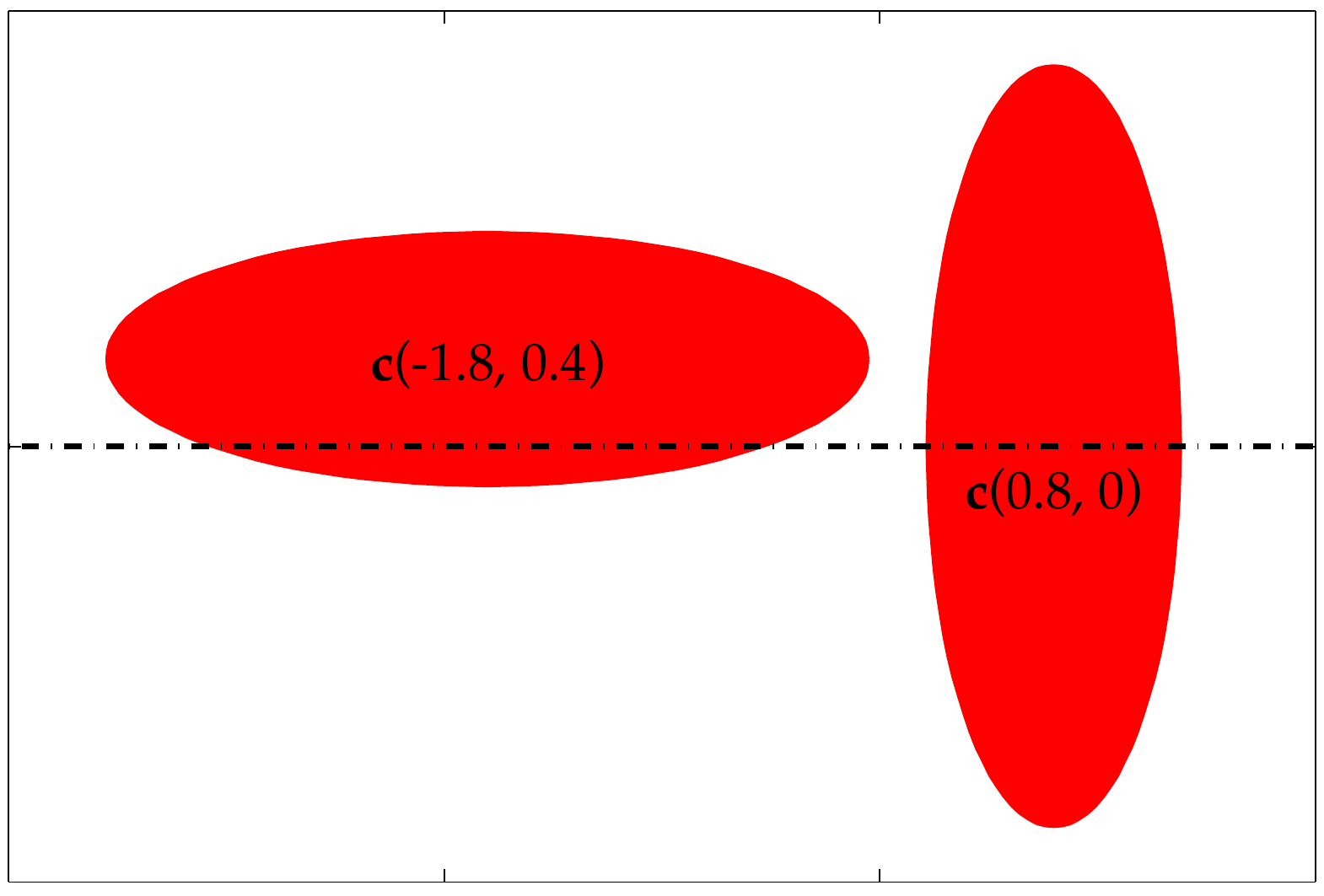}}}	
          \label{f:ICshearRA65}}
    \setcounter{subfigure}{0}
    \renewcommand*{\thesubfigure}{(a-2)} 
          \hspace{0cm}\subfigure[RA =
          0.99]{\scalebox{0.25}{{\includegraphics{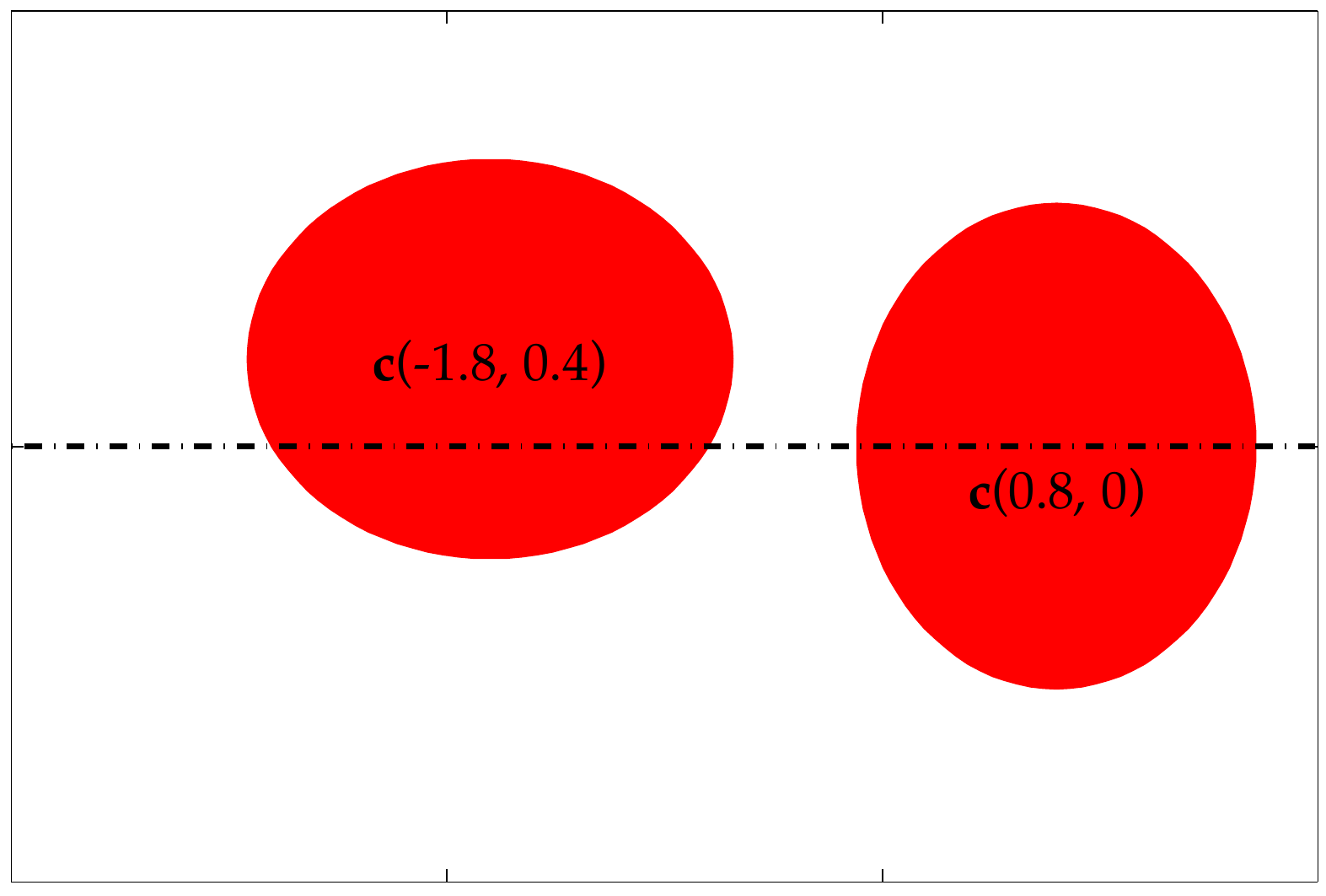}}}
          \label{f:ICshearRA99}}
    \mcaption{The initial configurations of two vesicles in an unbounded
    {\bf shear flow} $\uu = (y,0)$.  The left plot has vesicles of
    reduced area $0.65$ and the right plot has vesicles of reduced area
    $0.99$.  Both these simulations are run with viscosity contrasts
    $\nu = 1$ and $\nu = 10$.}{f:ICsShear}
  \end{minipage}
  \hfill
  \begin{minipage}[t]{0.46\textwidth}
    \centering
    \begin{tabular}{cc}\hline
        Parameter & Value \\ \hline
        Points on a vesicle $N$ & 96 \\
        Viscosity contrast $\nu$ & $\{1, 10\}$\\
        Number of SDC sweeps $n_{\text{sdc}}$ & 1 \\ 
        Time step size $\Delta t$ & $5 \times 10^{-4}$ \\ 
        CPU time ($\mathrm{RA}=0.65$, $\nu=1$) & 17 hours \\ 
        CPU time ($\mathrm{RA}=0.65$, $\nu=10$) & 61 hours \\ 
        CPU time ($\mathrm{RA}=0.99$, $\nu=1$) & 16 hours \\ 
        CPU time ($\mathrm{RA}=0.99$, $\nu=10$) & 36 hours \\ \hline
      \end{tabular}
      \captionof{table}{\textit{Parameters of the ground truth of a
      {\bf shear flow}}.}
      \label{t:shearGTparams}
    \end{minipage}
\end{minipage}
\end{figure}
\paragraph*{\textbf{Setup}} We consider two vesicles in an unbounded
shear flow $\mathbf{u} = (y,0)$. The initial configuration
(\figref{f:ICsShear}) results in the left vesicle traveling to the
right and over top of the right vesicle.  We consider reduced areas
$0.65$ and $0.99$ and viscosity contrasts $\nu=1$ and $\nu=10$.  We
simulate each of these cases with $N = 12, 16, 24, 32$ points per
vesicle and an error tolerance $\rho_{\mathrm{AL}} =10^{-2}, 10^{-3},
10^{-4}, 10^{-5}$ with and without the algorithms in \secref{s:algos}.
The time horizon is $T = 20$ so that the vesicles pass one another.  The
ground truth solutions are formed with the parameters in
\tabref{t:shearGTparams}.

\paragraph*{\textbf{Results}} We investigate the necessity of the LRCA to
maintain stability and we quantify their effect on the error in the
proximity of the vesicles, ${\epsilon}_{\mathrm{prox}}$.  This problem
is particularly difficult because the hydrodynamic force is inaccurate
at low resolutions, and this can cause vesicles to collide.  We report
the (self-) errors in proximity, the number of accepted and rejected time
steps, and the CPU times. 

In \tabref{t:ShearRA65VC1Errors}, we summarize the simulations of two
vesicles of reduce area $0.65$ with $\nu=1$ (top) and $\nu=10$
(bottom).  For almost all the simulations, the LRCA are not necessary
to maintain stability.  However, the error in the proximity of the
vesicles is decreased when the LRCA are used for all runs except $N =
32$ with $\nu=1$, and for the two highest resolutions with $\nu=10$.
In  these cases where the simulations with the LRCA  have greater
errors in the proximity than the original simulations, the  resolution
is sufficient for stability without the LRCA and  the effects of the
LRCA do not vanish yet, i.e. the decay of the repulsion length scale does not let the effects of the repulsion vanish yet at those high resolutions. That's why, the original simulations  are more
accurate than the ones with the LRCA. In addition, as expected, the
CPU time is increased when the algorithms are used, but the payoff is
additional stability and accuracy in almost all the examples. We also
increase the temporal resolution while keeping the spatial resolution
the same for $N = 12$ and $N = 16$. By doing so, the errors in the
proximity become less than the ones delivered by the two highest
spatial resolutions in shorter CPU times. However, lowering the
tolerances at the coarse spatial resolutions might significantly
increase the number of time steps taken and hence the CPU time because
it requires small time steps to keep the errors in area and
length below those low tolerances. Therefore, it
is not always efficient to refine the temporal resolution only. For
example, in the shear flow of two vesicles with RA = 0.65 and $\nu =
1$ decreasing the tolerance from $\rho_{\mathrm{AL}} = $ 1E-2 to
$\rho_{\mathrm{AL}} = $ 1E-3 with $N = 12$ leads to a sixfold increase
in the CPU time (see \tabref{t:ShearRA65VC1Errors}). Yet increasing
the spatial resolution from $N = 12$ to $N = 16$ while decreasing the
tolerance only triples the CPU time and results in a smaller error.

\figref{f:shearFramesRA65VC1} shows
snapshots of the simulation without viscosity contrast at one
resolution  both with and without our algorithms, and it is clear that
the LRCA are necessary to maintain physical vesicle shapes.  In
\figref{f:shearFramesRA65VC10} the two vesicles with $\nu=10$ are
illustrated at three different resolutions with the LRCA.  Here we see
convergence towards the ground truth and self-convergence within the 
low-resolution simulations when the spatio-temporal resolution is increased.

\begin{table}[H]
\mcaption{The (self-) errors in the proximity of two vesicles of RA = $0.65$
with viscosity contrast $\nu = 1$ (top) and $\nu=10$ (bottom) in a {\bf
shear flow} with and without the LRCA in \secref{s:algos} (see
Figures~\ref{f:shearFramesRA65VC1} and~\ref{f:shearFramesRA65VC10}).
The self-errors are computed with respect to the simulation in one row below. 
Also reported are the number of accepted and rejected time steps and
the CPU time. The dash "-" is put on the table for the simulations which break without the LRCA because the vesicle collisions cannot be handled. 
The ground truth simulations of $\nu = 1$ and $\nu = 10$ take 17 and 
61 hours, respectively. Both simulations have $N = 96$ and $\Delta t = $ 5E-4.}{t:ShearRA65VC1Errors}
\centering
\begin{tabular}{c c| c c c c c | c c c c}
 \multicolumn{11}{c}{$\boldsymbol{\nu=1}$} \\
 \hline
 \hline
 & & \multicolumn{5}{c|}{LRCA} &
 \multicolumn{4}{c}{Original} \\
\hline
$N$ & ${\rho}_{\mathrm{AL}}$ &  ${\epsilon}_{\mathrm{prox}}^{\mathrm{g}}$ & 
${\epsilon}_{\mathrm{prox}}^{\mathrm{s}}$ & Accepts & Rejects
& Time (sec) &  ${\epsilon}_{\mathrm{prox}}^{\mathrm{g}}$ & Accepts & Rejects & Time (sec) \\ 
 \hline
 
12 & 1E-2 & 1.2E-1 & 4.2E-2 & 94 & 4  & 64 & 2.5E-1 & 128 & 10  & 58 \\
 
16 & 1E-3 & 3.7E-2 & 4.7E-2 & 310 & 7  & 193  & 2.0E-1 & 345 & 14  & 132 \\
 
24 & 1E-4 & 4.1E-2 & 1.5E-2 & 998 & 11  & 826  & 4.1E-2 & 1026 & 14  & 400  \\
 
32 & 1E-5 & 2.8E-2 &        & 3156 & 15  & 1930 & 9.1E-3 & 3174 & 15  & 808  \\
\hline
12 & 1E-2 & 1.2E-1 & 3.3E-2 & 94 & 4  & 64 & 2.5E-1 & 128 & 10  & 58 \\
 
12 & 1E-3 & 8.0E-2 & 6.8E-2 & 354 & 10  & 348  & 8.4E-2 & 761 & 10  & 335 \\
 
12 & 1E-4 & 1.5E-2 &        & 1165 & 15  & 1150  & 1.0E-2 & 5226 & 16  & 1880  \\
 \hline
 16 & 1E-3 & 3.7E-2 & 3.5E-2 & 310 & 7  & 193  & 2.0E-1 & 345 & 14  & 132 \\

16 & 1E-4 & 2.6E-2 &        & 955 & 5  & 921  & 1.9E-1 & 2086 & 21  & 889 \\
\end{tabular} 

\begin{tabular}{c c| c c c c c | c c c c}
\multicolumn{11}{c}{$\boldsymbol{\nu=10}$} \\
\hline
\hline
 & & \multicolumn{5}{c|}{LRCA} &
 \multicolumn{4}{c}{Original} \\
\hline
$N$ & ${\rho}_{\mathrm{AL}}$ &  ${\epsilon}_{\mathrm{prox}}^{\mathrm{g}}$ & 
${\epsilon}_{\mathrm{prox}}^{\mathrm{s}}$ & Accepts & Rejects
& Time (sec) &  ${\epsilon}_{\mathrm{prox}}^{\mathrm{g}}$ & Accepts & Rejects & Time (sec) \\ 
 \hline
 
 12 & 1E-2 & 3.9E+0 & 1.9E+0 & 93 & 9  & 60 & 3.0E+0 & 98 & 11 & 41.4 \\
 
 16 & 1E-3 & 1.1E+0 & 2.1E-1 & 227 & 19  & 205 & - & - & - & -\\
 
 24 & 1E-4 & 3.1E-1 & 7.0E-1 & 786 & 32  & 843 & 1.6E-1 & 773 & 31 & 438  \\
 
 32 & 1E-5 & 1.4E-1 &       & 2567 & 37  & 2660  & 3.7E-2 & 2480 & 35 & 1120\\
 \hline
 12 & 1E-2 & 3.9E+0 & 2.2E+0 & 93 & 9  & 60 & 3.0E+0 & 98 & 11 & 41.4 \\

 12 & 1E-3 & 6.4E-1 & 7.3E-1 & 274 & 11  & 366 & - & - & - & - \\

 12 & 1E-4 & 7.4E-2 &  & 844 & 12  & 1220 & - & - & - & - \\
\hline
 16 & 1E-3 & 1.1E+0 & 2.1E-1 & 227 & 19  & 205 & - & - & - & -\\

 16 & 1E-4 & 4.7E-1 &  & 789 & 11  & 1530 & - & - & - & -\\
\end{tabular}

\end{table} 

\begin{figure}[H]
\begin{center}
\includegraphics[width=\textwidth]{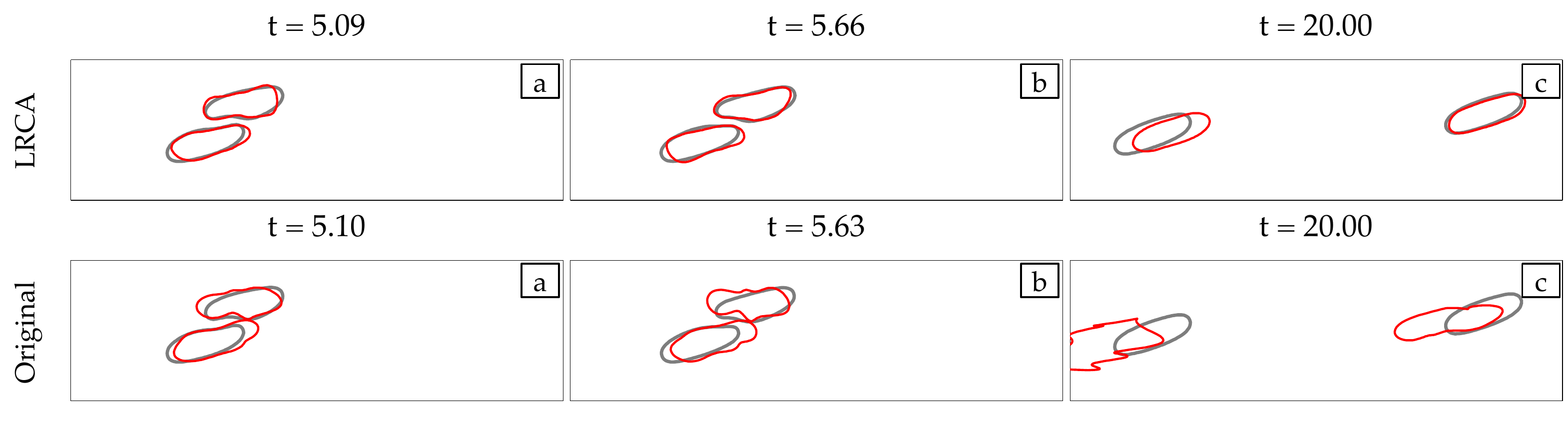}
\mcaption{Two vesicles with reduced area $0.65$, viscosity contrast
$\nu = 1$, and discretized with $N=12$ points in a {\bf shear flow}.
The error tolerance is $\rho_{\mathrm{AL}}=10^{-2}$ and the grey
vesicles are the ground truth. In the top row, the LRCA 
are used, and in the bottom row, they are not.
The error metric we are using seems to be underestimating the error.
Although the error in proximity can be considered reasonable,  
the original simulation has non-physical vesicles.}{f:shearFramesRA65VC1}
\end{center}
\end{figure}

\begin{figure}[H]
\begin{center}
\includegraphics[width=\textwidth]{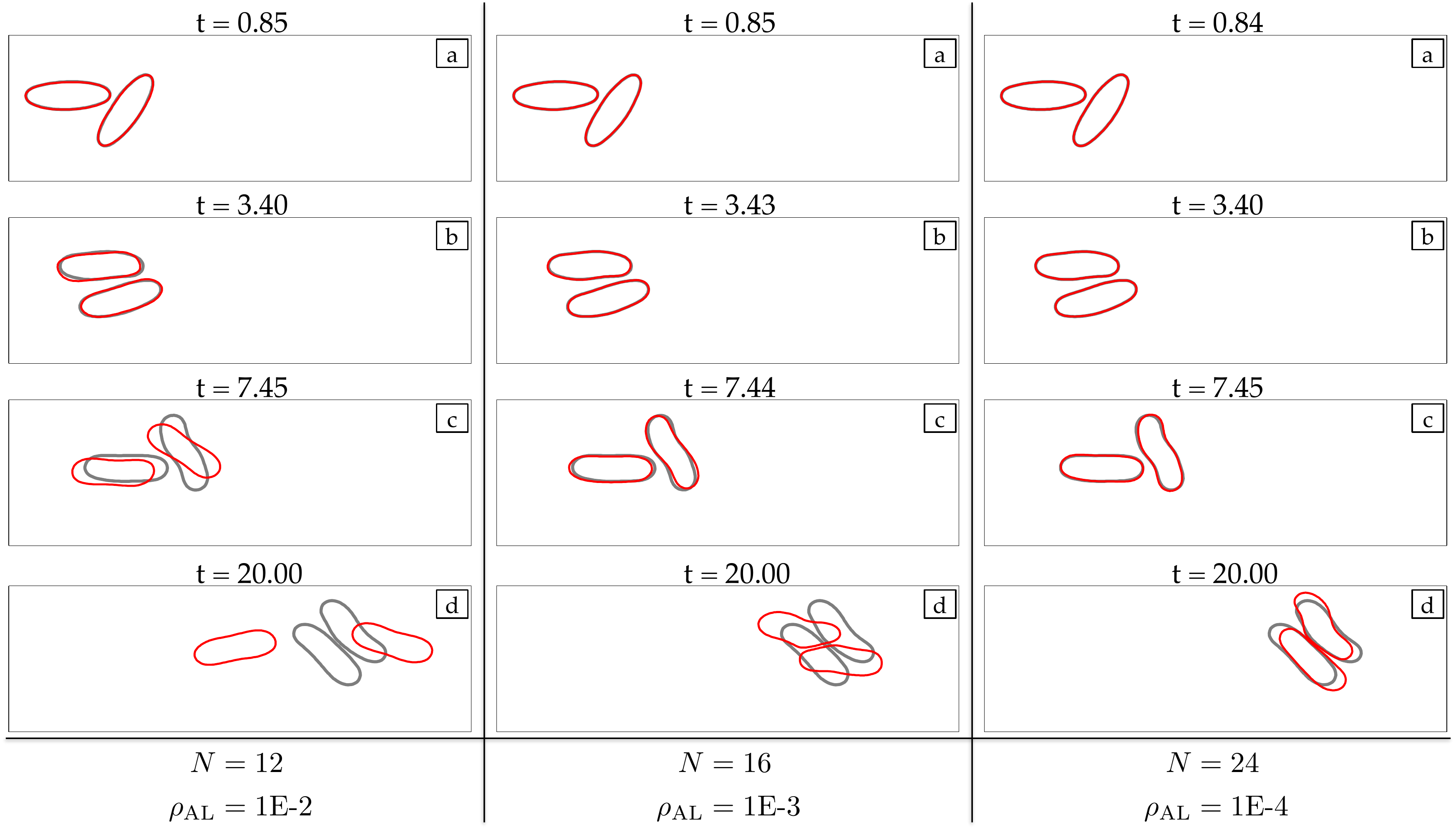}
\mcaption{Two vesicles with RA = 0.65 and ${\nu = 10}$ in a {\bf shear
flow} using the LRCA.  Vesicles are
discretized with $N = 12, 16, 24$ points and the error tolerances in
area and length are ${\rho}_{\mathrm{AL}} = 10^{-2}, 10^{-3}, 10^{-4}$,
respectively. The grey vesicles are from the ground
truth. Here, we show that as the resolution increases, the low-resolution 
simulations converge among themselves and to the ground truth.}{f:shearFramesRA65VC10}
\end{center}
\end{figure}

Finally, we present results for the vesicles of reduced area $0.99$
with the two different viscosity contrasts in
\tabref{t:ShearRA99VC1Errors}.  Here, vesicles do not come as close as
those of reduced area 0.65.  At all the resolutions we consider, not
using the LRCA delivers more accurate results in terms of the vesicles'
proximity with less CPU time.  However if the algorithms are not used
at the resolutions $N \leq 16$, the errors in area and length of the
vesicles are $\mathcal{O}(10^{-1})$.  This leads the vesicles to have
non-physical shapes at the time horizon (see
\figref{f:shearFramesRA99VC1} and \figref{f:shearFramesRA99VC10}).
\begin{table}[H]
\mcaption{The (self-) errors in the proximity of two vesicles of RA = $0.99$
with viscosity contrast $\nu = 1$ (top) and $\nu = 10$ (bottom) in a
{\bf shear flow} with and without the LRCA in \secref{s:algos} (see
\figref{f:shearFramesRA99VC1} and \figref{f:shearFramesRA99VC10} for
frames of the simulations at the coarsest resolution).  The self-errors are computed with respect to the simulation in one row below. 
Also reported are the number of accepted and rejected time steps and
the CPU time. The ground truth simulations of $\nu = 1$ and $\nu = 10$ take 16 and 36 hours, respectively. Both simulations have $N = 96$ and $\Delta t = $ 5E-4.}{t:ShearRA99VC1Errors}
\centering
\begin{tabular}{c c| c c c c c | c c c c}
\multicolumn{11}{c}{$\boldsymbol{\nu=1}$} \\
\hline
\hline
& & \multicolumn{5}{c|}{LRCA} &
\multicolumn{4}{c}{Original} \\
\hline
$N$ & ${\rho}_{\mathrm{AL}}$ &  ${\epsilon}_{\mathrm{prox}}^{\mathrm{g}}$ & 
${\epsilon}_{\mathrm{prox}}^{\mathrm{s}}$ & Accepts & Rejects
& Time (sec) &  ${\epsilon}_{\mathrm{prox}}^{\mathrm{g}}$ & Accepts & Rejects & Time (sec) \\ 
 \hline
 
 12 & 1E-2 & 5.3E-1 & 1.3E-1 & 110 & 2  & 57 & 5.5E-1 & 110 & 3  & 38 \\
 
 16 & 1E-3 & 3.4E-1 & 2.6E-1 & 373 & 3  & 199 & 1.1E-1 & 343 & 6  & 134 \\
 
 24 & 1E-4 & 6.5E-2 & 4.5E-2 & 1192 & 9  & 699 & 1.5E-2 & 1191 & 9  & 446 \\
 
 32 & 1E-5 & 1.8E-2 &       & 3766 & 13  & 2300 & 4.8E-3 & 3767 & 13  & 862 \\
 \hline
 12 & 1E-2 & 5.3E-1 & 1.9E-1 & 110 & 2  & 57 & 5.5E-1 & 110 & 3  & 38 \\

 12 & 1E-3 & 2.8E-1 & 3.4E-1 & 333  & 3  & 193 & 3.8E-1 & 370 & 5  & 177 \\

 12 & 1E-4 & 4.3E-2 &   & 1187 & 6  & 805 & 3.1E-1 & 1592 & 19  & 599 \\
 \hline
 16 & 1E-3 & 3.4E-1 & 3.5E-1 & 373 & 3  & 199 & 1.1E-1 & 343 & 6  & 134 \\

 16 & 1E-4 & 1.1E-2 &   & 1092 & 5  & 852 & 1.2E-1 & 1217 & 6  & 509 \\
\end{tabular} 

\begin{tabular}{c c| c c c c c | c c c c}
\multicolumn{11}{c}{$\boldsymbol{\nu=10}$} \\
\hline
\hline
& & \multicolumn{5}{c|}{LRCA} &
\multicolumn{4}{c}{Original} \\
\hline
$N$ & ${\rho}_{\mathrm{AL}}$ &  ${\epsilon}_{\mathrm{prox}}^{\mathrm{g}}$ & 
${\epsilon}_{\mathrm{prox}}^{\mathrm{s}}$ & Accepts & Rejects
& Time (sec) &  ${\epsilon}_{\mathrm{prox}}^{\mathrm{g}}$ & Accepts & Rejects & Time (sec) \\ 
 \hline
 
 12 & 1E-2 & 6.8E-1 & 3.0E-1 & 99 & 3  & 52 & 3.8E-1 & 93 & 4 & 38\\
 
 16 & 1E-3 & 3.0E-1 & 1.6E-1 & 318 & 7  & 181 & 7.5E-2 & 318 & 7 & 127\\
 
 24 & 1E-4 & 1.2E-1 & 1.0E-1 & 1030 & 7  & 668 & 9.4E-3 & 1030 & 7 & 384\\
 
 32 & 1E-5 & 1.1E-2 &      & 3160 & 8  & 2250 & 3.2E-3 & 3160 & 8 & 972\\
 \hline
 12 & 1E-2 & 6.8E-1 & 4.2E-1 & 99 & 3  & 52 & 3.8E-1 & 93 & 4 & 38\\

 12 & 1E-3 & 1.8E-1 & 1.5E-1 & 328 & 3  & 272 & 2.8E-1 & 294 & 3 & 194\\

 12 & 1E-4 & 2.4E-2 &  & 956 & 4  & 921 & 2.4E-1 & 958 & 4 & 563\\
 \hline
 16 & 1E-3 & 3.0E-1 & 2.0E-1 & 318 & 7  & 181 & 7.5E-2 & 318 & 7 & 127\\

 16 & 1E-4 & 7.5E-2 &  & 955 & 4  & 1230 & 6.6E-2 & 955 & 4 & 580\\
\end{tabular}

\end{table} 

\begin{figure}[H]
\begin{center}
\includegraphics[width=\textwidth]{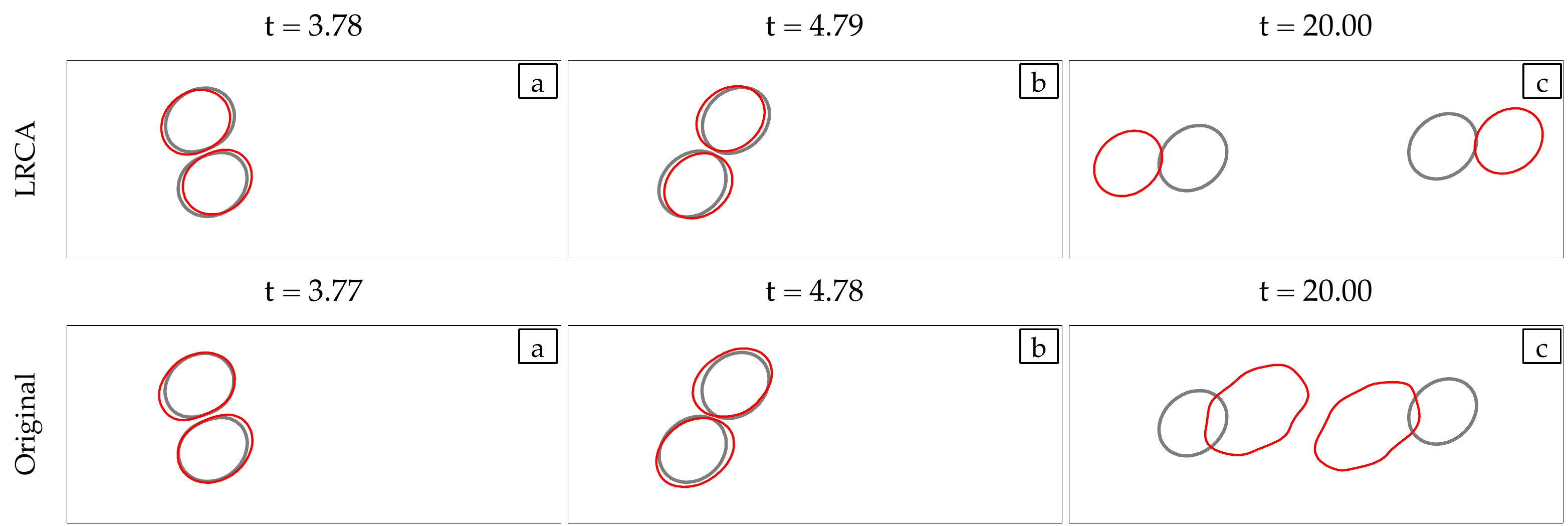}
\mcaption{Two vesicles with reduced area $0.99$, viscosity contrast
$\nu = 1$, and discretized with $N=12$ points in a {\bf shear flow}.
The error tolerance is $\rho_{\mathrm{AL}}=10^{-2}$ and the grey
vesicles are the ground truth. In the top row, the LRCA are used, 
and in the bottom row, they are not. Similar to simulations shown 
in \figref{f:shearFramesRA65VC1}, although the errors in the proximity 
are very close with and without the LRCA at these resolutions ($N = 12$ and 
${\rho}_{\mathrm{AL}} = 1\mathrm{E}-2$), the LRCA are necessary to maintain 
physical vesicle shapes.}{f:shearFramesRA99VC1}
\end{center}
\end{figure}

\begin{figure}[H]
\begin{center}
\includegraphics[width=\textwidth]{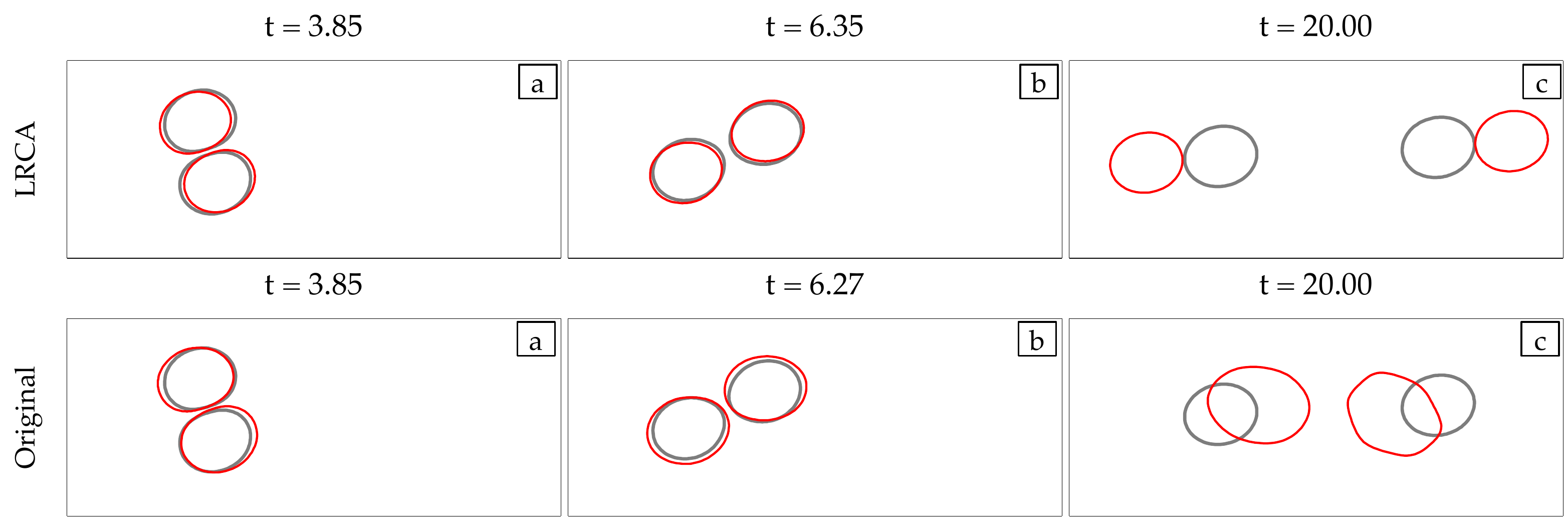}
\mcaption{Two vesicles with reduced area $0.99$, viscosity contrast
$\nu = 10$, and discretized with $N=12$ points in a {\bf shear flow}.
The error tolerance is $\rho_{\mathrm{AL}}=10^{-2}$ and the grey
vesicles are the ground truth. In the top row, the LRCA are used, 
and in the bottom row, they are
not.}{f:shearFramesRA99VC10}
\end{center}
\end{figure}

\subsection{Stenosis flow}\label{s:stenosis}
\begin{minipage}{\textwidth}
  \begin{minipage}[b]{0.49\textwidth}
    \centering
    \includegraphics[scale=0.4]{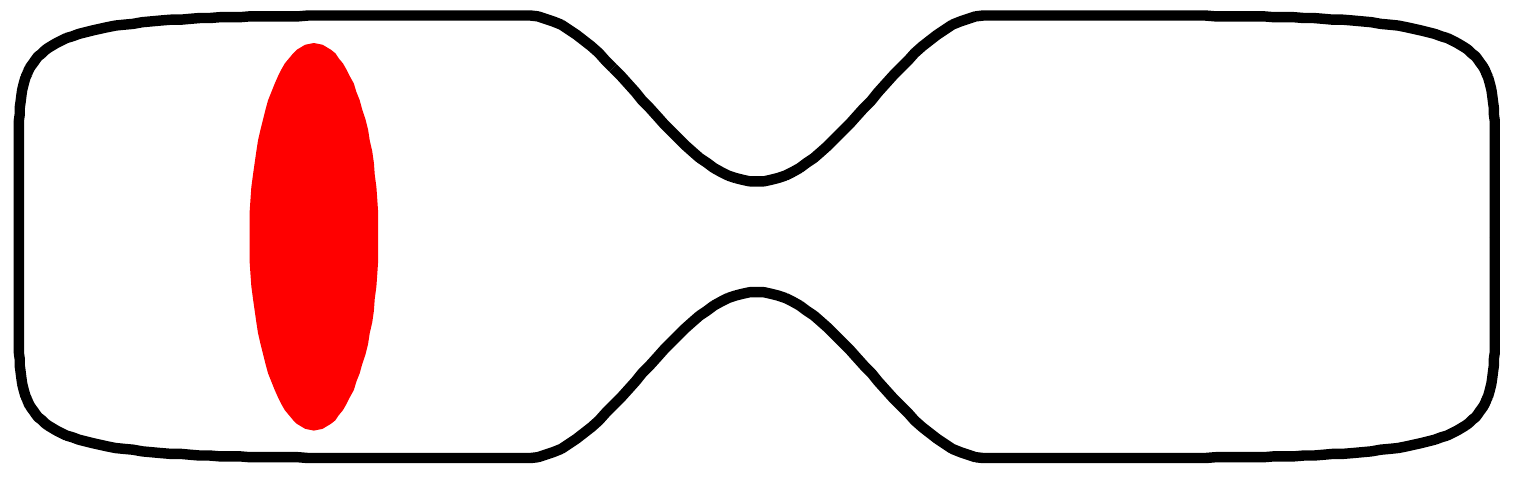}
    \captionof{figure}{\textit{The initial configuration of a {\bf stenosis flow}.}}
    \label{f:ICchoke1Ves}
  \end{minipage}
  \hfill
  \begin{minipage}[b]{0.49\textwidth}
    \centering
    \begin{tabular}{cc}\hline
        Parameter & Value \\ \hline
        Points on a vesicle $N$ & 128 \\
        Points on a wall $N_{\text{wall}}$ & 480 \\
        Number of SDC sweeps $n_{\text{sdc}}$ & 1 \\ 
        Time step size $\Delta t$ & $10^{-3}$ \\ 
        CPU time  & 22 hours \\ \hline
      \end{tabular}
      \captionof{table}{\textit{Parameters of the ground truth of a
      {\bf stenosis flow}}.}
      \label{t:stenosisGTparams}
    \end{minipage}
  \end{minipage}

\paragraph*{\textbf{Setup}} We consider a single vesicle of reduced area
0.65 passing through a constricted tube (stenosis) without viscosity
contrast (\figref{f:ICchoke1Ves}). The flow is driven by a parabolic
flow profile at the intake and the outtake and the vesicle's initial
height is 3.5 times larger than the size of the constriction.  We choose
a time horizon $T = 12$ so that the vesicle passes through the
constriction.  We simulate this example with $N = 12, 16, 24, 32$ points
on the vesicle and $N_{\mathrm{wall}} = 256$ points on the wall with and
without the LRCA.  The ground truth solution
is formed with the parameters in \tabref{t:stenosisGTparams}.

\paragraph*{\textbf{Results}} We again investigate the stability of our
scheme with and without the LRCA. In this example,
reparametrization is necessary since the vesicle becomes highly
deformed, which results in high frequencies in the shape that need to be
removed.  Time adaptivity and repulsion are necessary for the vesicle
to pass through the constriction without crossing the outer boundary.
In \figref{f:chokeSpaceTimeResFrames} plots of the vesicle passing
through the constriction at different resolutions with the LRCA are 
qualitatively compared with the ground truth (grey vesicle).  Even at the lowest
resolution, the vesicle passes through the constriction, and the
vesicle shape and center agree quite well with the ground truth. Whereas 
vesicle-wall collisions cannot be handled without the LRCA and the 
simulations break at these resolutions (i.e. $N \leq 24$).

We report the (self-) errors in the center, the number of accepted and rejected
time steps, and the CPU time, both with and without the LRCA in
\tabref{t:ChokeSpaceTimeErrors}.  We see that without the algorithms,
the low-resolution simulations are not stable with $N \leq 24$.  At
these resolutions, even with a very small time step, the dynamics when
the shape is close to the solid wall can not be resolved.  
However, with the help of the LRCA, 
the simulations are stable and deliver
acceptably accurate results in short CPU times. Even with $N = 32$
where the algorithms are unnecessary for stability, using them reduces the
total number of time steps resulting in a computationally faster method. 
Additionally, the self-error in vesicle's center decreases as the resolution increases.

\begin{table}[H]
\mcaption{The (self-) error in the center of the vesicle in a {\bf stenosis
flow} with and without the LRCA in \secref{s:algos} (see
\figref{f:chokeSpaceTimeResFrames} for frames of these simulations
using the algorithms). 
The self-errors are computed with respect to the simulation 
in one row below. 
Also reported are the number of accepted and rejected time steps and
the CPU time. The original simulations break
when $N\leq24$ because the vesicle-wall collisions cannot be
handled. The ground truth simulation takes 22 hours with $N = 128$ 
and $\Delta t $ = 1E-3.}{t:ChokeSpaceTimeErrors}
\centering
\begin{tabular}{c c| c c c c c | c c c c}
& & \multicolumn{5}{c|}{LRCA} &
\multicolumn{4}{c}{Original} \\
\hline
$N$ & ${\rho}_{\mathrm{AL}}$ &  ${\epsilon}_{\mathrm{center}}^{\mathrm{g}}$ & 
${\epsilon}_{\mathrm{center}}^{\mathrm{s}}$ & Accepts & Rejects
& Time (sec) &  ${\epsilon}_{\mathrm{center}}^{\mathrm{g}}$ & Accepts & Rejects & Time (sec) \\ 
 \hline
 
  12 & 1E-1 & 1.7E-1 & 4.6E-2 & 29 & 6  & 83  & - & - & - & -\\
 
  16 & 1E-2 & 8.1E-2 & 3.2E-2 & 64 & 12  & 116 & - & - & - & -\\
 
  24 & 1E-3 & 2.6E-2 & 7.5E-3 & 208 & 32  & 348 & - & - & - & -\\
 
  32 & 1E-4 & 1.2E-2 &        & 567 & 27  & 887 & 1.4E-2 & 1312 & 155 & 1950\\
 
\end{tabular} 
\end{table} 

\begin{figure}[H]
\begin{center}
\includegraphics[scale=0.53]{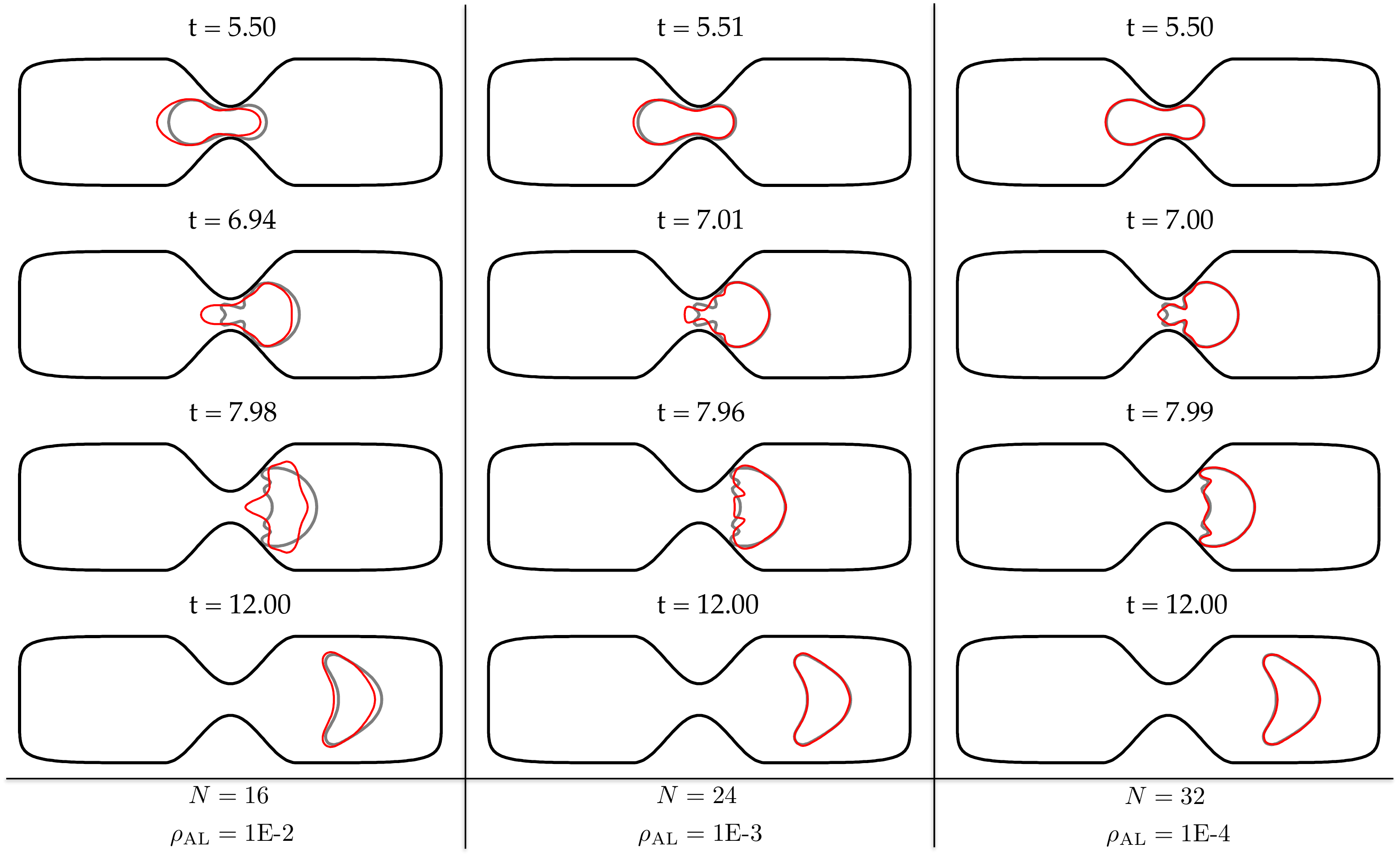}
\mcaption{A single vesicle passing through a constricted tube
({\bf stenosis}). Here we vary the temporal and spatial resolutions
simultaneously and use the LRCA.  The spatial resolution and
error tolerances are indicated at the bottom of each column.  The wall
is discretized with $N_{\text{wall}} = 256$ in all of the simulations.
The grey vesicle is the ground truth. While these low-resolution 
simulations are stable and accurate with the LRCA, vesicle-wall collisions 
cannot be handled in the original simulations.}{f:chokeSpaceTimeResFrames}
\end{center}
\end{figure}

\subsection{Taylor-Green flow}\label{s:taylorGreen}
\begin{minipage}{\textwidth}
  \begin{minipage}[b]{0.4\textwidth}
    \centering
    \includegraphics[scale=0.4]{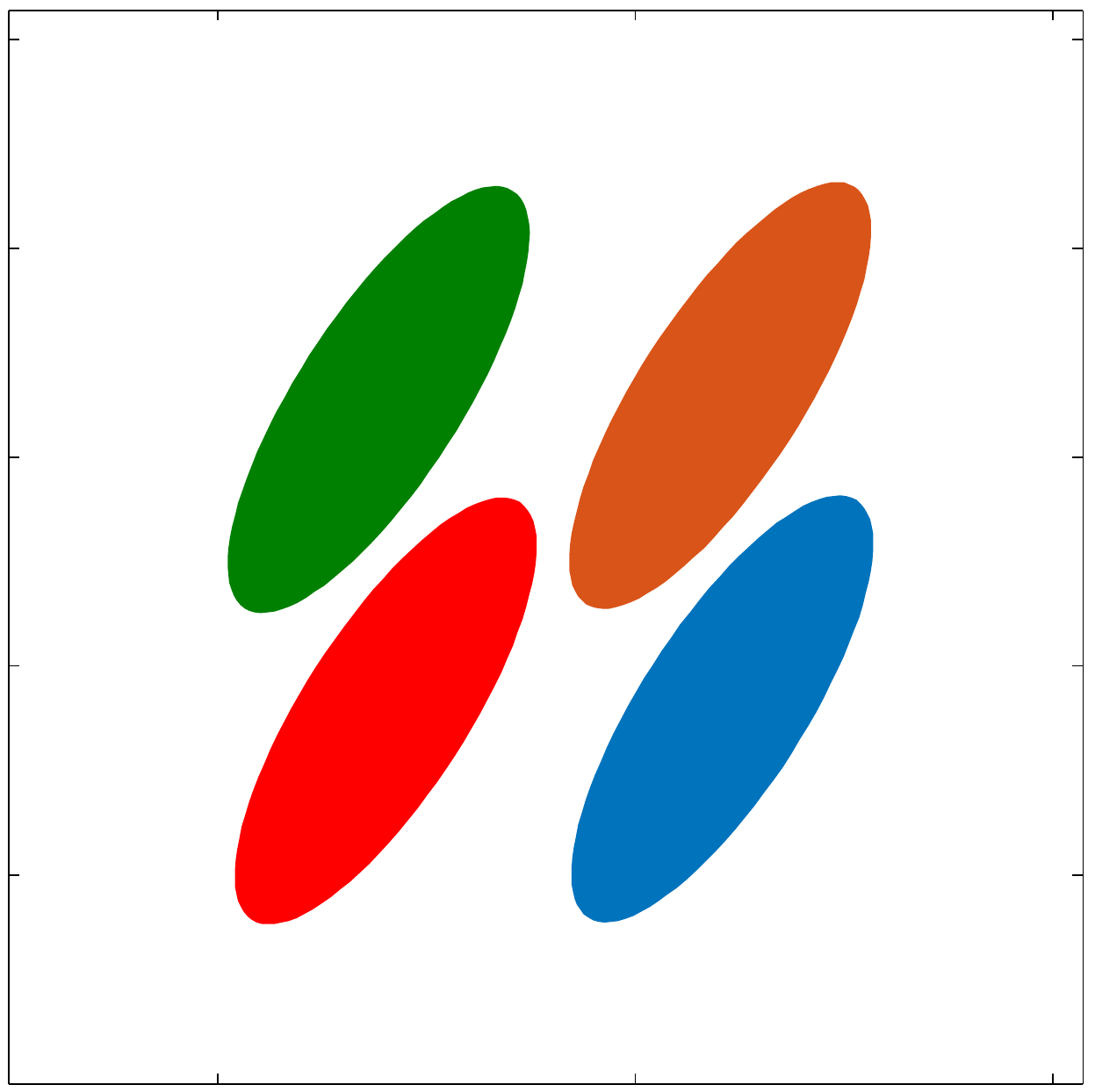}
    \captionof{figure}{\textit{Initial configuration of a {\bf Taylor-Green flow}.}}
    \label{f:ICtaylor}
  \end{minipage}
  \hfill
  \begin{minipage}[b]{0.59\textwidth}
    \centering
    \begin{tabular}{cc}\hline
        Parameter & Value \\ \hline
        Viscosity contrast $\nu$ & $\{1, 10\}$\\
        Points per vesicle $N$ ($\nu = 1$) & 96 \\
        Points per vesicle $N$ ($\nu = 10$) & 64 \\
        Time step size $\Delta t$ ($\nu = 1$) & $2 \times 10^{-4}$ \\ 
        Time step size $\Delta t$ ($\nu = 10$) & $10^{-3}$ \\ 
        Number of SDC sweeps $n_{\text{sdc}}$ & 1 \\
        CPU time ($\nu = 1$) & 71.1 hours \\
        CPU time ($\nu = 10$) & 76.4 hours \\ \hline
      \end{tabular}
      \captionof{table}{\textit{Parameters of the ground truth of a {\bf Taylor-Green flow}.}}
      \label{t:TGparams}
    \end{minipage}
  \end{minipage}

\paragraph*{\textbf{Setup}} We consider four large vesicles of reduced
area $0.65$ in the periodic cell $(0,\pi)^2$ with the background
Taylor-Green flow $\mathbf{u} = (\sin x \cos y, -\cos x \sin y)$.  The
vesicles occupy about 55\% of the periodic cell (see
\figref{f:ICtaylor}).  We color each vesicle for tracking purposes.
The time horizon is $T = 20$ and we perform simulations with viscosity
contrasts $\nu = 1$ and $\nu = 10$.  We simulate these examples with $N
= 12, 16, 24, 32, 48$ points per vesicle, the error tolerances
${\rho}_{\mathrm{AL}} = 10^{-1}, 10^{-2}, 10^{-3}, 10^{-4}, 10^{-5}$,
with and without the LRCA.  A ground truth solution
for these examples is formed with the parameters in \tabref{t:TGparams}. 
We also demonstrate the convergence of the ground truth solution for 
the example with no viscosity contrast in~\figref{f:TGVC1Converg}.

\begin{figure}[!htb]
 \begin{minipage}{\textwidth}
 \begin{center}
\setcounter{subfigure}{0}
\renewcommand*{\thesubfigure}{(a)} 
      \hspace{0cm}\subfigure[$t = 10$]{\scalebox{0.33}{{\includegraphics{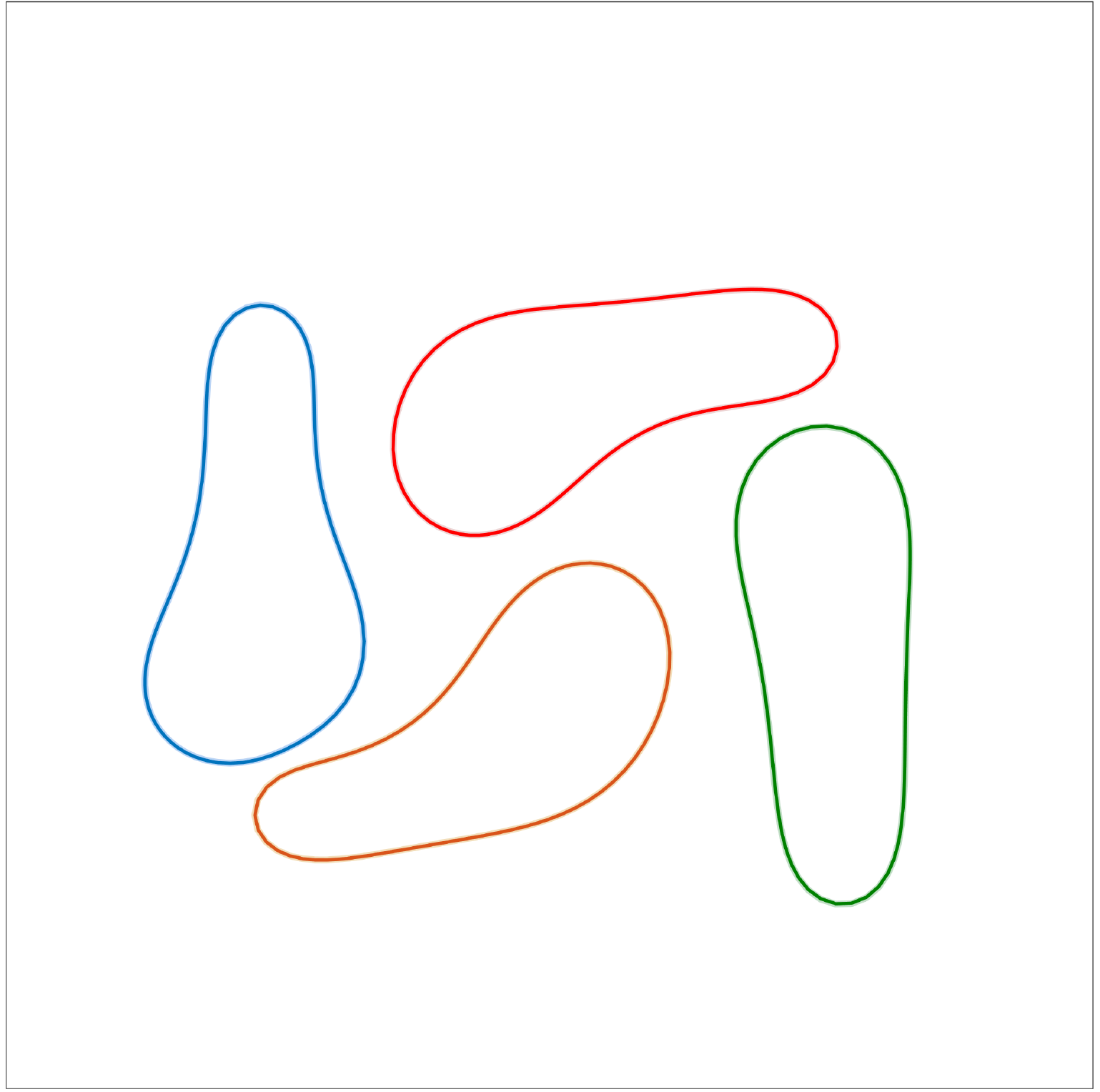}}} 
      \label{f:TGVC1GTt10}}
\setcounter{subfigure}{0}
\renewcommand*{\thesubfigure}{(b)} 
      \hspace{0cm}\subfigure[$t = 15$]{\scalebox{0.33}{{\includegraphics{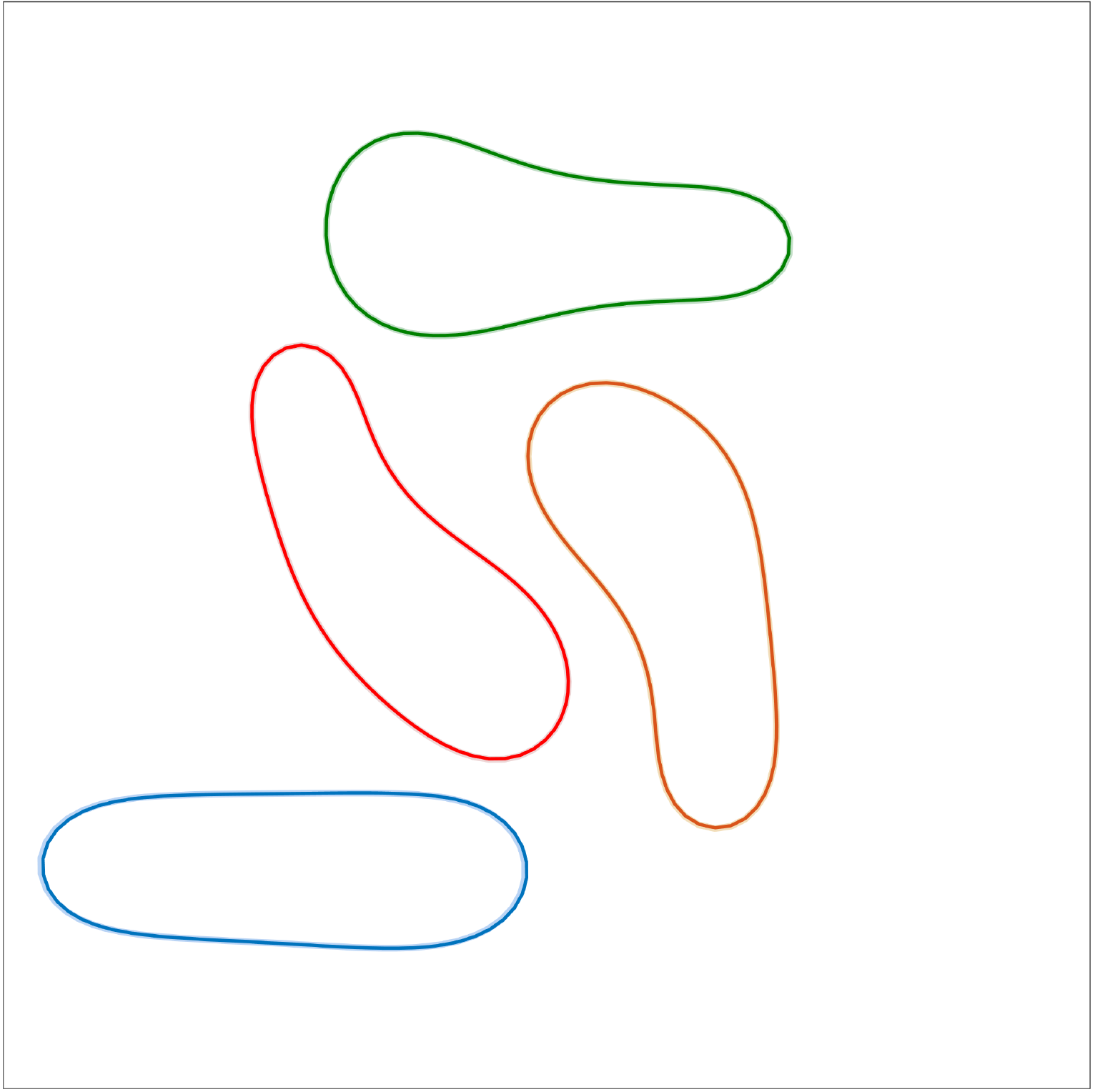}}}
      \label{f:TGVC1GTt15}}
\setcounter{subfigure}{0}
\renewcommand*{\thesubfigure}{(c)} 
      \hspace{0cm}\subfigure[$t = 20$]{\scalebox{0.33}{{\includegraphics{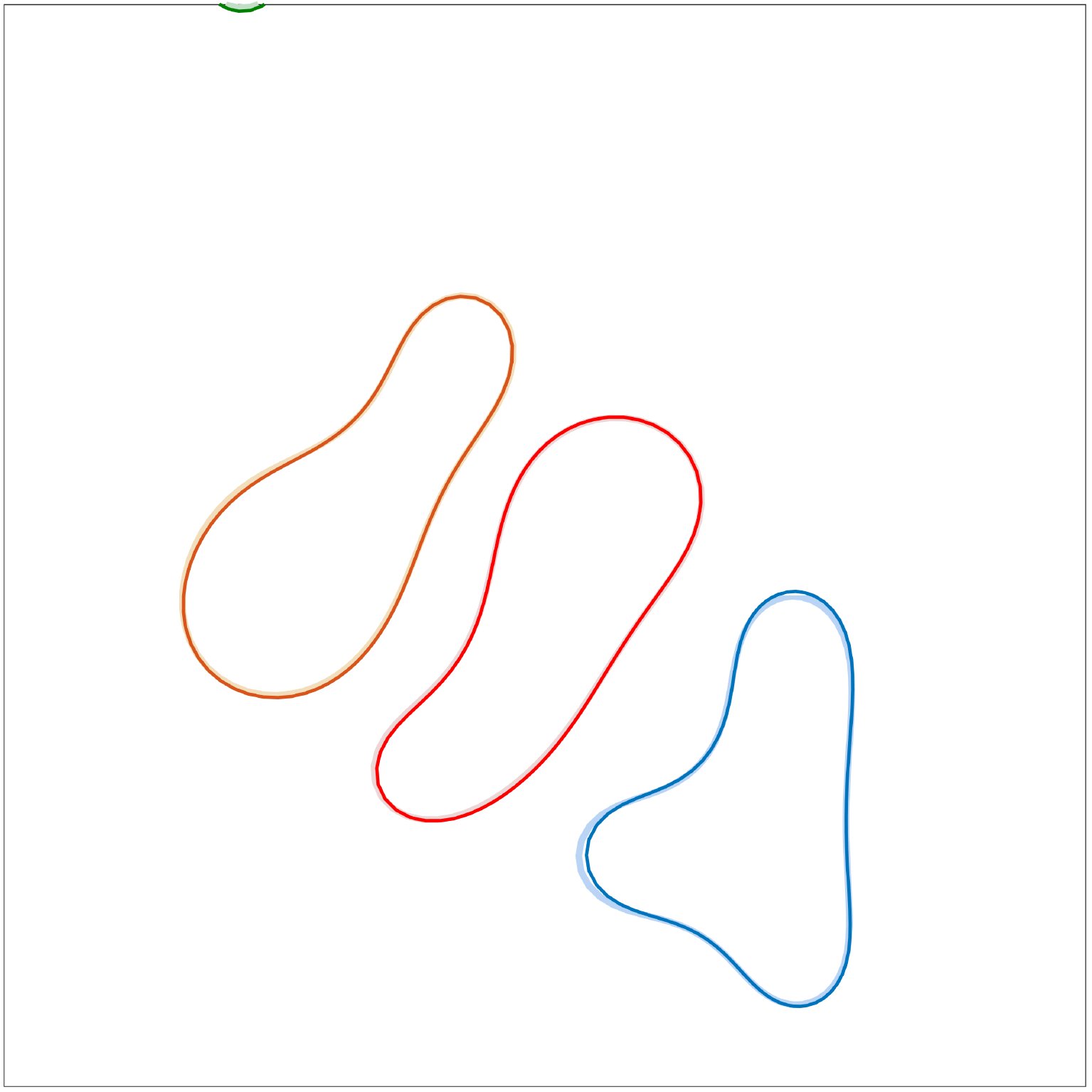}}}
      \label{f:TGVC1GTt20}}      
\end{center}
\end{minipage}
\mcaption{Convergence of the ground truth solution for the {\bf Taylor-Green flow} with no viscosity contrast. The ground truth solution for this example is formed with $N = 96$ points per vesicle and a constant time step size of $\Delta t$ = 2E-4, and shown with the faded colors. We superimpose the faded vesicles from the solution with a lower spatial and temporal resolution ($N = 64$ and $\Delta t$ = 1E-3). The errors in center and in inclination angle are $\epsilon_{\mathrm{center}} = $ 1E-2 and $\epsilon_{\mathrm{IA}} = $ 2E-2, respectively in the simulation with $N = 64$. Also see~\tabref{t:TG4VesVC1Errors} for the details of the simulation with $N = 64$.}{f:TGVC1Converg}
\end{figure}

\paragraph*{\textbf{Results}}  This example is more complex than the
previous examples since there are interactions between multiple
vesicles.  Therefore, we expect that the LRCA are essential for the stability at low resolutions.  We summarize the
results of the vesicles with $\nu=1$ in
\tabref{t:TG4VesVC1Errors} with and without the LRCA.  We report the
errors in the vesicles' centers, inclination angles, and effective
viscosity, as well as the number of accepted and rejected time steps,
and the total CPU time.  
The self-error is measured in terms of the effective viscosity only. 
Also, in \figref{f:TG4VesFrames} we plot
snapshots of the vesicle shapes at four different resolutions and
superimpose the ground truth solution.  We see that the LRCA result in
stability at much lower resolutions, but the errors in the center and 
inclination angle of the vesicles are large (i.e. $\bigO (1)$).  
The reason for that is this example has more vesicle-vesicle interactions 
than the previous two and the near collisions lead to more chaotic 
flows~\cite{narsimhan-shaqfeh-e13, aouane-misbah-e14}. Convergence in terms of the local error measures such as the error in center and inclination angle requires fine resolutions (i.e. at least $N = 64$ for the no viscosity contrast case, see~\figref{f:TGVC1Converg}). 
\figref{f:TG4VesFrames} shows that 
the centers and inclination angles of vesicles in 
the low-resolution simulations are close to those of the ground truth 
over a short time. As the vesicles interact more, the 
errors accumulate and result in diverging long-term 
behavior of an individual vesicle. 
However, the error of the effective
viscosity is satisfactory.  In
contrast, without the LRCA, stability is not achieved until $N=32$ due 
to vesicle-vesicle collisions which cannot be handled.
Smaller errors can be achieved without the LRCA, but this requires a
resolution of $N=48$. At the two lowest resolutions, we increase the temporal 
resolution  without changing the spatial resolution. While using the LRCA the errors decrease 
further with increasing temporal resolutions and the CPU times are still shorter 
than those with higher spatial resolutions, these simulations are not stable without the LRCA.

We repeat these experiments with viscosity contrast $\nu=10$ and 
we report the results in Tables~\ref{t:TG4VesVC10Errors}.  
Again, we see that with the 
LRCA, the errors in the center and inclination angle are large and
the error in the viscosity contrast is small. Without the
LRCA, stability requires $N=32$ points, and smaller errors than
than the those with our algorithms requires $N=48$ points.

\begin{table}[H]
\mcaption{The maximum errors in the vesicles' centers, inclination
angles, and the effective viscosity of four vesicles in a {\bf
Taylor-Green flow} with no viscosity contrast (\figref{f:TG4VesFrames})
and with the LRCA.  The self-errors in terms of the effective viscosity 
are computed with respect to the 
simulation in one row below. 
Also reported are the number of accepted and rejected time steps and
the CPU time. The dash "-" is put on the table for the simulations which break 
without the LRCA because the vesicle collisions cannot be avoided. The ground truth simulation takes 
71.1 hours with $N = 96$ and $\Delta t$ = 2E-4. 
}{t:TG4VesVC1Errors}
\centering
\begin{tabular}{c c| c c c c c c c}
\multicolumn{9}{c}{{\bf LRCA}} \\
\hline
\hline
$N$ & ${\rho}_{\mathrm{AL}}$ &  ${\epsilon}_{\mathrm{center}}^{\mathrm{g}}$ &
${\epsilon}_{\mathrm{IA}}^{\mathrm{g}}$ & ${\epsilon}_{{\mu}_{\mathrm{eff}}}^{\mathrm{g}}$ & 
${\epsilon}_{{\mu}_{\mathrm{eff}}}^{\mathrm{s}}$ & Accepts & Rejects 
& Time (sec)\\ 
 \hline
 
 12 & 1E-1 & 1.0E+0 & 9.7E-1 & 4.1E-1  & 3.7E-1 & 35 & 7  & 50 \\
 
 16 & 1E-2 & 1.8E+0 & 2.0E+0  & 5.0E-2 & 6.5E-3 & 92 & 8  & 106 \\
 
 24 & 1E-3 & 1.7E+0 & 2.1E+0  & 5.6E-2 & 2.3E-2 & 326 & 13  & 419 \\
 
 32 & 1E-4 & 1.6E+0 & 2.0E+0  & 3.4E-2 & 5.0E-3 & 1080 & 15  & 1390 \\
 
 48 & 1E-5 & 1.5E+0 & 4.2E-1 & 3.0E-2  &        & 3437 & 26  & 5990 \\

 64 & $\Delta t =$ 1E-3 & 1.0E-2 & 2.0E-2 & 1.2E-3  &        & 20001 & -  & 61200 \\

\hline
 12 & 1E-1 & 1.0E+0 & 9.7E-1 & 4.1E-1  & 3.3E-1 & 35 & 7  & 50 \\

 12 & 1E-2 & 7.4E-1 & 2.0E+0 & 1.1E-1  & 9.9E-2 & 104 & 15  & 143 \\

 12 & 1E-3 & 8.3E-1 & 2.0E+0 & 2.3E-2  & 9.1E-3 & 312 & 17  & 486 \\

 12 & 1E-4 & 5.1E-1 & 1.7E+0 & 1.4E-2  &        & 1458 & 28  & 2200 \\

 \hline
 16 & 1E-2 & 1.8E+0 & 2.0E+0  & 5.0E-2 & 3.9E-2 & 92 & 8  & 106 \\

 16 & 1E-3 & 1.7E+0 & 2.0E+0  & 2.5E-2 & 1.4E-2 & 333 & 18  & 359 \\

 16 & 1E-4 & 1.6E+0 & 1.8E+0  & 1.2E-2 &        & 1135 & 18  & 1610 \\

\end{tabular}

\begin{tabular}{c c| c c c c c c c}
\multicolumn{9}{c}{{\bf Original}} \\
\hline
\hline

$N$ & ${\rho}_{\mathrm{AL}}$ &  ${\epsilon}_{\mathrm{center}}^{\mathrm{g}}$ &
${\epsilon}_{\mathrm{IA}}^{\mathrm{g}}$ & ${\epsilon}_{{\mu}_{\mathrm{eff}}}^{\mathrm{g}}$ & 
${\epsilon}_{{\mu}_{\mathrm{eff}}}^{\mathrm{s}}$ & Accepts & Rejects 
& Time (sec)\\ 
 \hline
 
 12 & 1E-1 & - & - & - & - & - & - & - \\
 
 16 & 1E-2 & - & - & - & - & - & - & - \\
 
 24 & 1E-3 & - & - & - & - & - & - & - \\
 
 32 & 1E-4 & 1.5E+0 & 1.9E+0  & 2.2E-1 & 2.1E-1 & 1064 & 16  & 736 \\
 
 48 & 1E-5 & 5.4E-1 & 2.8E-1 & 1.5E-2 &         & 3306 & 29  & 3480 \\

 \hline
 12 & 1E-1 & - & - & - & - & - & - & - \\

 12 & 1E-2 & - & - & - & - & - & - & - \\

 12 & 1E-3 & - & - & - & - & - & - & - \\

 12 & 1E-4 & - & - & - & - & - & - & - \\

 \hline
 16 & 1E-2 & - & - & - & - & - & - & - \\

 16 & 1E-3 & - & - & - & - & - & - & - \\

 16 & 1E-4 & - & - & - & - & - & - & - \\

\end{tabular} 
\end{table} 

\begin{figure}[H]
\begin{center}
\includegraphics[scale=0.38]{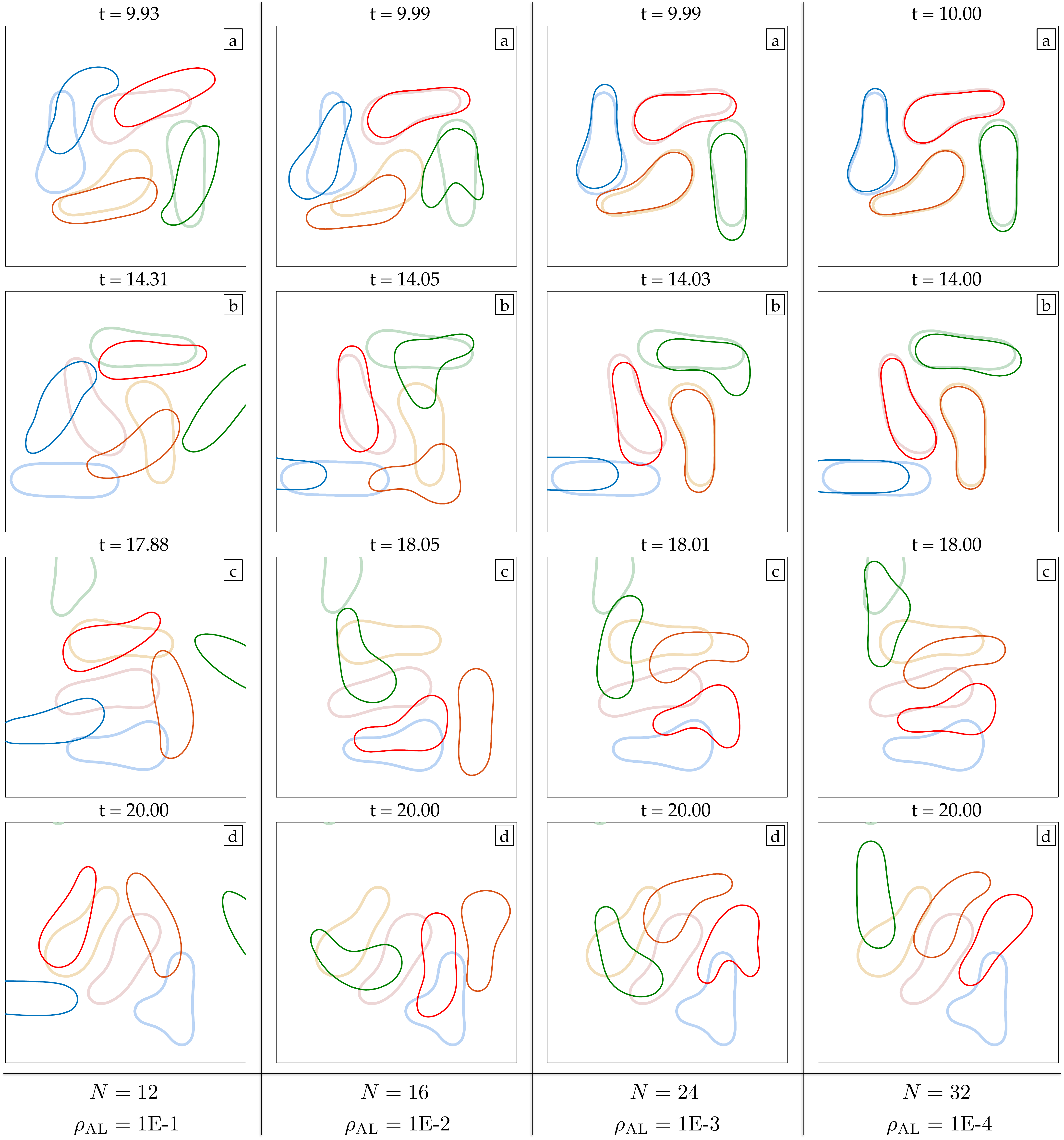}
\mcaption{Four vesicles in a {\bf Taylor-Green flow} with no viscosity
contrast and with the LRCA.  The spatial resolutions and the error
tolerances are given at the bottom of each column.  Faded vesicles
correspond to the ground truth and the low-resolution counterparts are
in bright colors.}{f:TG4VesFrames}
\end{center}
\end{figure}

\begin{table}[H]
\mcaption{The maximum errors in the vesicles' centers, inclination
angles, and the effective viscosity of four vesicles in a {\bf
Taylor-Green flow} with viscosity contrast $10$ and  with the 
LRCA. The self-errors in terms of the effective viscosity 
are computed with respect to the 
simulation in one row below. 
Also reported are the number of accepted and rejected time steps and
the CPU time. The dash "-" is put on the table for the simulations which break 
without the LRCA because the vesicle collisions cannot be avoided. The ground truth simulation takes 
76.4 hours with $N = 64$ and $\Delta t$ = 1E-3.}{t:TG4VesVC10Errors}
\centering
\begin{tabular}{c c| c c c c c c c}
\multicolumn{9}{c}{{\bf LRCA}} \\
\hline
\hline
$N$ & ${\rho}_{\mathrm{AL}}$ &  ${\epsilon}_{\mathrm{center}}^{\mathrm{g}}$ &
${\epsilon}_{\mathrm{IA}}^{\mathrm{g}}$ & ${\epsilon}_{{\mu}_{\mathrm{eff}}}^{\mathrm{g}}$ & 
${\epsilon}_{{\mu}_{\mathrm{eff}}}^{\mathrm{s}}$ & Accepts & Rejects 
& Time (sec)\\ 
 \hline
 
 12 & 1E-1 & 2.9E+0 & 2.0E+0  & 3.4E-1 & 2.4E-1 & 36 & 9  & 50 \\
 
 16 & 1E-2 & 1.6E+0 & 2.0E+0  & 1.3E-1 & 8.9E-2 & 91 & 7  & 87 \\
 
 24 & 1E-3 & 1.6E+0 & 5.9E-1  & 4.7E-2 & 9.2E-3 & 282 & 10  & 371 \\
 
 32 & 1E-4 & 3.4E-1 & 1.2E-1  & 3.8E-2 & 7.2E-2 & 881 & 10  & 1210 \\
 
 48 & 1E-5 & 1.1E-1 & 5.4E-2  & 3.6E-2 &        & 2835 & 15 & 8650 \\

 \hline
 12 & 1E-1 & 2.9E+0 & 2.0E+0  & 3.4E-1 & 2.5E-1 & 36 & 9  & 50 \\

 12 & 1E-2 & 1.7E+0 & 2.0E+0 & 1.2E-1  & 1.9E-2 & 97 & 8  & 150 \\

 12 & 1E-3 & 1.3E+0 & 2.0E+0 & 1.0E-1  & 2.3E-2 & 284 & 10  & 595 \\

 12 & 1E-4 & 1.1E-1 & 1.8E+0 & 8.1E-2  &        & 888 & 16  & 2090 \\

 \hline
 16 & 1E-2 & 1.6E+0 & 2.0E+0  & 1.3E-1 & 8.7E-2 & 91 & 7  & 87 \\

 16 & 1E-3 & 1.6E+0 & 2.0E+0  & 5.0E-2 & 7.8E-2 & 286 & 11  & 415 \\

 16 & 1E-4 & 2.5E-1 & 2.0E+0  & 3.1E-2 &        & 899 & 11  & 1670 \\

\end{tabular} 

\begin{tabular}{c c| c c c c c c c}
\multicolumn{9}{c}{{\bf Original}} \\
\hline
\hline
$N$ & ${\rho}_{\mathrm{AL}}$ &  ${\epsilon}_{\mathrm{center}}^{\mathrm{g}}$ &
${\epsilon}_{\mathrm{IA}}^{\mathrm{g}}$ & ${\epsilon}_{{\mu}_{\mathrm{eff}}}^{\mathrm{g}}$ & 
${\epsilon}_{{\mu}_{\mathrm{eff}}}^{\mathrm{s}}$ & Accepts & Rejects 
& Time (sec)\\  
 \hline
 
 12 & 1E-1 & - & - & - & - & - & - & - \\
 
 16 & 1E-2 & - & - & - & - & - & - & - \\
 
 24 & 1E-3 & - & - & - & - & - & - & - \\
 
 32 & 1E-4 & 3.8E-1 & 1.7E-1  & 2.7E-2 & 5.0E-2 & 894 & 11  & 940 \\
 
 48 & 1E-5 & 9.7E-2 & 5.8E-2 & 2.5E-2 &         & 2786 & 12 & 6380 \\

 \hline
 12 & 1E-1 & - & - & - & - & - & - & - \\

 12 & 1E-2 & - & - & - & - & - & - & - \\

 12 & 1E-3 & - & - & - & - & - & - & - \\

 12 & 1E-4 & - & - & - & - & - & - & - \\

 \hline
 16 & 1E-2 & - & - & - & - & - & - & - \\
 
 16 & 1E-3 & - & - & - & - & - & - & - \\

 16 & 1E-4 & - & - & - & - & - & - & - \\
\end{tabular} 
\end{table} 

\subsection{Couette flow}\label{s:couette}
\begin{figure}[H]
\begin{minipage}{\textwidth}
  \begin{minipage}[t]{0.55\textwidth}
    \centering
    \setcounter{subfigure}{0}
    \renewcommand*{\thesubfigure}{(a-1)} \hspace{0cm}\subfigure[Suspension 
    at $\phi = 20\%$]{\scalebox{0.3}{{\includegraphics{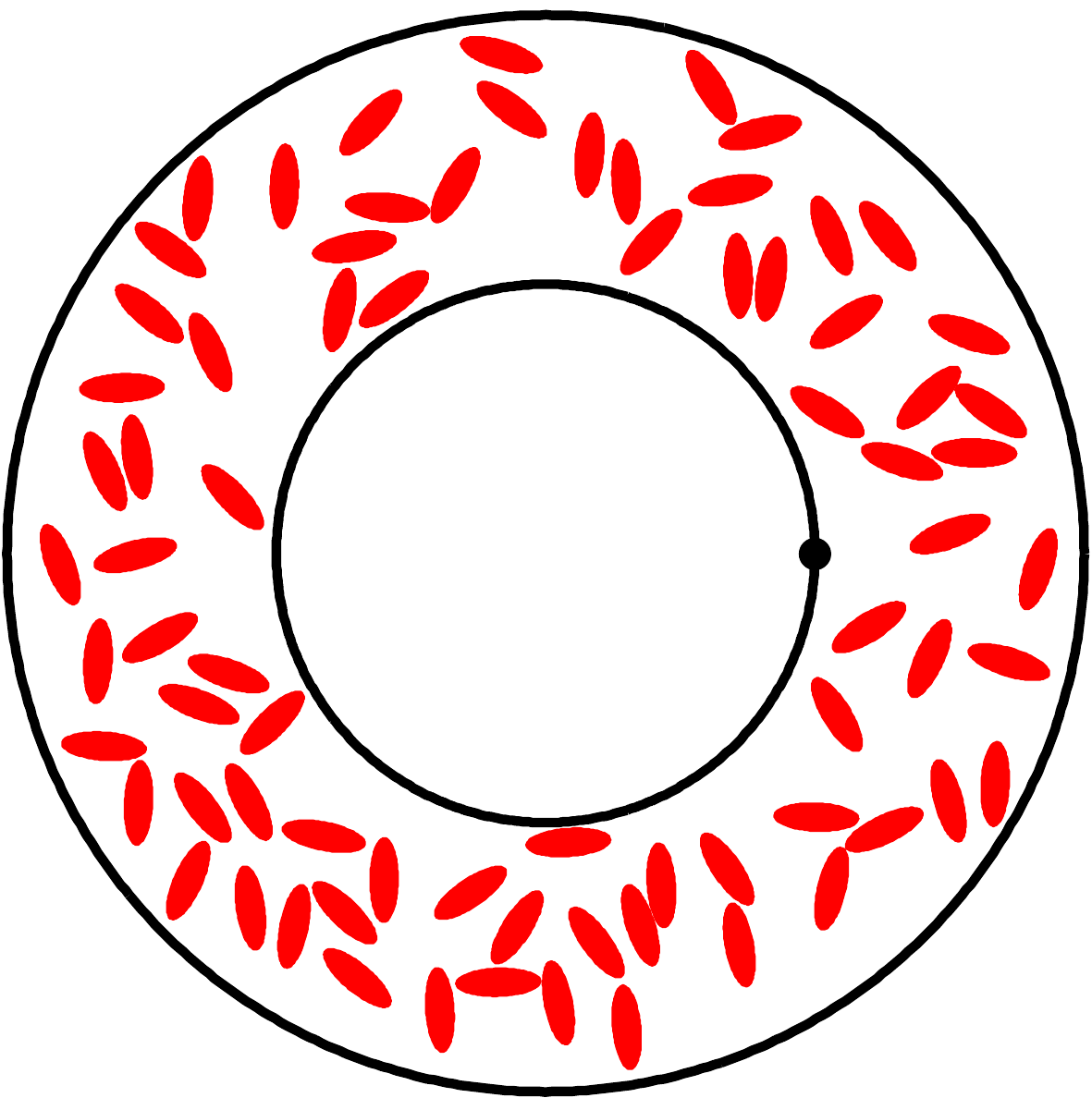}}}	
    \label{f:ICcouetteVF20}}
    \setcounter{subfigure}{0}
    \renewcommand*{\thesubfigure}{(a-2)} \hspace{0cm}\subfigure[Suspension 
    at $\phi = 40\%$]{\scalebox{0.3}{{\includegraphics{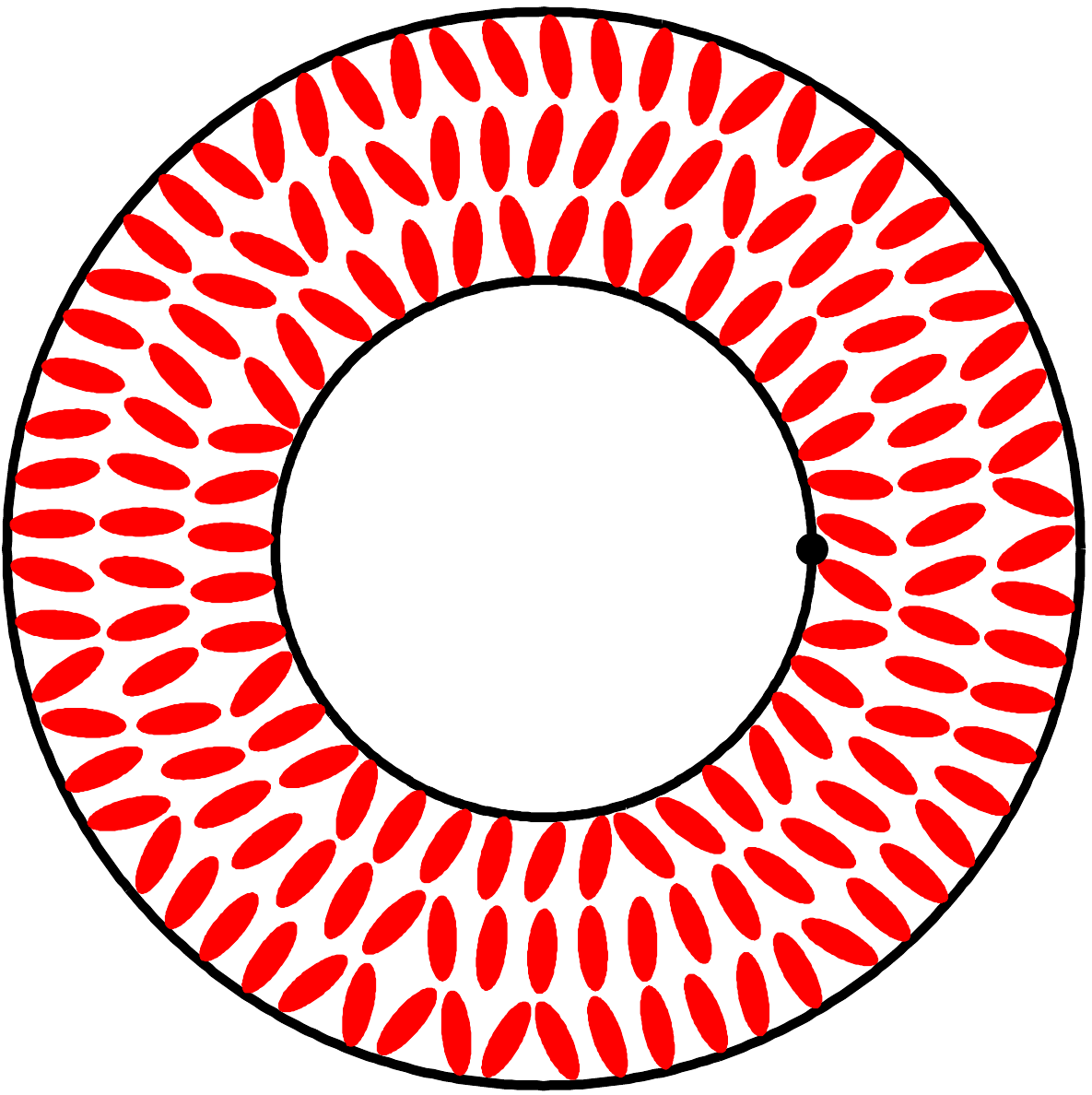}}}
    \label{f:ICcouetteVF40}}
    \mcaption{The initial configuration of two {\bf Couette}
    apparatuses with volume fractions $\phi = 20\%$ and $\phi = 40\%$.  For the
    ground truth solutions, we use the local area-length correction
    algorithm, but none of the other algorithms introduced in
    \secref{s:algos}.}{f:ICsCouette}
  \end{minipage}
  \hfill
  \begin{minipage}[t]{0.4\textwidth}
    \centering
    \begin{tabular}{cc}\hline
        Parameter & Value \\ \hline
        Points on a vesicle $N$ & 96 \\
        Points on a wall $N_{\text{wall}}$ & 256 \\
        Number of SDC sweeps $n_{\text{sdc}}$ & 1 \\ 
        Time step size $\Delta t$ & $10^{-2}$ \\ 
        CPU time($\phi=20\%$) & 3 weeks \\ 
        CPU time($\phi=40\%$) & 1 month \\ \hline
      \end{tabular}
      \captionof{table}{\textit{Parameters of the ground truth of a
      {\bf Couette flow}}.}
      \label{t:couetteGTparams}
    \end{minipage}
  \end{minipage}
\end{figure}

\paragraph*{\textbf{Setup}} We consider two Couette flows with volume 
fractions $\phi = 20\%$ (75 vesicles) and $\phi = 40\%$ (150 vesicles) 
without viscosity contrast (see \figref{f:ICsCouette}).  The inner boundary has radii
$R_{1}=10$ and is rotating with constant angular velocity while the
outer boundary has radii $R_{2}=20$ and is stationary.  We choose a
time horizon $T = 100$ which results in the inner cylinder completing
approximately 16 rotations.  We simulate these cases with $N = 16, 24$
points per vesicle, $N_{\text{wall}} = 128$ points per wall, error
tolerances ${\rho}_{\mathrm{AL}} = 10^{-2}, 10^{-3}$, and the
LRCA.  The ground truth solution for
these examples use the parameters in the caption of
\figref{f:ICsCouette}.  

\paragraph*{\textbf{Results}} A Couette apparatus is often used to
investigate properties of suspensions such as shear-induced
diffusion~\cite{podgorski-misbah-e11} and effective viscosity.  
High volume fraction
suspensions are of particular importance since red blood cells make up
approximately 45\% of human blood~\cite{goldsmith-skalak75}.  In
addition, long time horizons are required for statistical analysis.  Therefore, there
are a large number of interactions between vesicles and walls, near
collisions, and highly deformed vesicles.  The interactions and shapes
can be resolved with fine resolutions, but at a significant
computational cost.

We are interested in the errors of upscaled variables.  We report the
(self-) errors in effective viscosity, space-time average and time average of
the $L^2$ norm of a velocity field in \tabref{t:CouetteVF20Errors} for
$\phi = 20\%$ and in \tabref{t:CouetteVF40Errors} for $\phi = 40\%$.
We also present the frames from the simulations of the suspension at
$\phi = 20\%$ in \figref{f:CouetteVF20Frames} and the suspension at
$\phi = 40\%$ in \figref{f:CouetteVF40Frames} at various resolutions.
As in the previous example, the vesicle trajectories are not captured
by the simulations with the LRCA.  However, the errors in the upscaled
quantities are at an acceptable level of $\mathcal{O}(10^{-2})$ even
with $N=16$ points on each vesicle.  In addition, the computation
speedup is significant; the low-resolution runs required no more than a
little over a day ($\phi=20\%$) and less than a week $(\phi=40\%)$.  In
contrast, the ground truth simulations required 3 weeks ($\phi=20\%$)
and a month ($\phi=40\%$).

\begin{table}[H]
\mcaption{The (self-) errors in the effective viscosity
${\epsilon}_{\mu_{\mathrm{eff}}}$, time-space average of the velocity
${\epsilon}_{\langle v \rangle}$, and the time average of the $L^2$
norm of the velocity ${\epsilon}_{\langle V \rangle}$ of a suspension
at $\phi = 20\%$ in a {\bf Couette flow} with the LRCA
(\figref{f:CouetteVF20Frames}). The self-errors are computed with 
respect to the simulation in one row below. 
Also reported are the number of accepted and rejected time steps and
the CPU time.The ground truth simulation takes 
3 weeks with $N = 96$ and $\Delta t = $ 1E-2.}{t:CouetteVF20Errors}
\centering
\begin{tabular}{c c| c c c c c c c c c}

$N$ & ${\rho}_{\mathrm{AL}}$ &  ${\epsilon}_{{\mu}_{\mathrm{eff}}}^{\mathrm{g}}$ & 
${\epsilon}_{{\mu}_{\mathrm{eff}}}^{\mathrm{s}}$ & 
${\epsilon}_{\langle v \rangle}^{\mathrm{g}}$ 
& ${\epsilon}_{\langle v \rangle}^{\mathrm{s}}$ 
& ${\epsilon}_{\langle V \rangle}^{\mathrm{g}}$ 
& ${\epsilon}_{\langle V \rangle}^{\mathrm{s}} $ &
 Accepts & Rejects & Time (hours)\\ 
 \hline
 
 16 & 2E-2 & 5.5E-2 & 1.8E-2 & 7.9E-2 & 5.7E-2 & 2.3E-2 & 1.2E-2 & 507 & 74  & 9.3 \\

 24 & 1E-3 & 3.8E-2 &        & 3.6E-2 &        & 1.8E-2 &        &2160 & 55  & 32.2 \\
 
 \hline
 16 & 2E-2 & 5.5E-2 & 2.2E-2 & 7.9E-2 & 4.1E-2 & 2.3E-2 & 5.6E-3 & 507 & 74  & 9.3 \\

 16 & 1E-3 & 3.4E-2 &        &  4.0E-2 &       & 1.7E-2 &        & 2499 & 82 & 32.8 \\
\hline
 24 & 2E-2 & 4.2E-2 & 3.9E-3 & 8.8E-2 & 5.3E-2 & 2.9E-2 & 1.1E-2 & 402 & 31  & 10.1 \\

 24 & 1E-3 & 3.8E-2 &        & 3.6E-2 &        & 1.8E-2 &        &2160 & 55  & 32.2 \\

\end{tabular} 
\end{table} 

\begin{table}[H]
\mcaption{The (self-) errors in the effective viscosity
${\epsilon}_{\mu_{\mathrm{eff}}}$, time-space average of the velocity
${\epsilon}_{\langle v \rangle}$, and the time average of the $L^2$
norm of the velocity ${\epsilon}_{\langle V \rangle}$ of a suspension
at $\phi = 40\%$ in a {\bf Couette flow} with the LRCA
(\figref{f:CouetteVF40Frames}). The self-errors are computed with 
respect to the simulation in one row below. 
Also reported are the number of accepted and rejected time steps and
the CPU time. The ground truth simulation takes 
approximately a month with $N = 96$ and $\Delta t = $ 1E-2.}{t:CouetteVF40Errors}
\centering
\begin{tabular}{c c| c c c c c c c c c}

$N$ & ${\rho}_{\mathrm{AL}}$ &  ${\epsilon}_{{\mu}_{\mathrm{eff}}}^{\mathrm{g}}$ & 
${\epsilon}_{{\mu}_{\mathrm{eff}}}^{\mathrm{s}}$ & 
${\epsilon}_{\langle v \rangle}^{\mathrm{g}}$ 
& ${\epsilon}_{\langle v \rangle}^{\mathrm{s}}$ 
& ${\epsilon}_{\langle V \rangle}^{\mathrm{g}}$ 
& ${\epsilon}_{\langle V \rangle}^{\mathrm{s}} $ &
 Accepts & Rejects & Time (hours)\\ 
 \hline
 
 16 & 1E-2 & 2.7E-2 & 2.3E-2 & 1.3E-1 & 5.7E-2 & 5.9E-2 & 1.2E-2 & 1007 & 125  & 62.8 \\

 24 & 1E-3 & 3.1E-3 &        & 7.8E-2 &        & 4.7E-2 &        & 2894 & 67  & 138.6 \\
 
 \hline
 16 & 1E-2 & 2.7E-2 & 4.3E-2 & 1.3E-1 & 7.3E-2 & 5.9E-2 & 2.9E-2 & 1007 & 125  & 62.8 \\

 16 & 1E-3 & 1.6E-2 &        & 6.1E-2 &        & 3.1E-2 &        & 3517 & 110 & 156.1 \\

\hline 
 24 & 1E-2 & 1.9E-2 & 1.6E-2 & 1.5E-1 & 7.7E-2 & 7.2E-2 & 2.6E-2 & 806 & 37  & 42.2 \\
 
 24 & 1E-3 & 3.1E-3 &        & 7.8E-2 &        & 4.7E-2 &        & 2894 & 67  & 138.6 \\

\end{tabular} 
\end{table} 

\begin{figure}[H]
\centering
\includegraphics[scale=1.1]{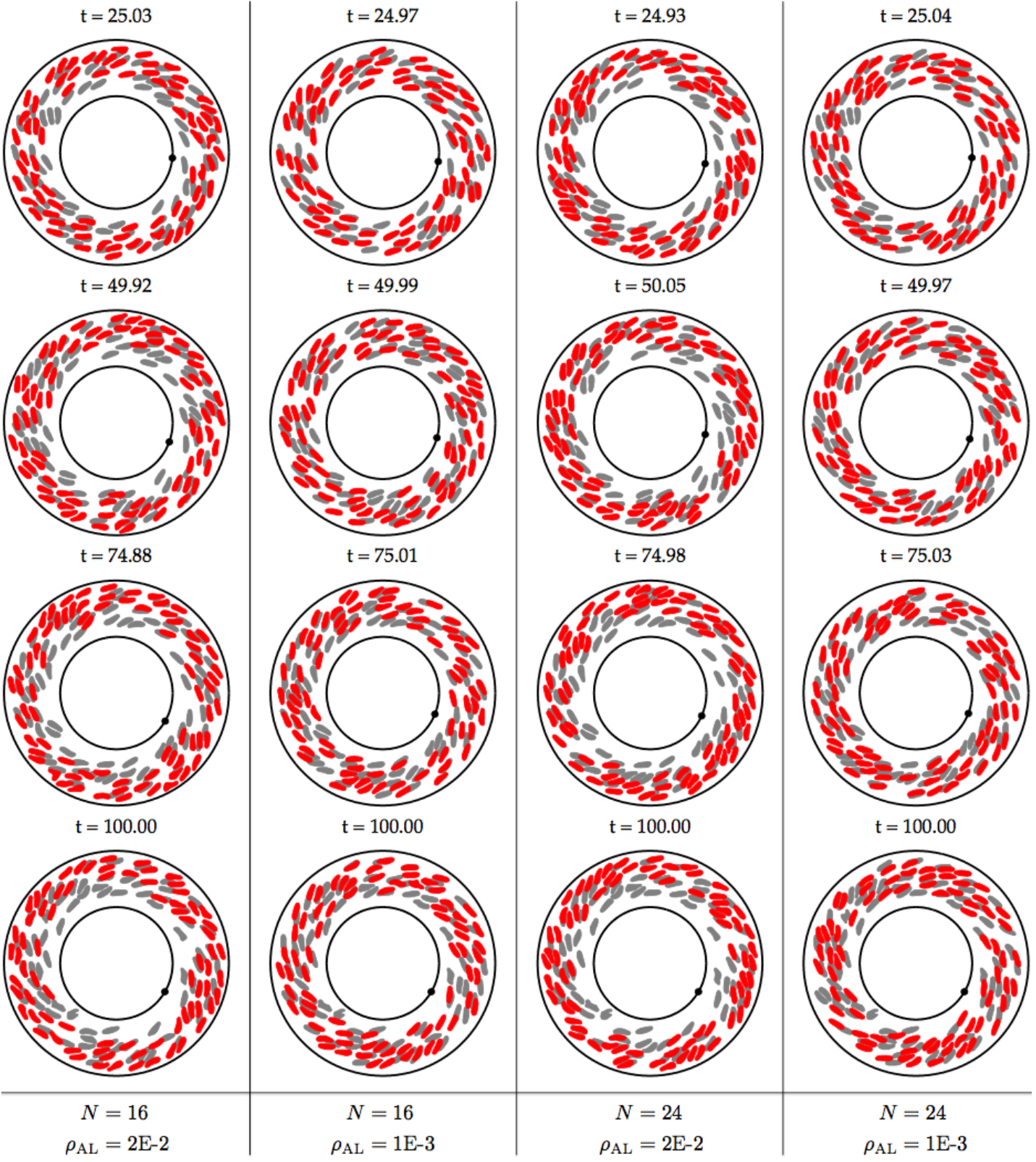}
\mcaption{75 vesicles in a {\bf Couette flow} which corresponds to a
volume fraction of $\phi = 20\%$. The ground truth is shown with grey
vesicles superimposed with the low-resolution counterpart (red
vesicles).}{f:CouetteVF20Frames}
\end{figure}

\begin{figure}[H]
\centering
\includegraphics[scale=1.1]{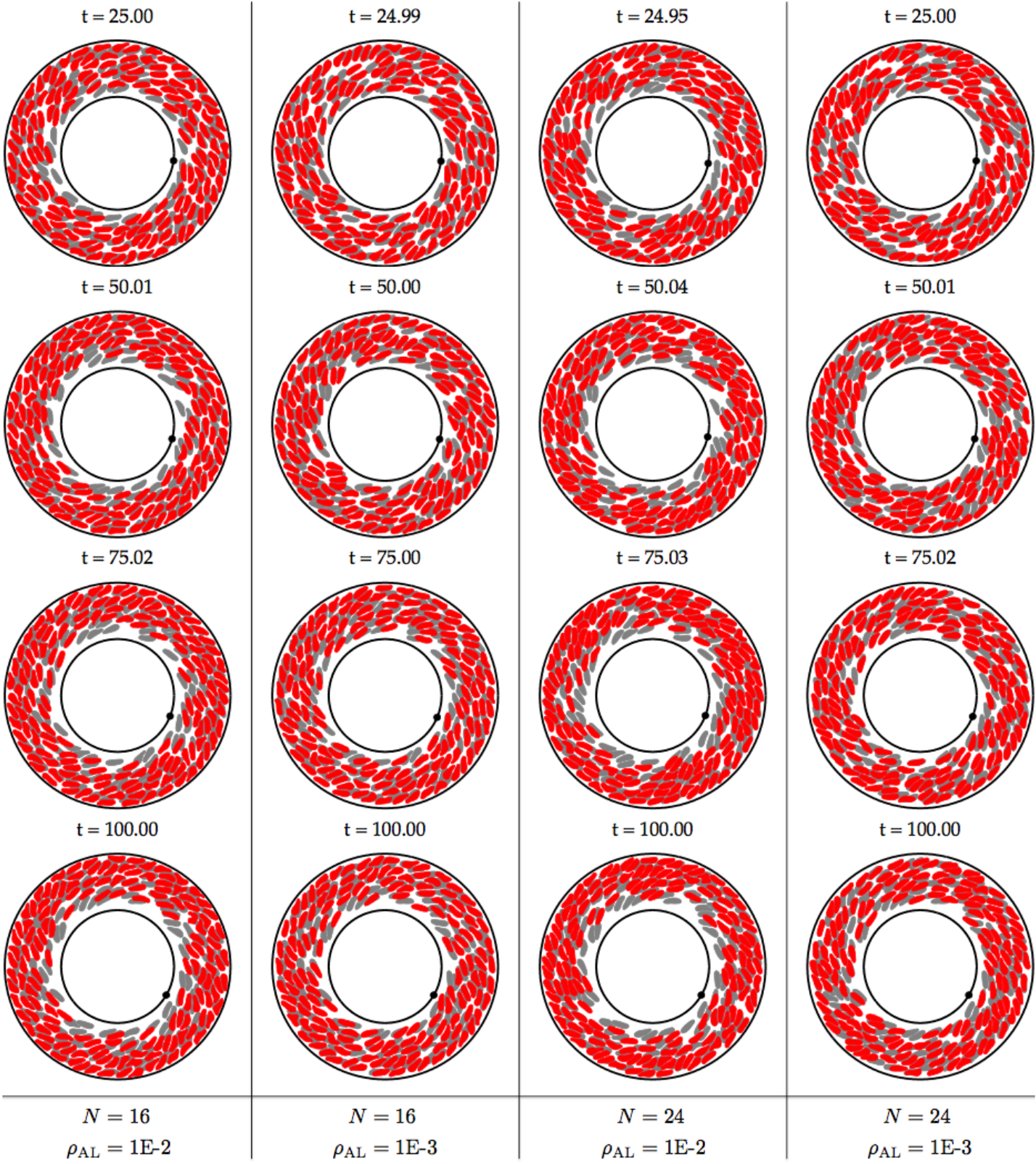}
\mcaption{150 vesicles in a {\bf Couette flow} which corresponds to a
volume fraction of $\phi = 40\%$. The ground truth is shown with grey
vesicles superimposed with the low-resolution counterpart (red
vesicles).}{f:CouetteVF40Frames}
\end{figure}
\paragraph*{{Cell-Free Layer}}Next, we investigate how accurately the
low-resolution simulations can capture the statistics of the vesicle
locations.  In this setup, vesicles are known to migrate away from the
walls resulting in a so-called cell-free layer near the
walls~\cite{leighton-acrivos87}.  This layer is captured by our
coarse spatial but fine temporal resolution simulations, i.e. low error tolerances ${\rho}_{\mathrm{AL}}$
(see Figures~\ref{f:CouetteVF20Frames}
and~\ref{f:CouetteVF40Frames}), but at the high error tolerances
the cell-free layer is thicker than the ground
truth (first and third columns).  To further demonstrate this point, we
plot the probability distribution functions of distances of the
vesicles' centers to the origin throughout the simulations in
\figref{f:CentersStatsCouette}.  The figure shows that the simulations
with the error tolerance ${\rho}_{\mathrm{AL}} = 10^{-3}$ estimate the
cell-free layer accurately at both spatial resolutions, while with
tolerance $\rho_{\mathrm{AL}} = 10^{-2}$, the cell-free layer is larger than the ground
truth. This suggests that although the local errors are too large in the simulations 
of dense suspensions at low resolutions, the upscaled quantities and statistics are rather insensitive to the local errors and can be accurately captured by the low-resolution simulations.

\begin{figure}[H]
 \begin{minipage}{\textwidth}
 \begin{center}
\setcounter{subfigure}{0}
\renewcommand*{\thesubfigure}{(a-1)} 
      \hspace{0cm}\subfigure[A Couette flow with $\phi = 20\%$]{\scalebox{0.32}{{\includegraphics{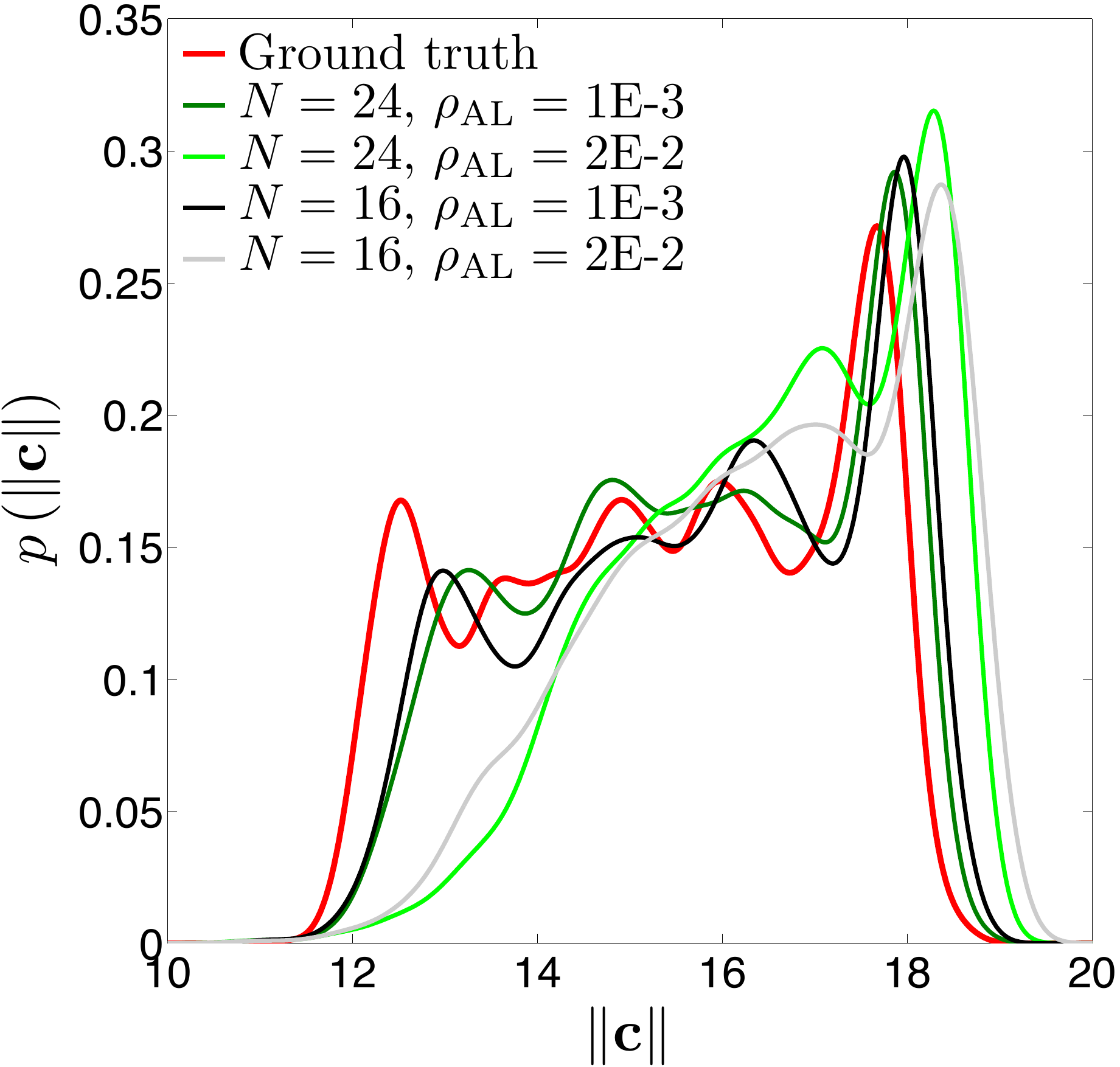}}}	
      \label{f:VF20Centers}}
\setcounter{subfigure}{0}
\renewcommand*{\thesubfigure}{(a-2)} 
      \hspace{0cm}\subfigure[A Couette flow with $\phi = 40\%$]{\scalebox{0.32}{{\includegraphics{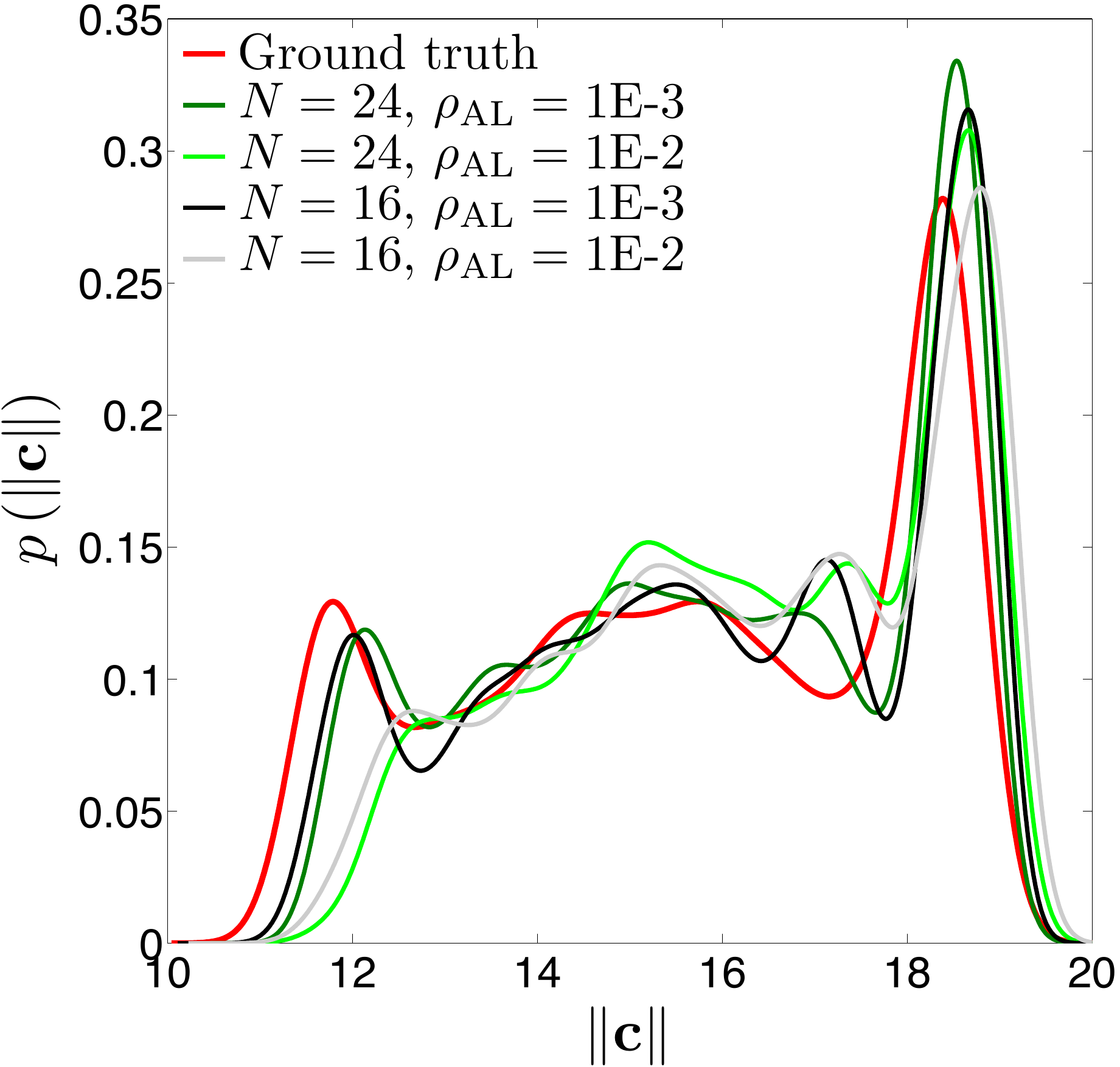}}}
      \label{f:VF40Centers}}
\end{center}
\mcaption{Statistics of the vesicles' concentration in a {\bf Couette
flow}. We plot the probability distributions of distances of the
vesicles' centers to the origin for the suspensions with volume fractions
$\phi = 20\%$ (left) and $\phi = 40\%$ (right).}{f:CentersStatsCouette}
\end{minipage}
\end{figure}

\paragraph*{Statistics of the velocity field} 
We also use simulations of vesicle suspensions in a Couette apparatus
to infer mixing properties of the
suspensions~\cite{kabacaoglu-biros-e16}.  For this reason, it is
important to estimate the velocity field accurately.  We compute the
error in the space-time averages of the velocity field discussed
above.  In \figref{f:VelocityStatsCouette} we present statistics of the
magnitude of the velocity field, $\|\mathbf{V}\|$, at points equally
distributed in the azimuthal direction at three different radii
($\frac{r - R_1}{R_2-R_1} = 0.2, 0.5, 0.8$).  Then we plot the
probability distribution function of $\|\mathbf{V}\|$ in
\figref{f:VelocityStatsCouette}. In the absence of vesicles,
$\mathbf{V}$ is only a function of the radial position in a Couette
flow. However, the presence of vesicles perturbs the velocity field.
The low-resolution simulations with the error tolerances
${\rho}_{\mathrm{AL}} = 10^{-3}$ estimate the statistics of the
velocity field closely. Similar to the statistics to capture 
the cell-free layer (\figref{f:CentersStatsCouette}), higher temporal 
resolutions provide more accurate velocity statistics while the 
spatial resolution does not significantly affect the results 
(see \figref{f:VelocityStatsCouette}).

\begin{figure}[H]
 \begin{minipage}{\textwidth}
 \begin{center}
\setcounter{subfigure}{0}
\renewcommand*{\thesubfigure}{(a-1)} 
      \hspace{0cm}\subfigure[at $\frac{r - R_1}{R_2-R_1} = 0.2$ ($\phi = 20\%$)]{\scalebox{0.28}{{\includegraphics{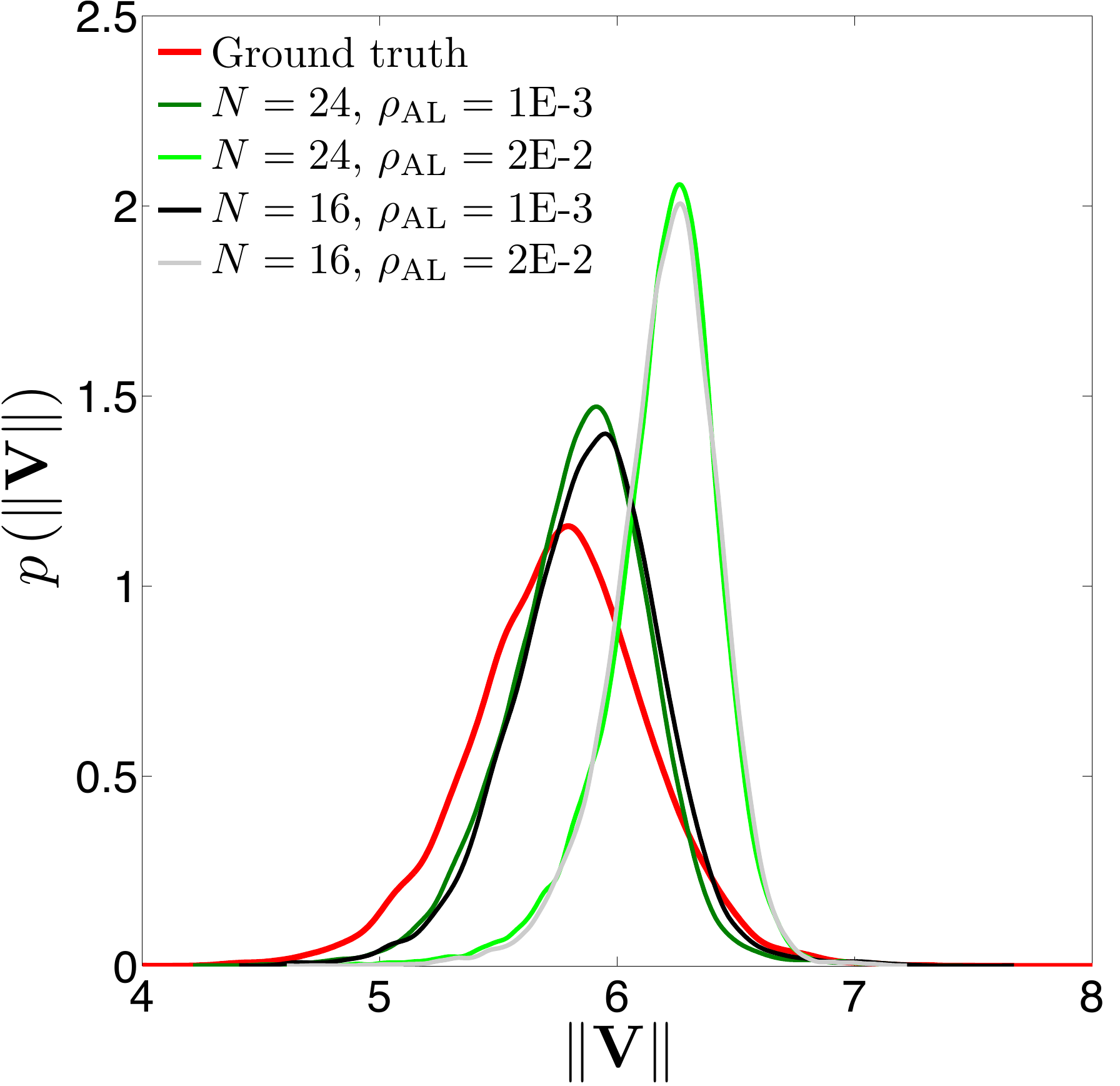}}}	
      \label{f:VF20atR1}}
\setcounter{subfigure}{0}
\renewcommand*{\thesubfigure}{(a-2)} 
      \hspace{0cm}\subfigure[at $\frac{r - R_1}{R_2-R_1} = 0.5$ ($\phi = 20\%$)]{\scalebox{0.28}{{\includegraphics{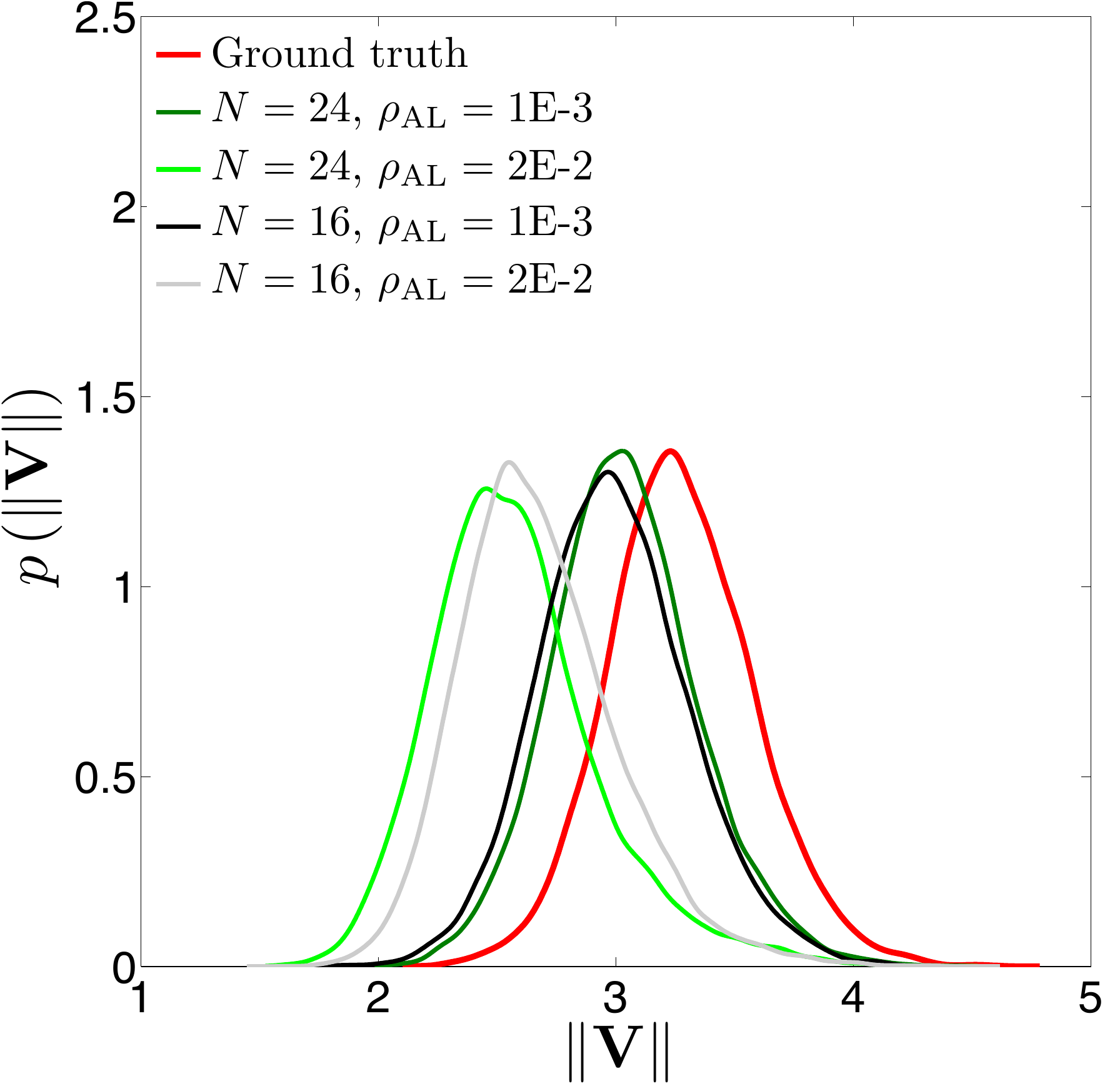}}}
      \label{f:VF20atR2}}
\setcounter{subfigure}{0}
\renewcommand*{\thesubfigure}{(a-3)} 
      \hspace{0cm}\subfigure[at $\frac{r - R_1}{R_2-R_1} = 0.8$ ($\phi = 20\%$)]{\scalebox{0.28}{{\includegraphics{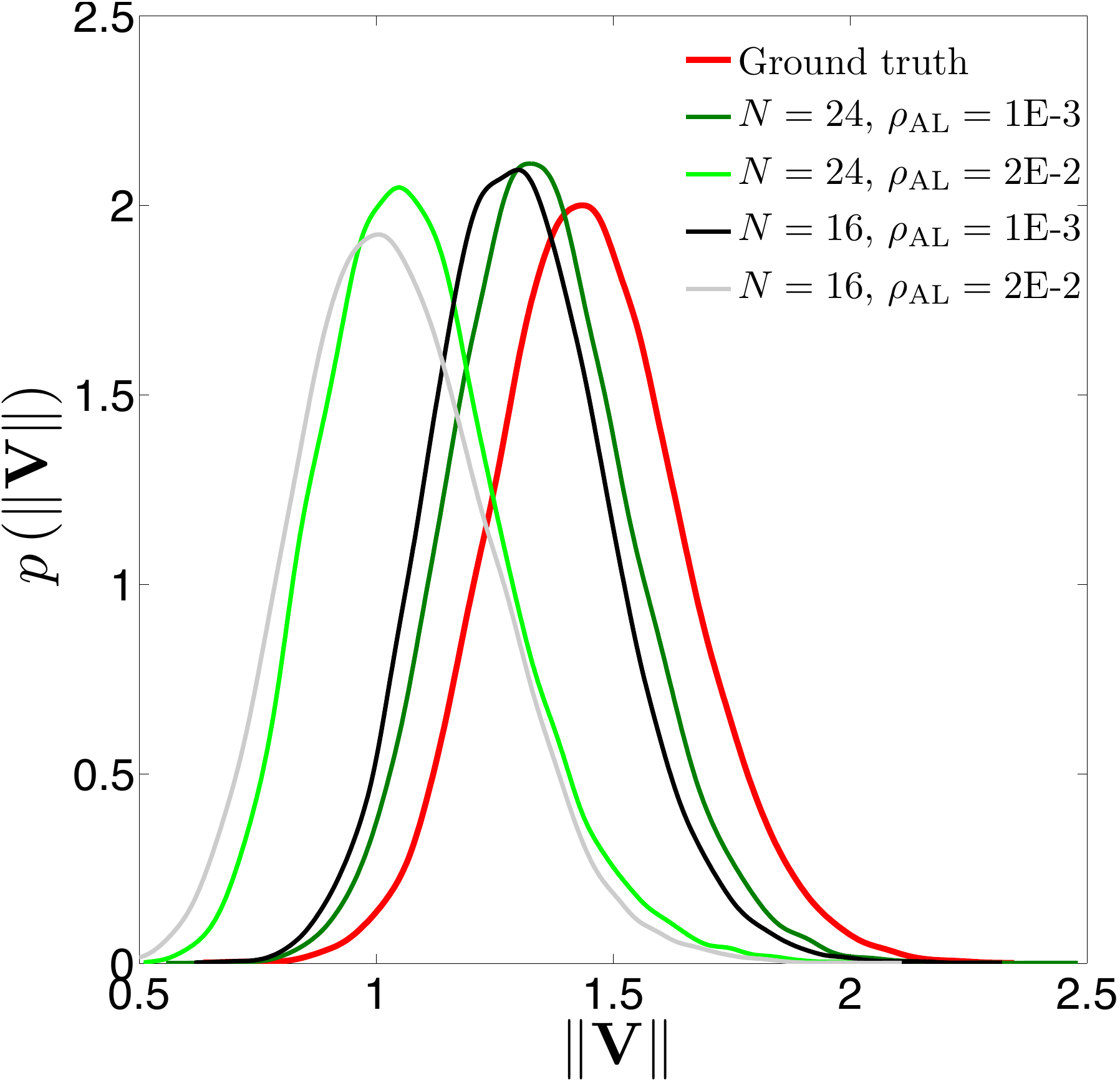}}}
      \label{f:VF20atR3}}      
\end{center}
\end{minipage}
 \begin{minipage}{\textwidth}
 \begin{center}
\setcounter{subfigure}{0}
\renewcommand*{\thesubfigure}{(a-1)} 
      \hspace{0cm}\subfigure[at $\frac{r - R_1}{R_2-R_1} = 0.2$ ($\phi = 40\%$)]{\scalebox{0.28}{{\includegraphics{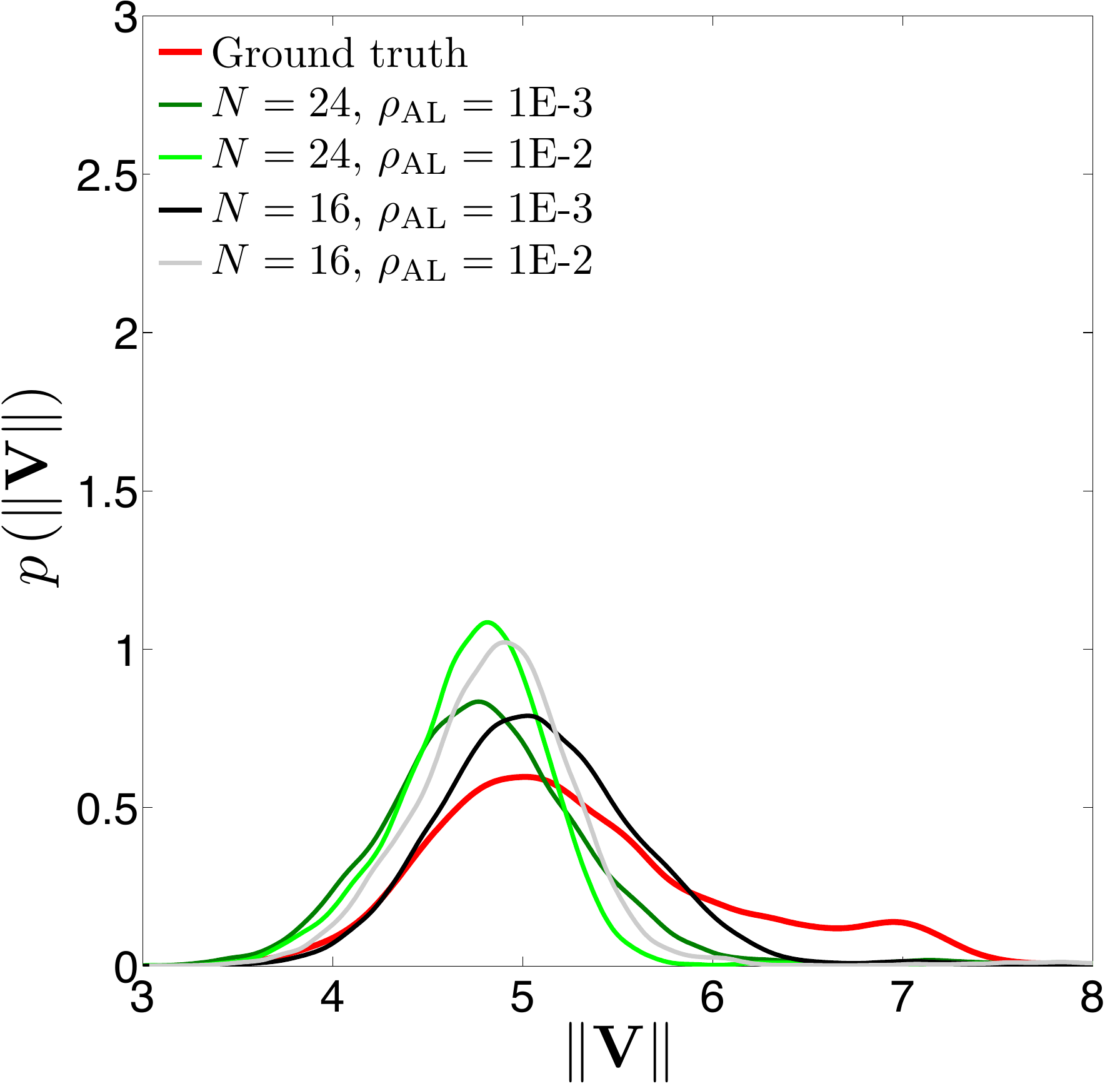}}}	
      \label{f:VF40atR1}}
\setcounter{subfigure}{0}
\renewcommand*{\thesubfigure}{(a-2)} 
      \hspace{0cm}\subfigure[at $\frac{r - R_1}{R_2-R_1} = 0.5$ ($\phi = 40\%$)]{\scalebox{0.28}{{\includegraphics{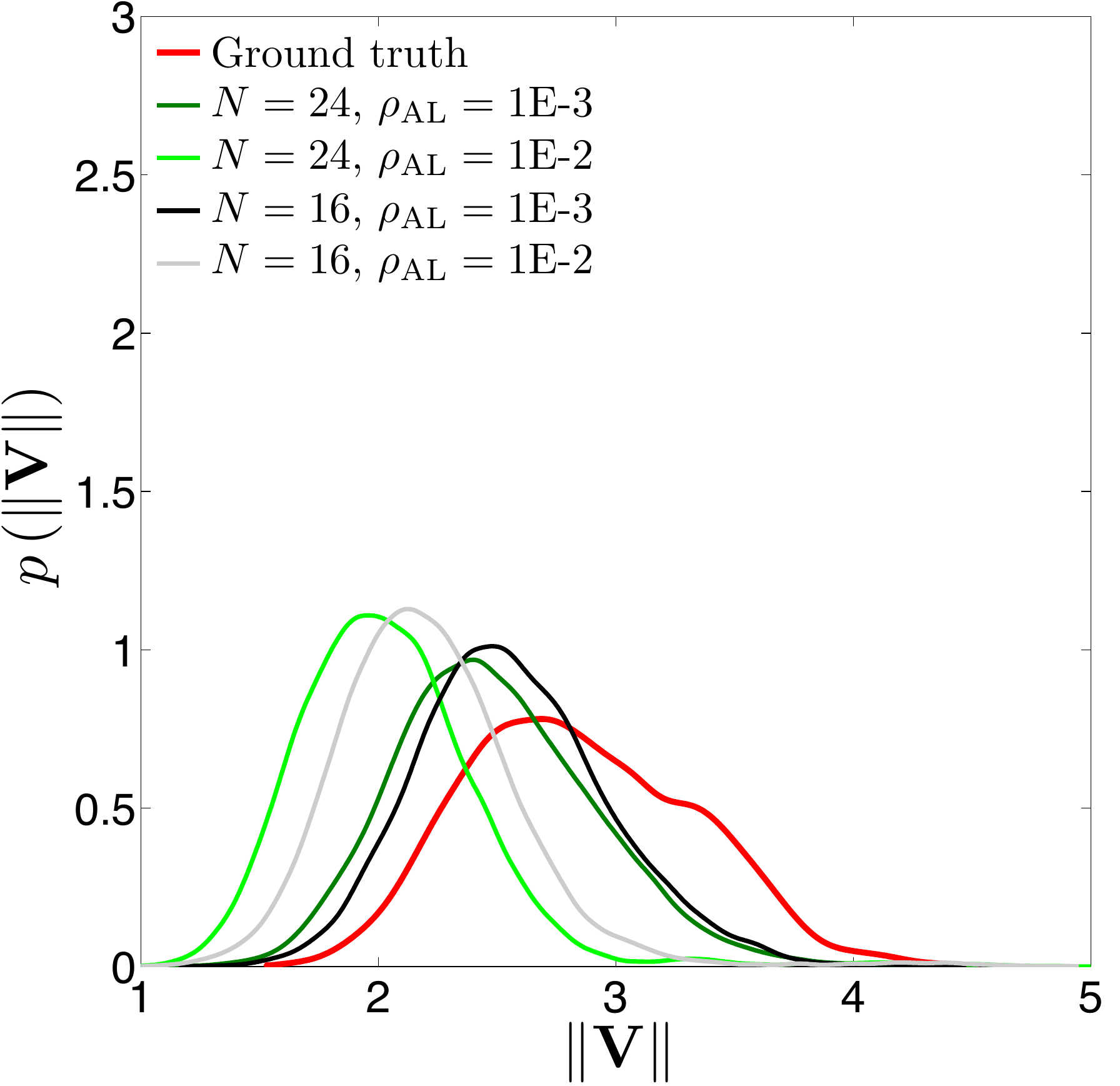}}}
      \label{f:VF40atR2}}
\setcounter{subfigure}{0}
\renewcommand*{\thesubfigure}{(a-3)} 
      \hspace{0cm}\subfigure[at $\frac{r - R_1}{R_2-R_1} = 0.8$ ($\phi = 40\%$)]{\scalebox{0.28}{{\includegraphics{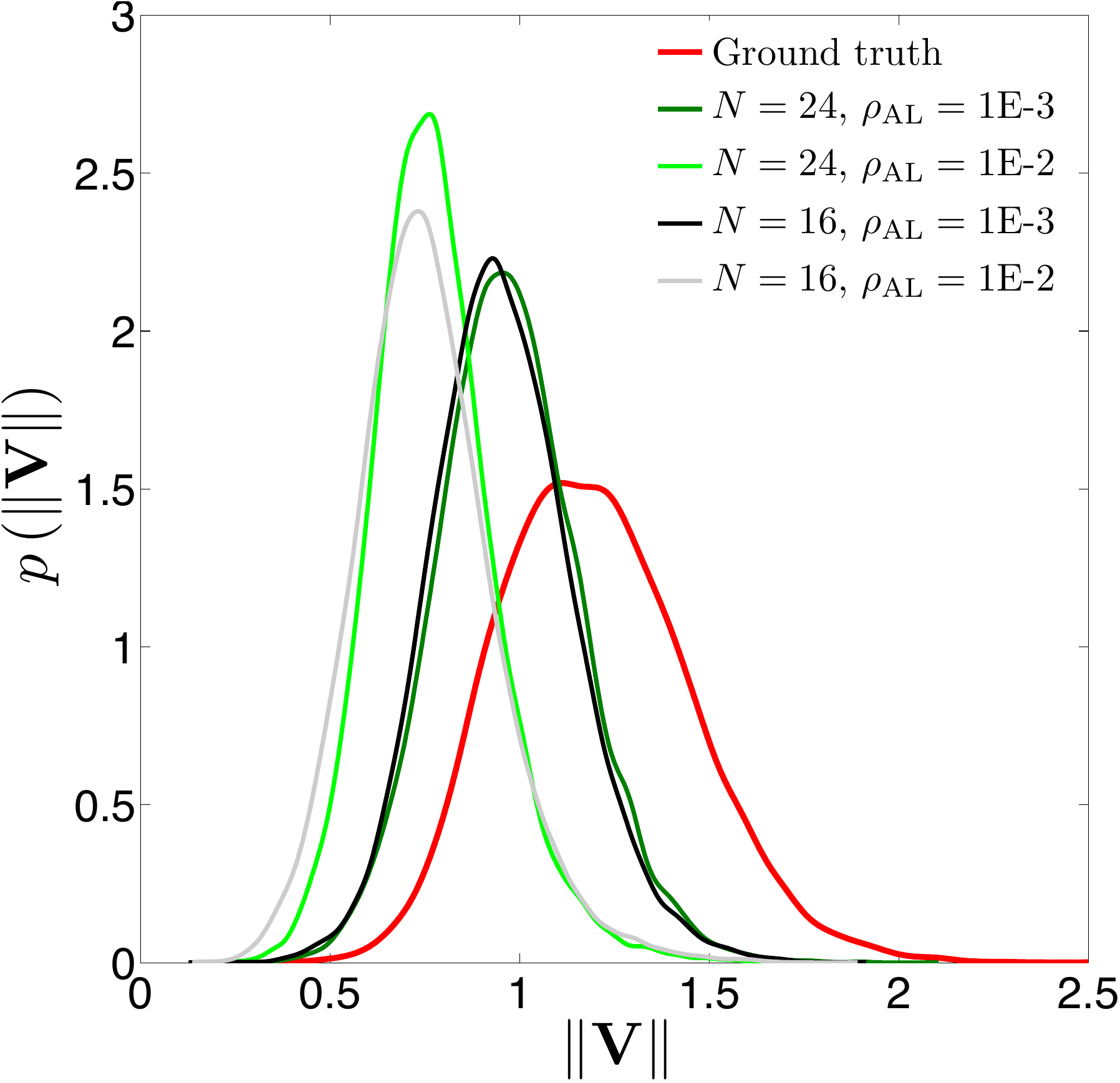}}}
      \label{f:VF40atR3}}      
\end{center}
\end{minipage}
\mcaption{Statistics of the velocity field of a {\bf Couette flow}. We
compute the probability distribution function of the velocity
magnitudes at points distributed equally in the azimuthal direction at
three different radii. The top plots correspond to the volume fraction
$\phi = 20\%$ and the bottom plots correspond to the volume fraction
$\phi = 40\%$.}{f:VelocityStatsCouette}
\end{figure}

\subsection{Microfluidic device}\label{s:dld}
\begin{figure}[H]
 \begin{minipage}{\textwidth}
 \begin{center}
\setcounter{subfigure}{0}
\renewcommand*{\thesubfigure}{(a)} 
      \hspace{0cm}\subfigure[$N = 8$ and $\rho_{\mathrm{AL}} = $ 1E-2]{\scalebox{0.45}{{\includegraphics{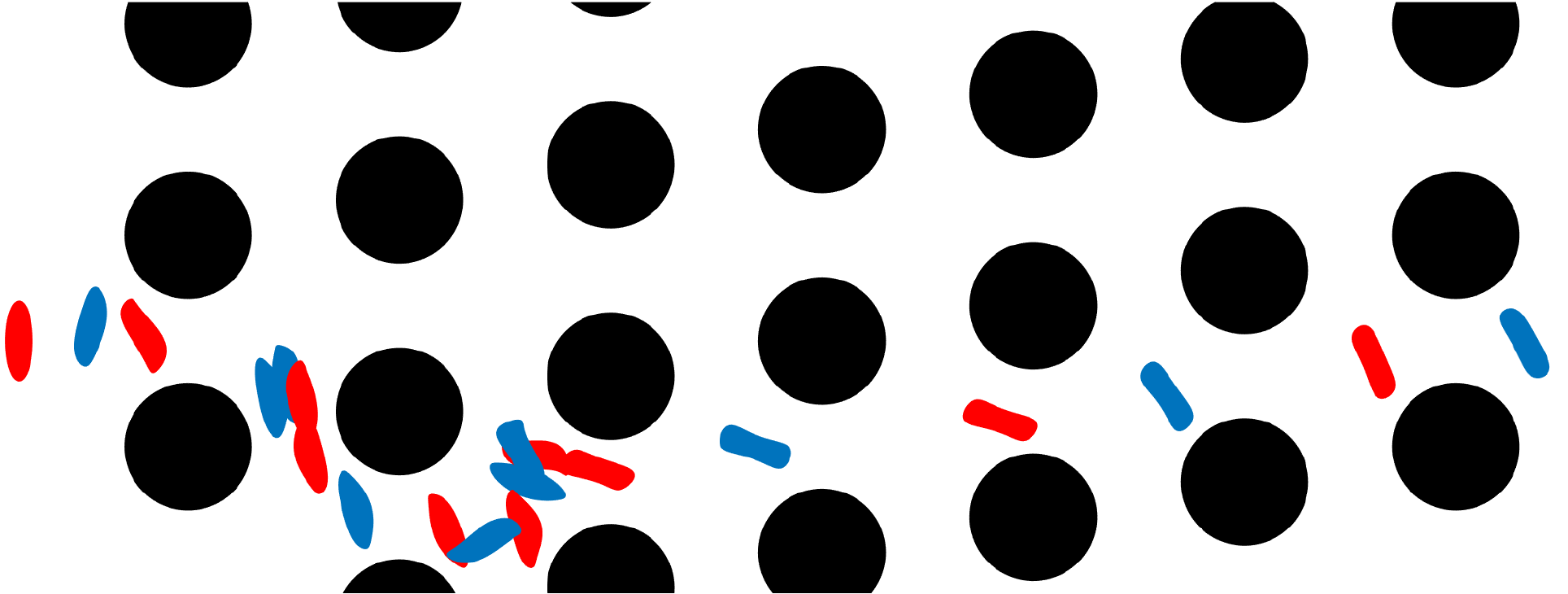}}}  
      \label{f:DLDN8}}    
\setcounter{subfigure}{0}
\renewcommand*{\thesubfigure}{(b)} 
      \hspace{0cm}\subfigure[$N = 16$ and $\rho_{\mathrm{AL}} = $ 1E-3]{\scalebox{0.45}{{\includegraphics{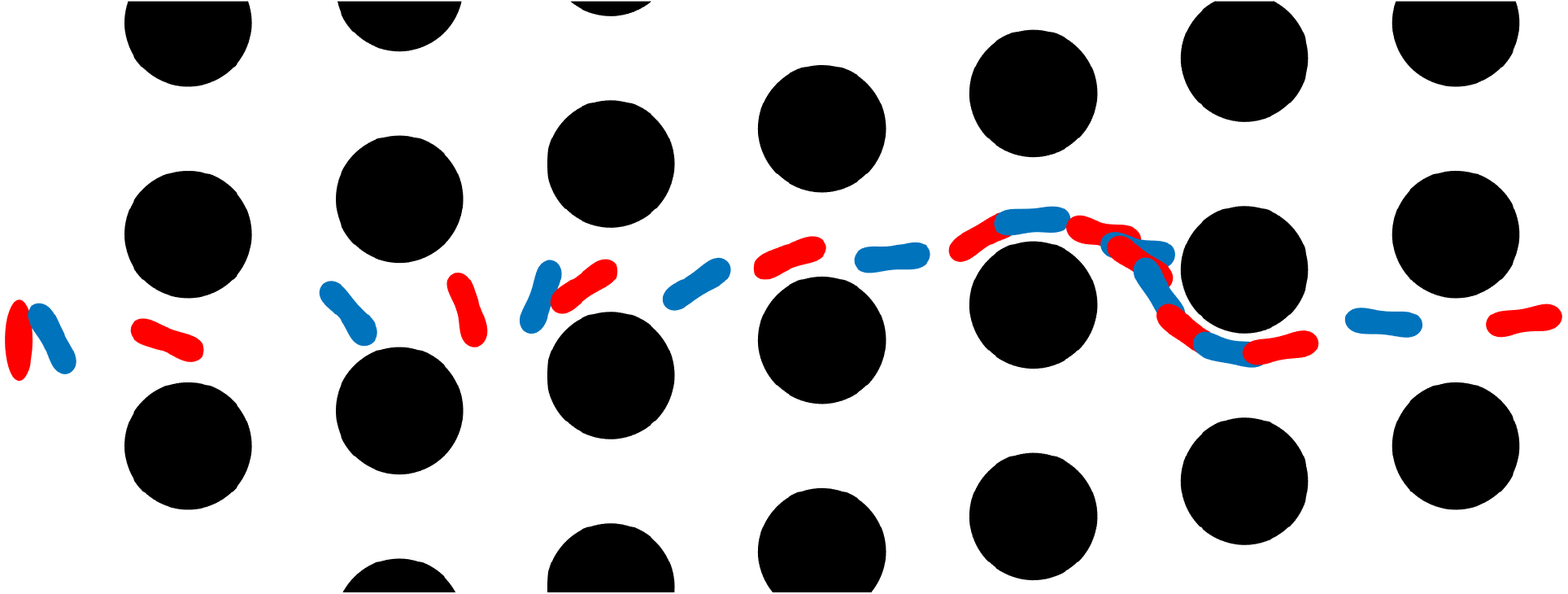}}}  
      \label{f:DLDN16}}    
\end{center}
\end{minipage}
 \begin{minipage}{\textwidth}
 \begin{center}
\setcounter{subfigure}{0}
\renewcommand*{\thesubfigure}{(c)} 
      \hspace{0cm}\subfigure[$N = 32$ and $\rho_{\mathrm{AL}} = $ 1E-4]{\scalebox{0.45}{{\includegraphics{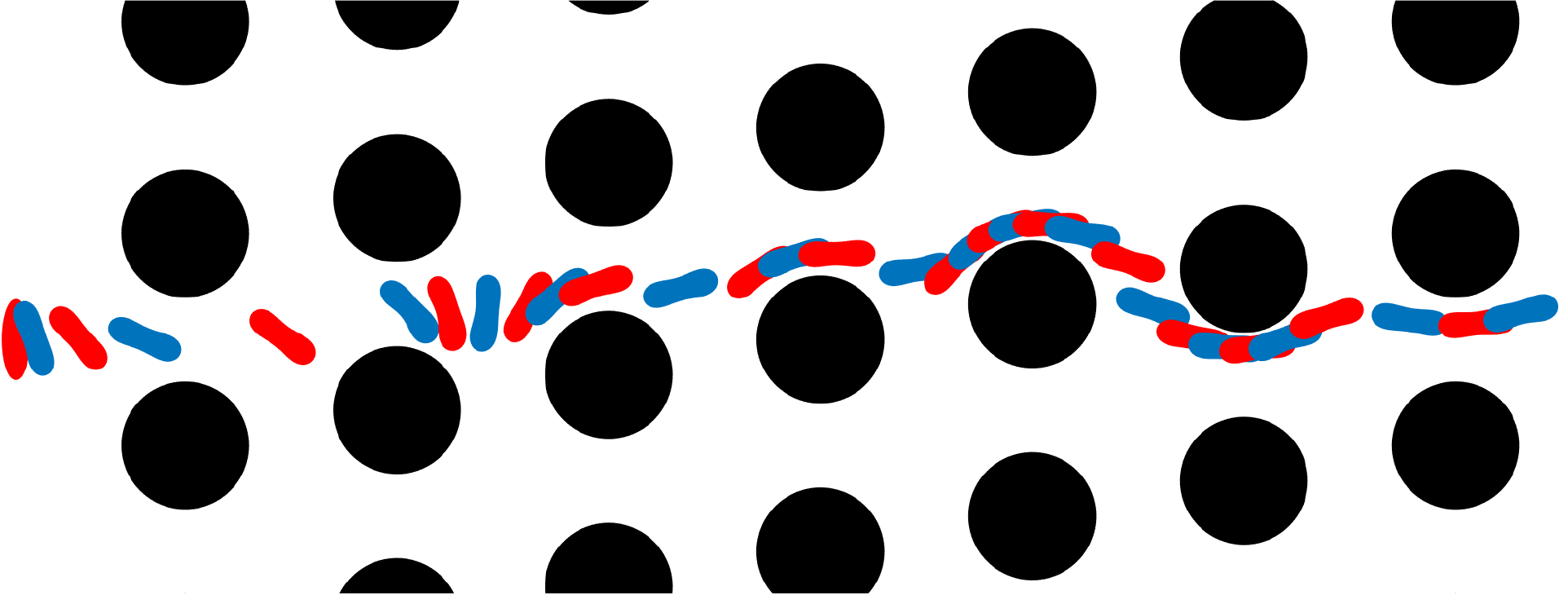}}}  
      \label{f:DLDN32}}    
\setcounter{subfigure}{0}
\renewcommand*{\thesubfigure}{(d)} 
      \hspace{0cm}\subfigure[$N = 64$ and $\rho_{\mathrm{AL}} = $ 1E-4]{\scalebox{0.45}{{\includegraphics{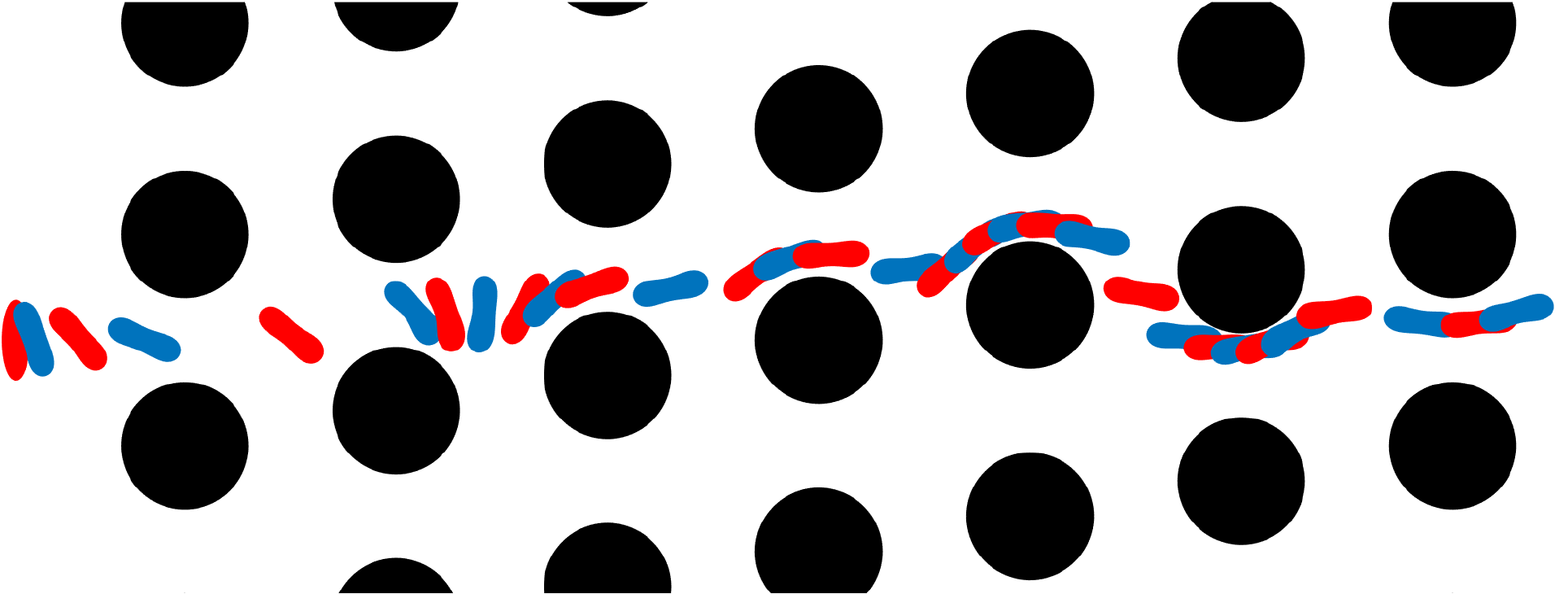}}}  
      \label{f:DLDN64}}    
\end{center}
\end{minipage}
\mcaption{Snapshots of zig-zagging RBCs from our low-resolution (the first three) and the ground truth (at the bottom) simulations of the microfluidic device. The regular alternation between blue and red RBCs represents sequential frames with variable time intervals. The device uses the technique called deterministic lateral displacement (DLD) to separate cells based on their deformability. Our DLD  device consists of arrays of circular pillars (shown in black) and 
  an exterior wall (not shown). The suspension flows from left to right (aligned with the horizontal axis). We impose a parabolic velocity at the intake and the outtake that causes a  healthy red blood cell to cross the inclined rows of pillars. This crossing is called \emph{``zig-zagging''} and has also been observed experimentally \cite{beech-tegenfeldt-e12}. 
In the ground truth simulation we can see that just before the last two columns of pillars, the cell
goes around the pillar and crosses rows, thus, it ``zig-zags''. If a cell does not zig-zag, we say that the cell displaces (laterally) along a row of pillars. }{f:dldEx}
\end{figure}

Deterministic lateral displacement (DLD) is a microfluidic technique to separate particles depending on their sizes and deformability without using any external force~\cite{huang-sturm-e04}. A DLD device consists of matrix of pillars, where the rows are arranged at an angle with the $x$-axis (horizontal) and the imposed velocity profile (or pressure difference) is aligned with the $x$-axis.  When a particle (e.g., rigid particles, vesicles, or red blood cells) enters the device it typically exhibits two modes of motion. Either it \emph{``displaces''} or it \emph{``zig-zags''}. These two terms are explained in~\figref{f:dldEx}. The basic idea is that if we want to separate particles, we design a DLD device in which one set of particles displaces and the other zig-zags. The experimental study \cite{beech-tegenfeldt-e12} shows that the technique can be used to separate red blood cells depending on their deformability. Follow up  numerical studies  \cite{quek-chiam-e11,ye-yu-e14,kruger-coveney-e14,zhang-fedosov-e15,vernekar-kruger15} systematically analyzed the separation of red blood cells using DLD and successfully reproduced the results of the experiments. Among these numerical studies~\cite{quek-chiam-e11,ye-yu-e14,zhang-fedosov-e15} are two-dimensional and~\cite{kruger-coveney-e14,vernekar-kruger15} are three-dimensional. Here, we want our 2D model to reproduce these numerical and experimental results using as coarse discretization as possible.

\paragraph*{\textbf{Setup}} The DLD device we consider here consists of circular pillars with a diameter of 15 $\mu m$ bounded by an exterior wall (not shown). We impose a Poiseuille flow as a velocity boundary condition at the intake and the outtake, and hence the velocity between two laterally adjacent pillars is parabolic. We consider a healthy red blood cell which has a reduced area of 0.65 and a viscosity contrast $\nu = 10$. The lengths of the long and short axes of the RBC are  $ 8\,\mu m$ and $3\,\mu m$. The inclination angle of the device is 0.17 rad and the center-to-center distance between the pillars is 25 $\mu m$. The setup of this DLD device (geometry and imposed velocity) are such so that the cell zig-zags (see \figref{f:dldEx}).

We discretize the exterior wall with $N_{\mathrm{wall}} = 3712$ points, the pillars with $N_{\mathrm{pillar}} = 64$ points. In our convergence study we do not change these resolutions. The repulsion length scale we use here is
$d_{\min} = 0.5h_{\max}$.  We start with $N = 8$ points per vesicle and $\rho_{\mathrm{AL}} = 1\mathrm{E}-2$. If the simulation can be completed within the allocated CPU time $T_{\mathrm{CPU}}$, we  perform a self-convergence test to determine the accuracy of the low-resolution solution. For this purpose we run another simulation of the example with a higher resolution $N = 16$ and $\rho_{\mathrm{AL}} = 1\mathrm{E}-3$. This \emph{``ground truth''} solution is performed using $N = 64$ points per vesicle and $\rho_{\mathrm{AL}} = 1\mathrm{E}-4$. For reference, the ground truth simulation requires 5.6 hours (on a single workstation) and our ground truth solution is in agreement with the experimental results reported in~\cite{beech-tegenfeldt-e12} (at the botton in~\figref{f:dldEx} we depict the trajectory of a cell using our ground-truth simulation).

\paragraph*{\textbf{Results}} We are interested in capturing the  true motion of the cells, i.e., displacement vs zig-zag and the correct point of zig-zagging so we can properly characterize the behavior of the device. We report a qualitative error metric (zig-zagging or not, and the pillar in which zig-zagging takes place). We also report two quantitative errors, one highly sensitive to the accuracy of the calculation and one less sensitive one. 
The first one (sensitive) is the error in the vesicle's center $\epsilon_{\mathrm{center}}$, specifically, its maximum over all time steps in \tabref{t:DLDResults}. The second error (less sensitive to numerical errors) is in the time it takes for the RBC to travel to the end of the device. We denote this error measure by ${\epsilon}_{\mathrm{T}}$. In terms of computational efficiency, we also report the number of accepted and rejected time steps, and the total CPU time. For both quantitative error metrics we report the ``self-error'' (as in a self-convergence study) without a ground truth using the superscript ``s'', and the error with respect the ground truth using the superscript ``g''. 

Our black-box solver took us to an accurate solution as follows: the simulation with $N = 8$ and $\rho_{\mathrm{AL}} =$ 1E-2 was completed within 3 hours. Then in order to estimate its accuracy we performed another simulation with $N = 16$ and $\rho_{\mathrm{AL}} =$ 1E-3, which took slightly longer than 2 hours. The self-error of the first simulation in the vesicle's center turned out to be ${\epsilon}_{\mathrm{center}}^{\mathrm{s}} = $ 2.8E+0, which is a close estimate of the error in the vesicle's center compared to the ground truth ${\epsilon}_{\mathrm{center}}^{\mathrm{g}} = $ 3.2E+0 and not acceptable. The self-error in the travel time is also large, i.e. $\epsilon_{\mathrm{T}}^{\mathrm{s}} =$ 2.5E-1. So another simulation with a higher resolution ($N = 24$, $\rho_{\mathrm{AL}} = $ 1E-4) was performed to measure the accuracy of the second simulation with $N = 16$ and $\rho_{\mathrm{AL}} $ = 1E-3. The self-error in the center still remains large but the self-error in the travel time decreases to $\bigO$(1E-2). Since this flow has several vesicle-wall interactions, the error in the center might be large at the low resolutions as in the Taylor-Green flow and the Couette apparatus examples. Therefore, if the quantity of interest is the travel time or the pillar in which zig-zagging takes place, $N =16$ and $\rho_{\mathrm{AL}} = $ 1E-3 seem to be sufficient for the accurate physics. We performed one more simulation with $N = 32$ and $\rho_{\mathrm{AL}} $ = 1E-4. This and the previous simulations had two times the CPU times of the first two runs. Additionally, the error in the vesicle's center or in the travel time did not improve further. So the self-convergence is achieved.
\begin{table}[H]
\mcaption{ The (self-) errors in the vesicle's center ${\epsilon}_{\mathrm{center}}$ and the vesicle's travel time to the end of the device ${\epsilon}_{\mathrm{T}}$ for the simulations of the {\bf microfluidic device} with the LRCA. The superscript ``$\mathrm{s}$'' indicates self-convergence errors (that is error with respect the next finer solution) and the superscript ``$\mathrm{g}$'' indicates errors with the ground truth. ``Accepts'' and ``Rejects'' refer to the steps accepted or rejected in the time marching algorithm.  The self-convergence errors are computed with respect to the simulation in one row below. Also reported are the number of accepted and rejected time steps and the CPU time. The ground truth simulation takes  5.6 hours with $N = 64$ and $\rho_{\mathrm{AL}} = $ 1E-4.}{t:DLDResults}
\centering
\begin{tabular}{c c| c c c c c c c }

$N$ & ${\rho}_{\mathrm{AL}}$ & ${\epsilon}_{\mathrm{center}}^{\mathrm{g}}$ & ${\epsilon}_{\mathrm{center}}^{\mathrm{s}}$ & ${\epsilon}_{\mathrm{T}}^{\mathrm{g}}$ & ${\epsilon}_{\mathrm{T}}^{\mathrm{s}}$ & Accepts & Rejects & Time (hours)\\ 
 \hline
 
  8 & 1E-2 & 3.2E+0 & 2.8E+0 & 3.3E-1 & 2.5E-1 & 506 & 27 & 2.92 \\

 16 & 1E-3 & 1.0E+0 & 1.1E+0 & 1.0E-1 & 1.7E-2 & 396 & 20 & 2.06 \\

 24 & 1E-4 & 6.7E-1 & 2.4E-1 & 8.5E-2 & 3.1E-2 & 1227 & 40 & 5.22 \\

 32 & 1E-4 & 4.5E-1 &        & 5.6E-2 &        & 1228 & 40 & 5.25 \\

 \hline

 8 & 1E-2 & 3.2E+0 & 2.5E+0 & 3.3E-1 & 1.3E-1 & 506 & 27 & 2.92 \\ 

 8 & 1E-4 & 3.3E+0 &        & 4.1E-1 &        & 2429 & 75 & 10.25 \\

 \hline 
 16 & 1E-3 & 1.0E+0 & 3.5E-1 & 1.0E-1 & 4.4E-2 & 396 & 20 & 2.06 \\

 16 & 1E-4 & 1.1E+0 &        & 1.4E-1 &        & 1195 & 45 & 5.14 \\
 
\end{tabular} 
\end{table} 

In conclusion, the scheme correctly identifies the necessary resolution to resolve the quantities of interest. In this example $N=16$ is sufficient to capture the correct zig-zagging behavior. All the simulations exhibit zig-zagging but the $N=8$ case is completely off (see the top figure in~\figref{f:dldEx}). As we discuss, the simulation was run without changing any parameters, other than $N$ and $\rho_\mathrm{AL}$.

\section{Conclusions\label{s:conclusions}}
We have addressed issues with simulations of vesicle suspensions at
low discretization resolutions. We have developed a robust method by
introducing new schemes and implementing some standard techniques.  An
efficient scheme to determine an upsampling rate is used for computing
the nonlinear terms without introducing spurious oscillations. A
surface reparametrization algorithm smooths out vesicles' boundaries
by penalizing their high-frequency components.  The area and arc-
length of the vesicles are corrected at each time step to allow for
long-time scale simulations without changing the governing equations.
A new reliable adaptive time-stepping scheme that works for all
resolutions is used to choose the optimal time step size.  Finally, a
repulsion force between vesicles eliminates any chance of an non-
physical collision.  All these algorithms require certain parameters,
and these were set heuristically. So that our solver can be used as a black-box. We show the capabilities of the solver in a real-world
example of a microfluidic cell sorting technique which is studied
experimentally and numerically. The solver leads to a solution with an accurate
physics. 

We have discussed separate error measures for dilute and dense 
suspensions, and performed a systematic error
analysis to investigate the accuracy of our low-resolution
simulations.  The low-resolution correction algorithms we have
presented are essential for stable simulations and dropping one of them results in failure. 
Furthermore, by using
these algorithms we are able to accurately capture the statistics of
the underlying flow accurately with a coarse discretization.  One of
the most impressive examples is the Couette flow.  Its low-resolution
simulation, which takes less than a week, estimates accurately the
upscaled quantities such as effective viscosity and statistics computed
by the high-fidelity simulation, which takes more than a month.

\section*{Acknowledgements} \label{s:acknow}
This material is based upon work supported by AFOSR grants
FA9550-12-10484; by NSF grant CCF-1337393; by the U.S.~Department of
Energy, Office of Science, Office of Advanced Scientific Computing
Research, Applied Mathematics program under Award Numbers DE-SC0010518
and DE-SC0009286; by NIH grant 10042242; by DARPA grant
W911NF-115-2-0121; and by the Technische Universit\"{a}t
M\"{u}nchen---Institute for Advanced Study, funded by the German
Excellence Initiative (and the European Union Seventh Framework
Programme under grant agreement 291763).  Any opinions, findings, and
conclusions or recommendations expressed herein are those of the
authors and do not necessarily reflect the views of the AFOSR, DOE,
NIH, DARPA, and NSF. Computing time on the Texas Advanced Computing
Centers Stampede system was provided by an allocation from TACC and the
NSF.


\bibliographystyle{plainnat} 
\bibliography{refs}
\biboptions{sort&compress}
\end{document}